\documentclass{aa}  
\usepackage{natbib}
\usepackage{graphicx}
\usepackage{txfonts}
\usepackage{xcolor}
\usepackage{pdflscape}
\usepackage{multirow}
\usepackage{booktabs}
\usepackage{ulem}
\def\civ{{\sc{Civ}}$\lambda$1549\/}
\def\civonly{{C\sc{iv}}\/}
\def\civbc{{\sc{Civ}}$_{\rm BC}$\/}
\def\aliii{{Al\sc{iii}}$\lambda$1860\/}
\def\aliiionly{{Al\sc{iii}}\/}
\def\siiii{{Si\sc{iii}]}$\lambda$1892\/}
\def\siiiionly{{Si\sc{iii}]}\/}
\def\ciii{{C\sc{iii}]}$\lambda$1909\/}
\def\ciiionly{{C\sc{iii}]}\/}
\def\siiv{{S\sc{iv}}$\lambda$1398\/}

\def\siii{{Si\sc{ii}}$\lambda$1816\/}
\def\niii{{N\sc{iii}]}$\lambda$1750\/}
\def\heii{{He\sc{ii}}$\lambda$1640\/}

\def\niv{{N\sc{iv}]}$\lambda$1486\/}

\def\siiv{Si{\sc iv}$\lambda$1397\/}

\def\oiv{O\sc{iv}$]\lambda$1402\/}

\def\cii{{C\sc{ii}}$\lambda$1335\/}
\def\oi{{O\sc{i}}$\lambda$1304\/}
\def\siiia{{Si\sc{ii}}$\lambda$1306\/}

\def\hb{{\sc{H}}$\beta$\ }
\def\mgii{{Mg\sc{ii}}$\lambda$2800\/}

\def\oiiiopt1{{\sc{[Oiii]}}$\lambda$5007\/}
\def\oiiiuv{{O\sc{iii]}}$\lambda$1663\/}
\def\alii{{Al\sc{ii}}$\lambda$1670\/}
\def\heiiuv{He{\sc{ii}}$\lambda$1640\/}
\def\oiv{{\sc{Oiv]}}$\lambda$1402\/}
\def\lledd{$L/L_{\rm Edd}$}
\def\hbbc{{\sc{H}}$\beta_{\rm BC}$\/}
\def\lya{{Ly}$\alpha$\ }
\def\nv{{N\sc{v}}$\lambda$1240}
\def\kms{\,km\,s$^{-1}$}
\def\feii{{Fe{\sc ii}}\/}
\def\rk{$R_\mathrm{K}$}

\def\feiii{Fe{\sc iii}\/}
\def\rfe{$R_{\rm FeII}$}

\def\cmp{$c(\frac{1}{2})$}

\def\mbh{${M_\mathrm{BH}}$\/}
\def\lledd{$L/L_\mathrm{Edd}$\/}
\def\ledd{$L_\mathrm{Edd}$\/}
\def\rfe{$R_{\rm FeII}$}

\def\hb{{\sc{H}}$\beta$\/}
\def\uflux{erg s$^{-1}$ cm$^{-2}$\/}
\def\nh{n$\mathrm{_H}$}
\def\cmc{cm$^{-3}$}
\def\mo{M$_{\odot}$}
\def\mdot{$\dot{M}$}
\def\ergs{erg\,s$^{-1}$}

\begin{document} 
\tabcolsep=1pt  
\citeindextrue
\title{Extreme quasars at high redshift}
\subtitle{}
\author{M. L. Mart\'{\i}nez-Aldama\inst{1} \and A. del Olmo\inst{1} \and P. Marziani\inst{2} \and J. W. Sulentic\inst{1}  \and C.A. Negrete\inst{3} \and D. Dultzin\inst{4} \and M. D'Onofrio\inst{5} \and J. Perea\inst{1}}
\institute{
{Instituto de Astrofis\'{\i}ca de Andaluc\'{\i}a, IAA-CSIC, Glorieta  de la Astronom\'{\i}a s/n, E-18008 Granada, Spain}
\and{Istituto Nazionale d'Astrofisica (INAF), Osservatorio Astronomico di Padova, IT 35122, Padova, Italy}
\and {CONACYT Research Fellow, Instituto de Astronom\'{\i}a, UNAM, M\'exico D.F. 04510, M\'exico} \and {Instituto de Astronom\'{\i}a, UNAM, M\'exico D.F. 04510, M\'exico}
\and {Dipartimento di Fisica \& Astronomia ``Galileo Galilei'', Universit\`a di Padova, Padova,  Italy}
}

\abstract
{Quasars radiating at extreme Eddington ratios (hereafter xA quasars) are likely a prime mover of galactic evolution and have been hailed as potential distance indicators. Their properties are still scarcely known.} 
{We test the effectiveness of the selection criteria defined on the ``4D Eigenvector 1'' (4DE1) for identifying xA sources.  We provide a quantitative description of their rest-frame UV spectra (1300 -- 2200 \AA) in the redshift range $2 \lesssim z \lesssim 2.9$, with a focus on major emission features. }
{Nineteen extreme quasar candidates were identified using 4DE1 selection criteria applied to SDSS spectra: \aliii/\siiii\ $ \gtrsim  0.5$ and \ciii/\siiii\ $\lesssim 1$. The emission line spectra  was studied using  multicomponent fits of deep spectroscopic observations (S/N$\gtrsim 40-50$; spectral resolution $\approx$ 250 \kms) obtained with the OSIRIS at Gran Telescopio Canarias (GTC). }
{GTC spectra confirm that almost all of these quasars are xA sources with very similar properties. We provide spectrophotometric and line profile measurements for the \siiv+\oiv, \civ+\heiiuv, and the 1900\AA\ blend. This last feature is found to be predominantly composed of \aliii, \siiii\ and \feiii\  emission features, with weak \ciii. The spectra can be  characterized as very low ionization (ionization parameter, $\log U\sim -3$), a condition that explains the significant \feiii\ emission observed in the spectra. xA quasars show extreme properties in terms of \civ\ equivalent width and blueshift amplitudes. \civ\ shows low equivalent width, with a median value of 15\AA\ ($\lesssim$ 30\,\AA\ for the most sources), and high or extreme blueshift amplitudes (--5000\,$\lesssim$\,\cmp\,$\lesssim$\,--1000\,\kms). Weak-lined quasars appear as extreme xA quasars and not as an independent class. The \civ\  high amplitude blueshifts coexists in all cases save one with symmetric and narrower \aliii\ and \siiii\  profiles.  
Estimates of the Eddington ratio using the \aliii\ FWHM as a virial broadening estimator are consistent with the ones of a previous xA sample.}
{{xA quasars show distinguishing properties that make them easily identifiable in large surveys and potential ``standard candles" for cosmological applications. It is now feasible to assemble large samples of xA quasars from the latest data releases of the SDSS. We provide evidence that \aliii\ could be associated with a low-ionization virialized sub-system, supporting previous suggestions that \aliiionly\ is a reliable virial broadening estimator.}}

\keywords{quasars: general -- quasars: emission lines -- quasars: supermassive black holes \\ \\ \small{ Manuscript accepted for publication in A\&A, July 2018 } }

\maketitle 

%
\defcitealias{marzianietal09}{M09}
\defcitealias{marzianisulentic14}{MS14}
\defcitealias{sulenticetal14}{S14}
\defcitealias{sulenticetal17}{S17}
\defcitealias{sulenticetal07}{S07}
\defcitealias{negreteetal12}{N12}

\section{Introduction}

Quasars are found over an enormous range of distances (z$\sim$0 -- 7.5) in the Universe. For this reason they have occasionally been cited as the ultimate possible standard candles for use in cosmology  { (See for example the Chapter by in \citet{donofrioburigana09} and the recent reviews by \citet{sulenticetal14a} and \citet{czernyetal18})}. The problem with such a use has been the lack of a clear definition of ``quasar" and a contextualization of their diversity. Since 2000 a clearer idea of their nature and diversity has emerged using the 4D Eigenvector 1 (4DE1) formalism. We are now able to identify a quasar main sequence  { \citep[See][for a recent review]{marzianietal18}}, and recognize an extreme accretor (xA) quasar population at the end of this sequence {\citep[e.g.,][hereafter \citetalias{marzianisulentic14}]{marzianisulentic14}. } This xA population radiating at \lledd$\sim$1 offer the best opportunity to use quasars for cosmology \citep[\citetalias{marzianisulentic14}, ][]{wangetal14c}. This paper describes a search for extreme quasars at $ z\sim$2.3 built upon an extension of low--redshift 4DE1 studies.  

Quasar spectra show diverse properties  in measures of line intensity ratios and line profiles. These measures offer multifold diagnostics of emitting region structure and physical conditions \citep{marzianietal18}.  Organizing the diversity of quasar properties  has been an ongoing effort for many years. Perhaps the first successful attempt  was carried out by \citet{borosongreen92}. They proposed an Eigenvector 1 scheme based on a principal component analysis of the Palomar-Green quasar sample \citep[See ][for reviews up to the late 1990s]{gaskelletal99,sulenticetal00a}. Trends between measures of [\ion{O}{iii}]$\lambda\lambda$4959,5007, optical \feii\ emission  and full width at half maximum (FWHM) \hb\ were found, and appreciation of their importance has grown with time \citep[][and references therein]{sulenticmarziani15}. \citet{sulenticetal00c} expanded upon this work and defined a new scheme called 4D Eigenvector 1 (4DE1)  with the addition of two new parameters. Principal 4DE1 measures for low-$z$\ quasars involve: 1) FWHM of broad line  \hb\ (excluding any narrow emission component)\footnote{In the following we understand for broad profile of a line the total broad profile excluding the narrow component}; 2) the strength of the optical \feii\ blend at 4570 \AA\ normalized by the intensity of \hb: \rfe=I(\feii)/I(\hb); 3) the velocity shift at half maximum (\cmp) of the high-ionization line (HIL) \civ\  profile relative to a rest-frame (usually defined by measures of the [\ion{O}{iii}]$\lambda$5007 and/or narrow \hb\ centroid) and  4) the soft X-ray photon index ($\Gamma_{\rm soft}$).

\begin{table*}[h!]
\setlength{\tabcolsep}{5pt} 
\small
 
\begin{center}
\caption{Source identification and basic properties of the GTC-xA quasars. \label{tab:properties}}
\begin{tabular}{c c c c c c c c c c c c c c c c}\\ 
\hline\hline\noalign{\vskip 0.1cm}         
SDSS identification &  \multicolumn{1}{c}{\it{z}} & $\Delta z$ & Line & $m_{v}$ & $M \mathrm{_B}$ & $g-r$ & Comments\\
(1) & (2) & (3) & (4) & (5) & (6) & (7) & (8) \\
\hline
\noalign{\vskip 0.07cm}
SDSSJ000807.27-103942.7	 &	2.4660	&	0.0010	&	{\sc	iii}	&	19.15	&	-26.2	&	0.08	&  \\
SDSSJ004241.95+002213.9&	2.0560	&	0.0053	&	{\sc	iii}	&	19.05	&	-25.6	&	0.24	&  \\
SDSSJ021606.41+011509.5	&	2.2236	&	0.0008	&	{\sc	i,ii}	&	19.36	&	-25.2	&	0.49	& BAL \\
SDSSJ024154.42-004757.5	&	2.3919	&	0.0015	&	{\sc	iii}	&	19.24	&	-26.0	&	0.24	&  \\
SDSSJ084036.16+235524.7	&	2.1879	&	0.0012	&	{\sc	iii}	&	19.46	&	-25.3	&	0.40	& \\
SDSSJ101822.96+203558.6	&	2.2502	&	0.0035	&	{\sc	iii}	&	19.27	&	-25.5	&	0.33	&  \\
SDSSJ103527.40+445435.6	&	2.2639	&	0.0060	&	{\sc	iii}	&	19.34	&	-25.6	&	0.20	& \\
SDSSJ105806.16+600826.9	&	2.9406	&	0.0001	&	{\sc	i,ii}	&	19.29	&	-26.7	&	0.16	&  \\
SDSSJ110022.53+484012.6	&	2.0884	&	0.0028	&	{\sc	iii}	&	18.90	&	-25.9	&	0.13	&  \\
SDSSJ125659.79-033813.8 &	2.9801	&	0.0004	&	{\sc	ii}	&	19.27	&	-26.7	&	0.26	& BAL\\
SDSSJ131132.92+052751.2	&	2.1234	&	0.0009	&	{\sc	iii}	&	19.06	&	-25.4	&	0.72	& BAL\\
SDSSJ143525.31+400112.2	&	2.2615	&	0.0006	&	{\sc	iii}	&	18.30  	&	-26.6	&	0.17	& \\
SDSSJ144412.37+582636.9	&	2.3455	&	0.0018	&	{\sc	iii}	&	19.21	&	-25.8	&	0.37	& Mini--BAL\\
SDSSJ151258.36+352533.2	&	2.2382	&	0.0012	&	{\sc	iii}	&	19.21	&	-25.8	&	0.04	& RL, $\log P_{\nu}  \approx 33.8$, $\log$\rk$\approx$2.65\\
SDSSJ214009.01-064403.9	&	2.0808	&	0.0038	&	{\sc	iii}	&	19.26	&	-25.3	&	0.56	& BAL\\
SDSSJ220119.62-083911.6	&	2.1840	&	0.0015	&	{\sc	iii}	&	18.68	&	-26.1	&	0.35	& BAL\\
SDSSJ222753.07-092951.7	&	2.1639	&	0.0010	&	{\sc	iii}	&	19.11	&	-25.6	&	0.10	& \\
SDSSJ233132.83+010620.9	&	2.6271	&	0.0038	&	{\sc	iii}	&	19.19	&	-26.1	&	0.33	& RL, $\log  P_{\nu}\approx 34.1$, $\log$\rk$\approx$2.93\\
SDSSJ234657.25+145736.0	&	2.1682	&	0.0007	&	{\sc	iii}	&	19.11	&	-25.7	&	0.29	& RI, $\log  P_{\nu}\approx 32.7$, $\log$\rk$\approx$1.67\\
\hline	
\end{tabular}
\end{center} 
\small {\sc Notes.} Columns are as follows: (1) SDSS coordinate name. (2) Redshift. (3) Redshift uncertainty. (4) Lines used for redshift determination:  {\sc i}: \cii, {\sc ii}: OI+SiII $\lambda$1305, {\sc iii}: \aliii+\siiii. (5) Apparent Johnson $V$\ magnitude as reported by \citet{veroncettyveron10}. (6) Absolute $B$\ magnitude according to \citet{veroncettyveron10}. (7) $g-r$ color in magnitudes. (8) Quasar classification: BAL QSO: Broad Absorption Line Quasar; Mini--BAL: Mini Broad Absorption Line Quasar, radio-loud (RL) and radio-intermediate (RI). $P_{\nu}$ is power per unit frequency at 1.4 GHz in units of erg s$^{-1}$ Hz$^{-1}$.  \\
\end{table*}

The 4DE1 optical plane (OP), defined by the measures of the FWHM of  \hb\ and \rfe,  shows a reasonably well-defined sequence  \citep[a quasar ``main sequence'', MS, ][]{sulenticetal00a,marzianietal01}. The nomenclature is motivated by an analogy  with the role of the stellar main sequence in the H-R diagram which connects observational measures to physical properties \citep{sulenticetal08}.  The stellar MS is driven by stellar mass, while the quasar sequence is thought to be driven by Eddington ratio \citep[$\propto$\mdot,][]{marzianietal01,marzianietal03b}. \citet{sulenticetal00a} noted  a change in all 4DE1 measures near FWHM \hb\ = 4000\kms\ in low-$z$ quasar samples  (L\,$\lesssim 10^{47} $ erg s$^{-1}$). This change motivated an empirical designation of two quasar populations: Population A with FWHM \hb\,$\le 4000$\kms, \rfe\,$>$\,0.5, frequent \civ\ profile blueshifts and sources with a soft X-ray excess, and Population B with FWHM \hb\, $>$ 4000\kms, \rfe\,$<$\,0.5, absence of \civ\ blueshift and little or no soft X-ray excess \citep{benschetal15}. Radio-loud (RL) quasars are strongly concentrated in the Population B domain along with 30\% of radio quiet (RQ) sources \citep[e.g.,][and references therein]{sulenticetal03,zamfiretal08}.  

Low-ionization emission lines (LILs - \hb\ best studied {feature}) in quasars frequently show asymmetric profiles. The H$\beta$\ broad component (\hbbc) in Population B sources is usually well described  by a double Gaussian profile with one of the components centered on  the rest-frame and the second one redshifted by  $\approx$ 1000 -- 3000\kms\, \citep{zamfiretal10}, the very broad component (VBC). Population A sources rarely show the redshifted component. High-ionization line (\civ\  best studied prototype) profiles usually show blueward shifts/asymmetries in Population A sources. Pop. B objects also show weak or moderate  strength LILs like \feii\ and the \ion{Ca}{\sc ii} IR triplet \citep{sulenticetal06,martinez-aldamaetal15}. Usually Population B quasars do not  show any strong soft X--ray excess \citep[][and references therein]{sulenticetal00a,benschetal15}. Largely radio-quiet (RQ) population A sources  usually show symmetric profiles well--modeled with a Lorentz function \citep{sulenticetal02,zamfiretal10,craccoetal16}. Population A sources with the narrowest \hb\ profiles ($<$2000 km/s)  are often called narrow line Seyfert 1 sources (NLSy1), but in no sense represent a distinct class of quasars \citep{zamfiretal08,sulenticetal15}.

All of the above description involves quasars with z$<$1.0 where moderate to high S/N ground-based spectra exist for significant numbers of sources \hb\ (ground based) and \civ\ (HST FOS archival data). Within each quasar population systematic trends are revealed  in composite spectra of \hb\ \citep{ sulenticetal02, zamfiretal10} and \civonly\, \citep{bachevetal04}. \citet{sulenticetal02} defined Pop. A subclasses A1, A2, A3 and A4 in order of increasing intervals of 0.5\rfe\  have been defined.  Extreme Population A sources in bins A3 and A4 (xA) involve quasars with very strong  \rfe$\gtrsim$1.0. This criterion  is applicable only  to sources at low redshift. This study involves searching for higher redshift analogues of such extreme quasars. At high redshift ($z>2$) the spectral range used to define sources in the 4DE1 (FWHM \hb, \rfe)  context are lost unless NIR spectra are available for \hb\ \citep{marzianietal09}. Many \civonly\, spectra in the intermediate/high-$z$\ range are available in the SDSS/BOSS archives although low S/N often precludes detailed analysis. 

The 4DE1 provides a consistent picture of quasar observational properties in low-$z$ samples: beginning with  the low-\rfe\ and broad FWHM \hb, involving high black hole mass  ``disk-dominated'' quasars. As we move along the sequence  we encounter sources whose spectra show narrower LIL profiles, lower ionization spectra, and blueshifted \civ\ profiles providing  evidence of strong HIL emitting outflows: ``wind-dominated'' quasars  \citep{richardsetal11}.   Eddington ratio (convolved with the effects of orientation) appears to be the physical parameter driving  systematic changes of observational properties along the quasar MS \citep{sulenticetal00a,marzianietal01,boroson02}. The main sequence observational trends can be interpreted as driven by  small \mbh, higher \lledd\ (young? high accretors) towards the high \rfe\ end of the sequence . The  xA sources cluster around \lledd $\approx 1$\footnote{The precise \lledd\ values depend  on the black hole mass \mbh\ scaling law and on the bolometric corrections. Following the 4DE1 based assumptions described above, the highest values along the MS are \lledd $\approx 1 - 2$}.   This would imply that the low-$z$\ xA sources are the ``youngest'' (less massive than Pop. B) quasar population radiating at the highest Eddington ratios \citep{fraixburnetetal17}. 

There is  a growing consensus that sources at the high \rfe\ end of the MS are accreting at the highest rates, and are expected to be close to the radiative limit per unit black hole mass \citep{sunshen15,duetal16,duetal16a,sniegowskaetal17}.  If this is the case, xA quasars acquire a special meaning. While Population A HILs are dominated by blue shifted emission associated with outflows, the presence of almost symmetric and unshifted LILs (Balmer lines, but also Paschen lines, \citealt{lafrancaetal14}) indicate the coexistence of a LIL emitting region that is virialized.  Since \lledd\ tends toward a constant limiting value \citep{mineshigeetal00},  xA quasars can be considered as ``Eddington standard candles.'' If so that, if \mbh\ can be retrieved under the virial assumption, an estimate of the luminosity becomes possible since \lledd$\propto L$/\mbh\ \citep[][\citetalias{marzianisulentic14}]{wangetal13,lafrancaetal14}. This approach  is conceptually analogous to the use of the link between the velocity dispersion in virialized systems (i.e. the rotational velocity of  {\sc Hi} disks in spiral galaxies \citealt{tullyfisher77}).  Initial computations for samples of 100-200 low-$z$ \ quasars ($\lesssim 1$) confirm the conceptual validity of the ``virial luminosity'' estimates \citep[submitted]{negreteetal17,negreteetal18}, although scatter in the distance modulus   is still too large to draw meaningful inferences for cosmology.
 
4DE1 trends can also be helpful for interpreting high-$L$, high-$z$ quasars, although there are  two  caveats.  At high redshift, $z \sim  \ $2, the majority of sources show large FWHM due to a bias in luminosity \citep{sulenticetal14,sulenticetal17}:  quasars with luminosities comparable to the  low-$z$ low-$L$\ sources are still too faint to be efficiently discovered. More fundamentally, there is a minimum possible FWHM \hb\ at fixed luminosity, if the line emitting region is virialized and its size follows a scaling law with luminosity. In practice this means that at $\log L \gtrsim 47$, all lines have to be broader than FWHM$~$2000 \kms. By the same token the FWHM limit for  Population A becomes luminosity dependent.  The limit  established at 4000 \kms\ is valid only for low-$z$, relatively low-$L$\, quasars.   Sources with larger FWHM and emission line properties similar to the ones of the low-$z$ xA quasars have been found at  high-$z$ and high-$L$\ \citep[][\citetalias{marzianisulentic14}]{negreteetal12}.  Another important issue at high-$L$\ concerns the HIL blueshifts. While at low-$L$\ large blueshifts ($v_\mathrm{r} \lesssim -1000$ \kms) are confined to Pop. A \citep{sulenticetal07,richardsetal11}, at high-$L$ they are ubiquitous \citep{coatmanetal16,bischettietal17,bisognietal17}, even if Pop. A sources still show the largest blueshift 
amplitudes among all quasars \citep[][hereafter \citetalias{sulenticetal17}]{sulenticetal17}. At any rate, several recent studies confirm that \hb, observed in NIR spectra in quasars at $z \gtrsim 1$  shows fairly symmetric and unshifted profiles suggesting that the broadening is mainly due to virial motions of the line emitting gas \citep[][\citetalias{sulenticetal17}]{marzianietal09,bisognietal17,shenetal16,vietrietal18}. We will show in this paper  that this is probably true also for high-$L$ xA sources. 

Goals of this paper include testing the effectiveness of 4DE1 selection criteria for identifying high \lledd\ xA sources at z$\sim $2.4. This will enable us to analyze spectral properties of the identified xA quasars in the rest-frame UV region. A sample of candidate xA sources (hereafter GTC-xA) was observed with the Gran Telescopio de Canarias (GTC) using the OSIRIS spectrograph. We apply the 4DE1 selection criterion defined by \citetalias{marzianisulentic14} using UV diagnostic ratios: \aliii/\siiii$\gtrsim$0.5 and  \ciii/\siiii$\lesssim$1.0. The selected sources are intended to represent an xA population for which \rfe\ is expected to be larger than 1, with no limitation on line FWHM. The sample selection is described in Sect.\,\S\ref{sec:sample}. The observations and the data reduction are  presented in Sect.\,\S\ref{sec:observa}. We perform a multicomponent fitting and build Monte Carlo (MC) simulations to estimate measurement uncertainties (Sect.\,\S\ref{sec:analysis} and Appendix \ref{sec:errors}). Spectra and line measures are presented for the 1900\AA\ blend, the blend \civ\ + \heiiuv, and the \siiv\ region in Sect.\,\S\ref{sec:results}. A comparison with control samples at low-$z$ and/or $L$ is described in the Section \S\ref{sec:ewdis}. A composite spectrum for the GTC-xA sample (Sect.\,\S\ref{sec:composite1}) allows us to carry out a comparison  with  low--$z$/low--$L$ samples (Sect.\,\S\ref{sec:composites}). We discuss  the low-ionization spectra and identify \feiii\ and \feii\ features that are  prominent in our spectra (Sect.\,\S\ref{sec:feiii} and Sect.\,\S\ref{sec:fluo}). After estimating the main accretion parameters (\S \ref{sec:acc}),  we briefly analyze the relation of xA sources to Weak Line Quasars (WLQs) --  a related class of quasars with extreme properties (Sect.\,\S\ref{sec:wlq}).   

xA sources are especially important because they are the quasars radiating at the highest luminosity per unit mass. The extreme radiative properties of xAs make them prime candidates for maximum feedback effects on host galaxies. We briefly analyze the possibility of significant feedback effects in Sect. \S\ref{sec:feed}.  We conclude the paper with some consideration on the possible cosmological exploitation of xA quasars at high-$z$\ (Sect.\,\S\ref{sec:concl}).

\section{Sample description}
\label{sec:sample}

\citetalias{marzianisulentic14} extracted  3000 quasar spectra from the SDSS DR6 archive which provided coverage of the 1900\AA\ blend for sources in the redshift range $2.0   < z  < 2.9$, with $g < 19.5$. Intensity measures of \aliii, \siiii\ and \ciii\ were carried out with an automatic {\tt SPLOT} procedure within the {\tt IRAF} reduction package. The majority of SDSS spectra are quite noisy making them suitable for identifying samples of candidate sources, but not providing accurate spectroscopic measures \citep{sulenticmarziani15}. A preliminary selection of extreme Eddington candidates was made and sources were then vetted according to the xA selection criteria (\aliii/\siiii$\gtrsim$0.5 and  \ciii/\siiii$\lesssim$1.0). \citetalias{marzianisulentic14} considered the brightest candidates with moderate S/N ($\ge$15) spectra, leaving 58 candidate xA quasars  whose relative \aliii, \siiii\ and \ciii\ intensities satisfied the selection criteria, but whose S/N was too poor to make an accurate measurement of individual lines in the 1900\AA\ blend. Hence the need for new GTC spectroscopic observations . This paper present an analysis of 19 of the sources that constitute our GTC-xA sample.
 
Table \ref{tab:properties} gives source identifications and basic properties including: redshift and uncertainty (Col. 2 \& Col. 3), emission line used for the redshift estimate (Col. 4), apparent $V$ magnitude and absolute $B$ magnitude $M_{\rm B}$\ (Col. 5 \& 6) as given in \citet{veroncettyveron10}, and $(g - r)$ color index from the SDSS photometry (Col. 7). Column 8 identifies other observed features like its classification as BAL (Broad Absorption Line) or mini-BAL quasar, and the radio properties, if they are detected.  Low-$z$ studies \citep{zamfiretal08} define a radio-loud (RL) quasar with a radio/optical flux ratio $\log$\,\rk\,$\gtrsim$\,1.8 \citep{kellermannetal89}, or better a radio power measure $\log P_{\nu}\,>\,31.6$ [erg s$^{-1}$ Hz$^{1}$], independent of uncertainties in optical flux measures. \rk\ was obtained  normalizing the $k-$corrected radio flux at 1.4GHz by the $k-$corrected $B$\ magnitude. Two radio-detected sources exceed the \rk\ limit with the third one close to the limit. Only for SDSSJ233132.83+010620.9 was possible to estimate the radio spectral index from one observation at 8.4 GHz \citep{ceglowskietal15}: $\alpha_\mathrm{r}\approx 0.44$, that places it in the compact steep-spectrum radio domain. SDSSJ234657.25+145736.0 has only a low-resolution  NRAO VLA Sky Survey  (NVSS) map available. In this case, $P_{\nu}$ and \rk\ have been computed assuming $\alpha_\mathrm{r}=0 $. All of them exceed the $\log\ P_{\nu}$ limit, then we could have three RL quasars in GTC-xA sample. Since low-$z$ RL are almost never found in the xA domain in 4DE1, it is possible either that the three radio detected sources are not xA extreme accretors or that high-$z$ Pop. A quasars are more frequently RL. At this point the three RL sources in the sample must be treated with caution.

\subsection{The FOS-A, S14-A and FOS-xA ``control'' samples}
\label{sec:control}

The quasars in the GTC-xA sample are  thought to be highly accreting, with an average bolometric luminosity of $\log L\sim$ 47 [\ergs]. Absolute magnitude $M_\mathrm{B} \approx -26$, before extinction correction, corresponds to a comoving space density of  $\sim 10^{-6}$ mag$^{-1}$ Mpc$^{-3}$\ just beyond the turnover at the high luminosity end of the 2dF luminosity function \citep{boyleetal00}. 

In order to compare the behavior of the GTC-xA sample, we consider the FOS sample from \citet{sulenticetal07} as a control sample at low-$L$ and low-$z$. For the sake of the present paper, we restrict the control FOS sample to 28 Pop. A RQ sources covering the \civonly\ and 1900\AA\ blend spectral range and with previous measures for the \hb\ profile and \rfe\ \citep{marzianietal03a}. 23 objects are classified as Pop. A1-A2 sources (henceforth FOS-A sample) and 5 as xA sources (hereafter FOS-xA sample) including I Zw 1. The FOS sample has a typical bolometric luminosity  $\log L \sim 45.2$ [erg s$^{-1}$] and a redshift $z\lesssim$ 0.5.

The sample presented in \citet[hereafter S14]{sulenticetal14} covers a similar range in redshift ($z\sim2.3$, corresponding to a lookback time of $\approx$\,10 Gyr), and are in turn  a factor $\sim$10 less luminous ($\log L\sim\,46$) than the GTC-xA sample. S14 is representative of a general population of faint, moderately accreting quasars that are also found at intermediate redshift \citep{fraixburnetetal17}. S14 includes both Pop. A  and B sources, but no xA sources. Restriction to the 11 Pop. A quasars  (hereafter S14-A) offers a high-$z$ counterpart to the FOS-A sample, and therefore  suitable for a comparison between xAs  and a sample of Pop. A quasars of moderate $L$\ at $z \sim 2 - 2.5$, that represents a population expected to be relatively common  {($\Phi \sim  10^{-6}$ mag$^{-1}$ Mpc$^{-3}$).}

\section{Observations and Data Reduction}
\label{sec:observa}
\newcommand{\ang}{\mbox{\normalfont\AA}}

Long slit observations were carried out in service mode using the OSIRIS spectrograph at the 10.4m GTC telescope of the Roque de los Muchachos Observatory. Grisms R1000B and R1000R with 2x2 CCD binning were used for the observations, depending on the redshift of the source. The majority of the observations employed R1000B that covered the wavelength range from 3650--7400 \AA\ with a reciprocal dispersion of 2.1 \AA\ per pixel (R$\approx$1000). In our highest $z$\ sources, SDSS J105806.16+600826.9, SDSS J125659.79-033813.8, and SDSS J233132.83+010620.9 with $z\approx$ 2.94, 2.98 and 2.63 respectively, we used the R1000R grism with reciprocal dispersion of 2.6 \AA/pixel and  spectral coverage 5100 - 10000\AA. These two spectral ranges correspond the rest-frame region covering the  UV spectral features of interest such as the  \siiv, \civ\ and the 1900\AA\ blend. The spectra were obtained with an 0.6 arcsec slit width oriented at the parallactic angle in order to minimize atmospheric differential refraction. Table \ref{tab:obs} contains a summary of the observations including: SDSS identification, observation date, grism employed, total exposure time for the 3 individual exposures on each source, seeing estimated from the FWHM of field stars in the acquisition image, and the estimated S/N in the 1450\AA\ continuum region on the blue side of \civ.

\begin{table}[h!]
\setlength{\tabcolsep}{4pt} 
\caption{Log of Observations} 
\scriptsize
\begin{center}
\begin{tabular}{c c c c c c c}        
\hline\hline\noalign{\vskip 0.1cm}         
SDSS Identification &Obs. Date & Grism & Exp. Time & Seeing & S/N  \\
(1) & (2) & (3) & (4)& (5) & (6) \\
\hline
\noalign{\vskip 0.07cm}
SDSSJ000807.27-103942.7  & \ 22/07/2015 \ & R1000B & 1800 & 1.13 & 29 \\
SDSSJ004241.95+002213.9 & \ 22/07/2015 \ & R1000B & 1440 & 1.18 & 39 \\
SDSSJ021606.41+011509.5 & \ 14/08/2015 \ & R1000B & 1440 & 0.80 & 70 \\
SDSSJ024154.42-004757.5  & \ 14/08/2015 \ & R1000B & 2700 & 1.17 & 40 \\
SDSSJ084036.16+235524.7 & \ 18/04/2015 \ & R1000B & 1800 & 1.13 & 31 \\
SDSSJ101822.96+203558.6 & \ 12/06/2015 \ & R1000B & 1440 & 1.40 & 33 \\
SDSSJ103527.40+445435.6 & \ 23/03/2015 \ & R1000B & 1800 & 1.33 & 32 \\
SDSSJ105806.16+600826.9 & \ 12/06/2015 \ & R1000R & 2700 & 1.22 & 22 \\
SDSSJ110022.53+484012.6 & \ 18/04/2015 \ & R1000B & 1890 & 0.74 & 38 \\
SDSSJ125659.79-033813.8 & \  23/05/2015 \ & R1000R & 1440 & 1.29 & 12 \\
SDSSJ131132.92+052751.2 & \ 23/05/2015 \ & R1000B & 1800 & 1.42 & 31 \\
SDSSJ143525.31+400112.2 & \ 11/04/2015 \ & R1000B & 2340 & 0.98 & 60 \\
SDSSJ144412.37+582636.9 & \ 25/06/2015 \ & R1000B & 1980 & 0.89 & 23 \\
SDSSJ151258.36+352533.2 & \ 11/04/2015 \ & R1000B & 2340 & 0.86 & 37 \\
SDSSJ214009.01-064403.9 & \ 05/08/2015 \ & R1000B & 1800 & 1.13 & 60  \\ 
SDSSJ220119.62-083911.6 & \ 17/06/2015 \ & R1000B & 2340 & 0.83 & 75 \\
SDSSJ222753.07-092951.7 & \ 26/06/2015 \ & R1000B & 1800 & 0.91 & 62 \\
SDSSJ233132.83+010620.9 & \ 05/08/2015 \ & R1000R & 1800 & 1.29 & 22 \\
SDSSJ234657.25+145736.0 & \ 16/07/2015 \ & R1000B & 1440 & 0.96 & 54 \\[0.05cm]
\hline
 \end{tabular}
\end{center}
\small {\sc Notes.} Columns are as follows: (1) SDSS identification. (2) Observation Date. (3) Grism. (4) Exposure time in seconds. (5) Seeing in arcseconds.  (6) S/N measured in the continuum at 1450 {\AA}. \label{tab:obs} \\
\end{table}

Data reduction was carried out in a standard way using the {\tt IRAF} package. Bias subtraction and flat-fielding correction were performed nightly. Wavelength calibration was obtained using Hg+Ar and Ne lamps observed with the same configuration and slit width used for source observations. Wavelength  calibration rms was  less than 0.1 \AA. We checked the wavelength calibration for individual exposures with sky lines before source extraction, background substraction, and final combination. Scatter of the sky line wavelength peaks was $\lesssim$20 \kms. This value provides a realistic estimate of the wavelength scale uncertainty including zero point error. Spectral resolution estimated from FWHM of the skylines is $\sim$ {230 \kms and 250 \kms} for grisms R1000B and R1000R, respectively.

Instrumental response and flux calibration were obtained nightly with observations of the spectrophotometric standard stars Ross 640, GD24-9, Feige 110, Hiltner 600, and G158-100. In order to improve flux calibration, we also included  two additional flux standard stars, LDS749B and HZ21, as target objects. They were observed with both grisms and two slits: 0.6 arcsec (as used for quasar observations) and 5 arcsecs. A comparison between the different slits gives a change in the absolute flux calibration $\sim$10\%. Spectra were corrected for light losses due to the narrow slit width employed and also to differential light loss as a function  of wavelength. Taking into account the ratio between the slit width and the seeing during an observation allowed correction for the scale factor and for the wavelength dependence of the seeing following the method described by \citet{Bellazzini07}. Telluric absorptions, that affect our spectra mainly beyond 7600 \AA, were also corrected  using the  standard stars to obtain a normalized template of the absorption features. The template was  shifted if needed, and scaled for each individual source in an iterative and interactive procedure, until the residuals in the  telluric correction were negligible. Spectra were finally deredshifted as explained in  Section \S\ref{redshiftdet}.

\subsection{Extinction estimation}
\label{sec:ext}

It is visually apparent from examining our spectra that some of them (e.g., SDSSJ021606.41+011509.5, Fig. \ref{fig:specSDSSJ021606.41+011509.5}) show a flatter continuum than  cannot be modeled with a single power-law over the observed spectral range. This effect, as well as the presence of BALs, has been interpreted in the literature as indicating  the presence of dust or internal reddening. In order to assess the importance of internal reddening on the observed fluxes and derived magnitudes in our xA sample, we have estimated the reddening in each source by fitting its UV continuum with quasar templates excluding spectral regions with broad emission lines (e.g., \lya, \siiv, \civ\ and the 1900\AA\ blend). We used four QSO templates: 1) a median composite spectrum representative of the xA quasar population  and built with extreme accretor sources identified in the SDSS DataBase by \citetalias{marzianisulentic14} excluding BALs; 2) a template involving the composite FIRST Bright Quasar Survey spectrum (FBQS; \citet{brothertonetal01}); 3) the composite spectrum provided by \citet{harrisetal16} with BOSS spectra of quasars in the redshift range 2.1\,$<$\,z\,$<$\,3.5, and 4) the SDSS composite quasar spectra  \citep{vandenberketal01}. We reddened the templates using an SMC extinction law \citep{gordonclayton98} which appears to be the most appropriate reddening law for modeling quasar spectra \citep{yorketal06, galleranietal10}. We assumed a $R_V$ coefficient of 3.07 for the extinction law.

 \begin{table}[h!t]
\setlength{\tabcolsep}{4pt} 
\begin{center}
\caption{Extinction measures for the xA sources\label{tab:extinction}}
\begin{tabular}{c c c}\\ 
\hline\hline\noalign{\vskip 0.1cm}         
SDSS identification & $A_V$ (mag.)\\
\hline
\noalign{\vskip 0.07cm}
SDSSJ021606.41+011509.5	 &	0.270 	 \\
SDSSJ103527.40+445435.6	 &	0.178	 \\
SDSSJ131132.92+052751.2	 &	0.325	 \\
SDSSJ144412.37+582636.9	 &	0.106	 \\
SDSSJ214009.01-064403.9	&	0.250	 \\
SDSSJ233132.83+010620.9	&	0.150	\\
\hline	
\end{tabular}
\end{center}
\end{table}

In general, best fits were obtained with the xA composite although there were no  appreciable differences and all the fittings yielded  similar results. For the majority of sources (12) no additional extinction was needed and the continuum was well represented by  the templates. In 6 cases the reddening has a significant effect on the spectrum, amounting to $A_V$=0.1 to $A_V$=0.32. Table \ref{tab:extinction} reports reddening estimates parametrized by the A$_V$ value.  In the case of SDSSJ220119.62-083911.6, classified as BAL, the spectrum shows a broad and deep absorption in the blue wings of \civ, \siiv\ and \lya\ with an evident flattening of the continuum at wavelengths shorter than 1600 \AA. However, from 1700 \AA\ the spectrum shows a similar slope to the templates and does not show evidence for reddening. Hence no internal reddening was applied to this source  and the continuum at 1350 \AA\ needed for estimation of the luminosity was obtained by extrapolating the power law of the fit applied to the red spectral region.

\section{Data Analysis}
\label{sec:analysis}

\subsection{Redshift determination}
\label{redshiftdet}

Accurate redshift estimates are very important because redshift defines the quasar rest-frame from which emission line shifts can be measured. Shifts are particularly important for HILs like \civ\  \citep[e.g.,][]{gaskell82,espeyetal89,carswelletal91,marzianietal96}. Redshift estimates are chiefly obtained from narrow emission lines (or the narrow core of \hb) for low-$z$\ sources \citep[e.g.,][]{eracleoushalpern03,huetal08}. In the UV region covered by our spectra of $\rm z\approx\rm{2.3}$, narrow LILs are not present. We must resort to {broad} LILs and estimate the redshifts using three features: \aliii, \cii\ and \oi+\siiia. The strongest and hence most often detected LIL involves \aliii, which emerged in lower $z$\ studies as a kind of UV surrogate \hb\ \citep{bachevetal04}. While often detected, it is always part of the 1900\AA\ blend albeit on the blue end of it. Although \aliii\ and \siiii\ are blended lines, they are stronger than \cii\ and \oi+\siiia\ and in many of the spectra their peaks are clearly seen. We performed a multicomponent fit for each source, considering all the lines in the region of the 1900\AA\ blend (See Section \S\ref{sec:fitting}). The peaks of \aliii\ and \siiii\ were unconstrained in intensity in the fitting and the adopted model of the blend was the one with minimum $\chi^2$. \cii\ is the only isolated LIL in the observed UV range and is detected in only 5 sources (e. g.  SDSSJ021606.41+011509.5, See Fig. \ref{fig:specSDSSJ021606.41+011509.5}). The \oi+\siiia\ blend is also well seen in a few sources where \cii\ is detected. In the rest of the sample these lines are too weak to be useful and are often affected by absorption features. 

We constructed synthetic Gaussian profiles for \cii\ and \oi+\siiia\ lines using the {\tt IRAF} task {\tt MK1DSPEC} assuming FWHM $\sim$\,4000 \kms. We computed a redshift from the peak of the synthetic features. When the \aliii\ redshift was compared with the one of \cii\ and \oi+\siiia, we find {three} cases where the difference is less than 100 \kms. In these cases, we kept the redshift value determined by the \cii\ and \oi+\siiia\ (label {\sc{i}} and {\sc{ii}} in Table \ref{tab:properties}). In the cases where the difference is larger than 100 \kms, we considered the redshift given by \aliii\ and \siiii\ (label {\sc iii} in Table \ref{tab:properties}), since the peak of these lines is clearly observed. The uncertainty reported in Column 4 of  Table \ref{tab:properties} has been computed from the redshift difference between {  \cii, \oi+\siiia, and \aliii+\siiii. } 

\begin{figure*}[h!t]
	\begin{center}
		\includegraphics[width=18.cm]{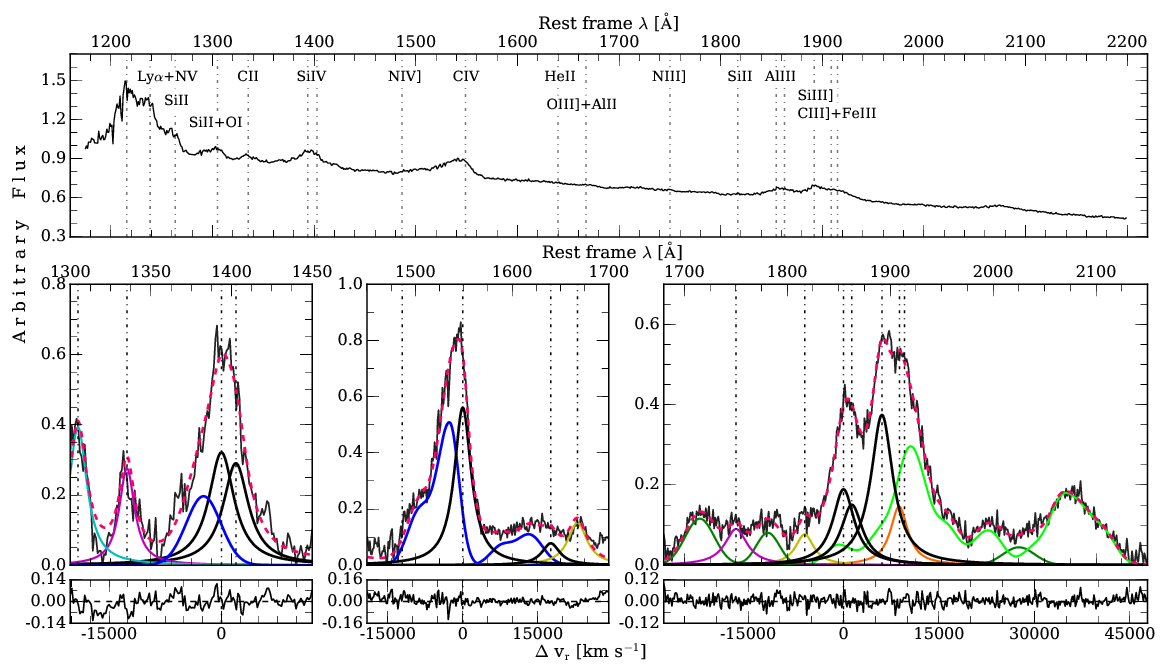}
	\end{center}
\caption{{Top panel: rest-frame composite spectrum (See Sect.\,\S\ref{sec:composite1}). Abscissa corresponds to vacuum rest-frame wavelength in \AA, while ordinate is in arbitrary intensity. } Dot-dashed vertical lines identify the position at rest-frame of the strongest emission lines. Bottom panels: multicomponent fits after continuum subtraction for the 1400\AA\,blend, \civ\ and 1900\AA\,blend spectral regions (Sect.\,\S\ref{sec:fitting}). In all the panels continuous black line marks the broad component at rest-frame associated to \siiv, \civ, \aliii\ and \siiii\ respectively, while the blue one corresponds to the blueshifted component associated to each emission. Dashed pink line marks the fit to the whole spectrum. Dot-dashed vertical lines correspond to the rest-frame wavelength of each emission line. In the \siiv\ spectral range, the cyan line marks the contribution of  \oi+\siiia\ blend, while the magenta line corresponds to the \cii\ emission line. In the \civ\ region, the yellow one corresponds to the \oiiiuv\ + \alii\ blend. In the 1900\AA\ blend range, \feiii\ and \feii\ contributions are traced by dark and pale green lines respectively, magenta line marks the \niii\ and the yellow one corresponds to the \siii. Lower panels correspond to the residuals, abscissa is in radial velocity units km s$^{-1}$.}
	\label{fig:restframespectra}
\end{figure*}

\subsection{Multicomponent fitting}
\label{sec:fitting}

Emission lines can be distinguished by their ionization potential (IP). The UV range covered in our spectra is populated by intermediate (IP $\sim$ 20--40 eV) and high ionization lines (IP $>$ 40 eV). They offer an opportunity to characterize the behavior of different ionic species at the same time. In order to analyze the emission lines in our spectra, we carried out multicomponent fits using the {\tt SPECFIT} routine from {\tt IRAF} \citep{kriss94}. This routine simultaneously fits the continuum, and emission/absorption line components. The best model is indicated by the minimum $\chi^2$ over a spectral range where all components are included.

The main continuum source in the UV region is thought to arise from the accretion disk \citep{malkansargent82}. In the absence of extinction the continuum can be modeled by a single power-law over the full observed spectral range. However, due to the presence of absorptions (BAL sources) or dust extinction the continuum is flattened out in several sources (Sect. \S\ref{sec:ext}). Whenever possible, we fit the entire spectra range with a single power-law or a linear continuum.  Otherwise we estimate locally the continuum. We  divide the observed spectral range in three parts, which are centered on the most important emission lines relevant to our work.

{\sc Region 1}: 1700--2200\AA. This range is dominated by the 1900\AA\ blend which includes \aliii, \siiii, \ciii\ and \feiii\ lines (See Sect.\,\S\ref{sec:feii-iii} and Appendix\,\ref{sec:errors} for \feiii\ line identification). On the blue side of the blend \siii\ and \niii\ are also detected. \aliiionly, \siiiionly\ and \ciiionly\ are intermediate--ionization lines (IIL) and according to \citet[][hereafter \citetalias{negreteetal12}]{negreteetal12} can be well-modeled with Lorentzian profiles. The strengths of the three lines are allowed to vary freely in our {\tt SPECFIT} model. FWHM \aliiionly\ and \siiiionly\ were assumed equal, while FWHM \ciiionly\ was unconstrained. \siii\ and \niii\ were also modeled with Lorentzian profiles with flux and FWHM allowed to vary freely. All Lorentzian profile peaks were fixed at rest-frame. \feii\ makes an important contribution in the range 1715--1785 \AA. We tried to use templates available in the literature \citep{bruhweilerverner08,mejia-restrepoetal16}, but we could not reproduce the observed contribution. If that template is scaled to reproduce these features, the \feii\ emission around the \mgii\ line would be overestimated by a large factor. On the converse if the template is normalized to \feii\ in the proximity of \mgii, the \feii\ emission in the spectral region 1700 -- 2200 \AA\, is negligible. We therefore decided to fit isolated Gaussian profiles for \feii\ at 1715 and 1785 \AA. Their flux and FWHM vary freely. \feiii\ emission makes a larger contribution than \feii\ and appears especially strong on the red side of the 1900\AA\ blend. We modeled the emission of this ion with the \citet{vestergaardwilkes01} template and included an extra component at 1914 \AA\ following \citetalias{negreteetal12} (the motivation for this choice is discussed in Sect. \S\ref{sec:feiii}). Around 2020--2080 \AA\ we were forced to include extra Gaussians in order to obtain a good fit (we found excess emission with respect to the \feiii\, template). The flux and FWHM of the features at 2020--2080 \AA\ were allowed also to vary freely.

{\sc Region 2}: 1450--1700 \AA. The \civ\  emission line dominates this region and is accompanied by \heii, \oiiiuv\, and \alii. The broad component (BC) of \civonly\ is modeled by a Lorentzian profile fixed at the rest-frame. The flux of the \civbc\ is free and FWHM is assumed to be the same as \aliii\ and \siiii. All the \civonly\ profiles in our sample show a blueshift/blueward asymmetry. In order to model it with {\tt SPECFIT}, we used one or two blueshifted skewed Gaussian profiles. The flux, FWHM, asymmetry and shift were unconstrained. \heii\ was modeled assuming components similar to those of \civ: Lorentzian and skewed Gaussian profiles for the BC and blueshifted components, respectively. The FWHM, shift and asymmetry were assumed equal to those of \civ, but the flux varies freely. \oiiiuv\ and \alii\ were also modeled with unshifted Lorentzian profiles with their fluxed and FWHM varying freely.

{\sc Region 3}: 1300--1450 \AA. The dominant emission in this region involves \siiv+\oiv\ (the 1400\AA\  blend), and is accompanied by weaker \siiia, \oi\ and \cii\ lines. The underlying assumption of the blend modeling is that the BC emission is dominated by \siiv\ due to collisional deexcitation of the inter combination \oiv\ multiplet \citep{willsnetzer79}, while the blue shifted component is due to an inextricable contribution of both \oiv+\siiv. The \siiv+\oiv\ feature is a high-ionization blend, and shows a blueshifted, asymmetric profile not unlike \civ. The broad component was modeled with the same emission components of \civonly, whenever possible (in several cases the blue side of the \siiv+\oiv\ blend was strongly contaminated by absorption features making a good fit impossible), with only the flux varying freely. In the case of strong absorption the blueshifted emission was modeled independently from the one of \civonly. 

For lines that are composed of more than one component (\civ, \heii\ and \siiv+\oiv) the total profile parameters were also computed: FWHM, centroid at half maximum (\cmp), and asymmetry index (AI) defined by \citet{zamfiretal10}. Unlike the {\tt SPECFIT} components, these parameters provide a description of the profile that is not dependent on the model decomposition of the profile.

Figure \ref{fig:restframespectra} shows the multicomponent fitting made to the composite spectrum obtained by combining the normalized spectra of the GTC-xA sample (See Section \S\ref{sec:composite1} for a complete description). The bottom panels of the figure present the fits performed on the \siiv, \civ\ and \aliii\ spectral regions of the continuum subtracted  spectrum. Residuals of the fits are shown in the lower part of the bottom  panels. Spectra and multicomponent fits for the individual 19 quasars analyzed in this paper are shown from Fig. \ref{fig:specSDSSJ000807.27-103942.7}  to \ref{fig:specSDSSJ234657.25+145736.0} in the Appendix A. {Error estimations of blended emission lines were evaluated by building Monte-Carlo (MC) simulations as explained in Appendix B.}

\section{Results}
\label{sec:results}

\subsection{Composite spectrum}
\label{sec:composite1}

We constructed a median composite spectrum from the individual observations in order to emphasize the main emission features (composites efficiently remove narrow absorption lines) and carry out a comparison with other samples (See Sect. \S\ref{sec:composites}). The composite GTC-xA spectrum corresponds to the median of all the normalized individual spectra including BALs.  This simple approach produced a spectrum that reflects the behavior of the xA objects (See Fig. \ref{fig:restframespectra}).

In xA sources, the 1900\AA\ blend of the composite spectrum shows a high contribution of \aliii\ and \siiii\ compared to \ciii. Most of the emission on the red side of \siiii\ can be attributed to \feiii\ excess in addition to the template, which is required to minimize the fit $\chi^{2}$. A fit with \ciii\ only (with no \feiii\ $\lambda$1914)  would leave a large residual at 1915--1920\,\AA.  Strong \feiii\ emission is confirmed also by the prominent bump at 2080\,\AA\ predominantly ascribed to the \feiii\ multiplet \#48. 

On the blue side of \civ, a smooth and shallow through is due to the combined effect of broad absorption lines that are frequent shortwards of \civ. The emission profile of \civ\ is in any case almost fully blueshifted, as observed in the majority of sources. Also the \siiv\ median profile shows a blueshift asymmetry even if it is affected by the heavy absorptions frequently observed on the blue side of this line. The net effect is that \siiv\,+\oiv\ appear more symmetric because their blue wings are truncated by narrow and broad absorptions.  The composite also clearly shows  the prominent low-ionization features associated with \cii, and \oi\ blended with \siiia. \civ\,+\heiiuv\ shows low equivalent width, close to the boundary of WLQs (See Sect. \S\ref{sec:wlq}). The composite spectrum is especially helpful for the increase in S/N that allows to trace the broad and faint \heiiuv\ profile, which shows a flat topped appearance. It is interpreted in the multicomponent fits as due to a strong blueshifted component, blended with a faint BC and with \oiiiuv\ and Al{\sc ii}$\lambda$1670.

\subsection{{Consistency of selection criteria for xA sources}}

\begin{figure}[htp!]
\begin{center}
\includegraphics[width=9cm]{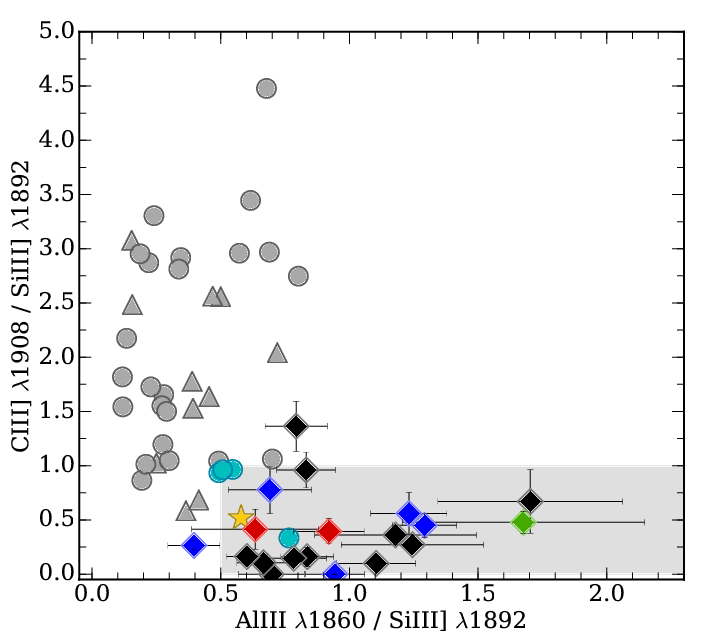}
\end{center}
\caption{Relation between intensity ratios \aliii/\siiii\ and \ciii/\siiii. Black, blue, green and red diamonds correspond to normal, BAL, mini--BAL and Radio--Loud quasars respectively for the present GTC-xA sample. Gray and cyan dots correspond to the FOS-A and FOS-xA sample respectively. Gray triangles correspond to S14-A sample. Yellow star marks the position of I Zw 1. Gray area represents the parameter space occupied by xA sources. 
\label{fig:critsel}}
\end{figure}

Using lines in the 1900\AA\ blend, \citetalias{marzianisulentic14} proposed that xA sources with \rfe$>$\,1 show  flux ratios \aliii/\siiii\,$\gtrsim$\,0.5 and \ciii/\siiii\,$\lesssim$\,1. Our sample was selected considering these criteria applied to SDSS noisy spectra previously excluded by \citetalias{marzianisulentic14}. The high S/N spectra of the GTC sample presented in this work confirm the defining criteria for the identification of the highly accreting sources in the UV range. Figure \ref{fig:critsel} shows the location of the 19 sources of our sample  in the plane defined by \ciii/\siiii\ vs \aliii/\siiii. In order to compared the behavior of the xA with the rest of the Pop. A, in Figure \ref{fig:critsel} are also represented control samples (FOS-xA, FOS-A and S14-A) described in Section \S\ref{sec:control}. 

Comparing the flux ratios shown by the low and high-$z$ Pop. A and xA samples, like \citetalias{marzianisulentic14} have done, we observe a clear difference between the two kind of populations. The FOS-xA and GTC-xA samples show flux ratios different to the ones of FOS-A and S14-A samples. The GTC-xA sources occupy a very well defined region in the bottom right side of the panel, while the Pop. A sample occupy the left top space. Two of the objects (SDSSJ222753.07-092951.7 and SDSSJ125659.79-033813.8, $\sim$10 \%\ of our sample) do not rigorously satisfy the selection criteria, although they present similar spectral properties to ones observed in the rest of the sample. The measured line ratios are actually borderline: the \aliii/\siiii\ ratio of the BAL  \object{SDSSJ125659.79-033813.8} is $\approx0.40 \pm 0.10$, the \ciii/\siiii\ ratio of \object{SDSSJ222753.07-092951.7} is $\approx1.36 \pm 0.23$. Therefore the two quasars will be considered along with all other GTC-xA sources.

\subsection{1900\AA\ Blend}
\label{sec:blend}

Table \ref{tab:blend} reports emission lines measurements corresponding to the 1900\AA\ blend region: \aliii, \siiii, \ciii, \niii\ and \siii.  The first column lists the SDSS name. The equivalent width ($W$), flux ($F$) and FWHM of \aliii\  are reported in Columns 2, 3, and 4 respectively. Equivalent width and flux are reported for \siiii\ in Col. 5 and 6, since its FWHM is assumed equal to the one of \aliii.  $W$, $F$ and FWHM of \ciii\ are reported in Col. 7, 8 and 9 respectively. They are followed by the flux of the two faint features outside of the blend: \niii\ and \siii\ (Col. 10 and 11). All the parameters have uncertainties at 1$\sigma$\ confidence level, which were obtained from the MC simulations. {The MC method followed to estimate the uncertainties is explained in Appendix \ref{sec:errors}.}

In several instances the \aliii\ doublet is partly resolved and is assumed to be at the rest-frame along with \siiii. In general, the \aliii\ flux, FWHM and $W$ can be estimated with good accuracy as the doublet can be efficiently deblended save for the cases that the lines are very broad (FWHM $\gtrsim $ 6000 \kms). \aliii\  is well fit with a symmetric and unshifted Lorentzian. Typical uncertainties in FWHM are $\lesssim 10$\%.  Only in one case (\object{SDSSJ084036.16+235524.7}) there is evidence for a blue asymmetry in the 1900\AA\ blend although the \aliii\ BC can be isolated with good accuracy. The fraction of  the blueshifted to the total \aliiionly\ emission would of $\approx 0.25$, small compared to a median value of $\approx 0.46$ for \civonly. The presence of a blue shifted component in \aliii\ is a rare occurrence. Only in one source of the \citetalias{sulenticetal17}  sample shows evidence of a blueward \aliii\ excess \citep[\object{HE0359-3959}, ][]{martinez-aldamaetal17}.

The peak of the 1900\AA\ blend is often not in correspondence of the \ciii\ wavelength.  If the blend of \siiiionly, \ciiionly\ and \feiii\ is peaked, the peak is preferentially redshifted beyond 1909 \AA. The 1900\AA\ blend profile  may show  excess emission on its red side (See e.g., \object{SDSSJ021606.41+011509.5}, Fig. \ref{fig:specSDSSJ021606.41+011509.5}).  {In these cases,} we suggest that  the main contribution is coming from the \feiii\ $\lambda$1914 line, modeled with a  Lorentzian in excess to the \feiii\ template (Sections \S\ref{sec:feii-iii} and \S\ref{sec:fluo}), or by unresolved \feiii\ emission.  

The weaker lines \niii\ and \siii\ are detected in about $2/3$\ of the sample, e.g.,  they are both prominent in the case of the quasar SDSSJ021606.41+011509.5.  The \niii\ line shows a wide range in $W$. It is not detected in four sources but the detections show  $W$\ in the range 0.5 -- 5 \AA.

\begin{table*}[h!t]
\setlength{\tabcolsep}{3pt} 
\small
\begin{center}
\caption{Measurements in the 1900\AA\ blend region \label{tab:blend}}
\scriptsize
\begin{tabular}{ccccccccccc}\\ 
\hline\hline\noalign{\vskip 0.1cm}         
\multirow{2}{*}{SDSS Identification} 
 &  \multicolumn{3}{c}{Al{\sc iii} 1860} & \multicolumn{2}{c}{Si{\sc iii]} 1892} & \multicolumn{3}{c}{C{\sc iii]} 1909} & \multicolumn{1}{c}{N{\sc iii}] 1750} & \multicolumn{1}{c}{Si{\sc ii} 1816}\\
    \cmidrule(lr){2-4} \cmidrule(lr){5-6} \cmidrule(lr){7-9} \cmidrule(lr){10-10} \cmidrule(lr){11-11}\noalign{\vskip 0.05cm}         
& $W$ & $F$ & FWHM & $W$ & $F$ & $W$ & $F$ & FWHM & $F$ & $F$ \\
(1) & (2) & (3) & (4) & (5) & (6) & (7) & (8) & (9) & (10) & (11) \\
\hline \noalign{\vskip 0.07cm}
SDSSJ000807.27-103942.7	&	9.6	$\pm$	1.2	&	11	$\pm$	0.8	&	5410	$\pm$	630	&	12.1	$\pm$	1.9	&	13.2	$\pm$	1.6	&	11.9	$\pm$	1.9	&	12.7	$\pm$	1.6	&	4810	$\pm$	790	&	6.5	$\pm$	0.9	&	3.1	$\pm$	1.0	\\	
SDSSJ004241.95+002213.9	&	9.6	$\pm$	1.3	&	25.2	$\pm$	2.2	&	5060	$\pm$	340	&	11.8	$\pm$	1.5	&	30.2	$\pm$	2.5	&	1.9	$\pm$	1.0	&	4.9	$\pm$	2.6	&	3060	$\pm$	1620	&	4.8	$\pm$	1.5	&	2.2	u	\\		
SDSSJ021606.41+011509.5	&	11.3	$\pm$	1.3	&	33.2	$\pm$	1.7	&	3190	$\pm$	220	&	9.0	$\pm$	1.2	&	25.7	$\pm$	2.1	&	3.7	$\pm$	1.0	&	11.6	$\pm$	2.8	&	2690	$\pm$	850	&	7.4	$\pm$	1.8	&	8.1	$\pm$	1.5$^\mathrm{a}$	\\	
SDSSJ024154.42-004757.5	&	5.1	$\pm$	1.1	&	9.5	$\pm$	1.9	&	3340	$\pm$	320	&	4.2	$\pm$	0.6	&	7.7	$\pm$	0.8	&	1.2	u		&	2.1	u	&	$\ldots$	&	$\ldots$	&	6.6	$\pm$	0.8	\\					
SDSSJ084036.16+235524.7	&	6.1	$\pm$	1.2	&	11.5	$\pm$	2	&	3420	$\pm$	380	&	5.3	$\pm$	1.2	&	9.8	$\pm$	2	&	2.0	$\pm$ 0.9	&	3.5	$\pm$ 1.6	&	2850	$\pm$ 420	&	$\ldots$	&	3.3	$\pm$ 0.6$^\mathrm{a}$	\\
SDSSJ101822.96+203558.6	&	8.5	$\pm$	1.2	&	26.7	$\pm$	2.8	&	4870	$\pm$	340	&	11.1	$\pm$	1.8	&	33.4	$\pm$	4.2	&	1.7	u		&	4.9	u	&	$\ldots$	&	3	u	&	6	$\pm$	2.4	\\				
SDSSJ103527.40+445435.6	&	7.9	$\pm$	1.4	&	19	$\pm$	2.8	&	7060	$\pm$	480	&	11.5	$\pm$	1.7	&	27.3	$\pm$	2.9	&	$\ldots$		&	$\ldots$	&	$\ldots$	&	$\ldots$	&	$\ldots$	\\									
SDSSJ105806.16+600826.9	&	5.1	$\pm$	0.8	&	12.8	$\pm$	1.6	&	4180	$\pm$	650	&	7.9	$\pm$	1.1	&	19.2	$\pm$	1.7	&	0.8	u		&	2	u	&	$\ldots$	&	4.2	$\pm$	2.1	&	6.7	$\pm$	1.3	\\			
SDSSJ110022.53+484012.6	&	9.1	$\pm$	0.8	&	21.8	$\pm$	2.2	&	4900	$\pm$	390	&	15.6	$\pm$	2.0	&	36.3	$\pm$	2.9	&	2.6	$\pm$	1.2	&	5.9	$\pm$	2.7	&	3510	$\pm$	1970	&	5.3	$\pm$	1.3	&	2.4	$\pm$	1.7	\\	
SDSSJ125659.79-033813.8	&	13.4	$\pm$	3.0$^\mathrm{a}$	&	24.1	$\pm$	4.9$^\mathrm{a}$	&	5950	$\pm$	600	&	34.9	$\pm$	6.4	&	60.9	$\pm$	9.4	&	9.5 $\pm$	4.3	&	16.2	$\pm$	7.2	&	4000 $\pm$ 2090		&	7.1 $\pm$ 4.5$^\mathrm{a}$	&	7.8  $\pm$  3.7$^\mathrm{a}$	\\					
SDSSJ131132.92+052751.2	&	5.8	$\pm$	0.7	&	16.4	$\pm$	1.3	&	3760	$\pm$	250	&	6.2	$\pm$	0.8	&	17.3	$\pm$	1.6	&	$\ldots$		&	$\ldots$	&	$\ldots$	&	$\ldots$	&	$\ldots$	\\								
SDSSJ143525.31+400112.2	&	4	$\pm$	0.5	&	25.8	$\pm$	2.3	&	3350	$\pm$	270	&	3.8	$\pm$	0.6	&	23.4	$\pm$	2.5	&	0.4	u		&	2.3	u	&	$\ldots$	&	5.3	$\pm$ 2.8	&	4.4	$\pm$	1.9	\\				
SDSSJ144412.37+582636.9	&	13	$\pm$	2.7	&	12.5	$\pm$	2.3	&	8080	$\pm$	1560	&	7.9	$\pm$	1.9	&	7.5	$\pm$	1.6	&	3.8	u		&	3.6	u	&	$\ldots$	&	5.8	$\pm$	0.5	&	1.6	$\pm$  1.0	\\			
SDSSJ151258.36+352533.2	&	4.1	$\pm$	0.6	&	10.8	$\pm$	1.1	&	2780	$\pm$	330	&	4.6	$\pm$	0.7	&	11.8	$\pm$	1.2	&	1.9	$\pm$	0.6	&	4.6	$\pm$	1.3	&	2110	$\pm$	700	&	5.8	$\pm$	1.2	&	5.3	$\pm$	1.2	\\	
SDSSJ214009.01-064403.9	&	7.7	$\pm$	0.9	&	13.9	$\pm$	0.9	&	2620	$\pm$	160	&	6.3	$\pm$	0.9	&	11.3	$\pm$	1.1	&	3.5	$\pm$	1.2	&	7.4	$\pm$	2.1	&	3500	$\pm$	1150	&	0.8	u	&	1.5	$\pm$	0.7	\\		
SDSSJ220119.62-083911.6	&	6.3	$\pm$	1.1	&	26.7	$\pm$	4	&	7250	$\pm$	1050	&	9.3	$\pm$	1.9	&	38.6	$\pm$	6.9	&	7.4	$\pm$	1.8	&	30	$\pm$	6.5	&	6110	$\pm$	1670	&	6.7	$\pm$	2.4	&	$\ldots$	\\				
SDSSJ222753.07-092951.7	&	6	$\pm$	0.8	&	11.2	$\pm$	1	&	4500	$\pm$	420	&	7.8	$\pm$	1.2	&	14.1	$\pm$	1.7	&	10.8	$\pm$	1.6	&	19.2	$\pm$	2.2	&	4550	$\pm$	590	&	2.9	$\pm$	1.2	&	0.5	u	\\		
SDSSJ233132.83+010620.9	&	5.5	$\pm$	1.5$^\mathrm{a}$	&	38.5	$\pm$	10.2$^\mathrm{a}$	&	6310	$\pm$	1640$^\mathrm{a}$	&	8.8	$\pm$	2.7	&	60.7	$\pm$	17.4	&	3.7	$\pm$	1.3	&	25.2	$\pm$	8.6	&	5530	$\pm$ 4700	&	10.2	$\pm$	5.5$^\mathrm{a}$	&	6.6	u$^\mathrm{a}$	\\		
SDSSJ234657.25+145736.0	&	6.9	$\pm$	0.8	&	21.1	$\pm$	1.3	&	3660	$\pm$	560	&	4.1	$\pm$	0.9	&	12.4	$\pm$	2.5	&	2.8	$\pm$	1.1	&	8.3	$\pm$	3.2	&	3040	$\pm$	1450	&	2.8	$\pm$	1.4$^\mathrm{a}	$	&	5.3	$\pm$	1.4$^\mathrm{a}	$\\
\hline	
\end{tabular}
\end{center}
\small {\sc Notes.} Columns are as follows: (1) SDSS name.  (2), (5) and (7) report the equivalent width in units of \AA. (3), (6), (8), (10) and (11) list fluxes in units of \uflux. (4) and (9) correspond to the FWHM of \aliiionly\ and \ciiionly in units of \kms.  $^\mathrm{a}$ means that measurement is contaminated by absorption lines. A $u$ letter marks an upper limit to the measurement.

\end{table*} 

\subsection{\civ\ and \heiiuv}
\label{sec:civheii}

Measurements associated with the emission lines of the \civ\ spectral region are reported in Table \ref{tab:c4}. The first column lists the SDSS name. Column 2 reports the rest-frame specific continuum flux ($f_\lambda$) at 1350 \AA. No error estimate is provided; the inter calibration with different standard star spectra suggests an uncertainty around $10 $\%, which should include all main source of errors in the absolute flux scale.  Cols. 3 to 7 tabulate $W$, flux, FWHM, centroid at half intensity (\cmp) and asymmetry index (A.I.) of the \civ\ total profile (broad plus blueshifted components). Col. 8 and 9 list the equivalent width and $F$ of the broad component of \civonly. Its FWHM is not reported, because it is assumed to be the same as FWHM  \aliii, due to the BC have to be emitted by the same zone. The equivalent width, $F$, FWHM and \cmp\ of the blueshifted component of \civonly\ are given in columns 10 to 13,  respectively. 

Table \ref{tab:he2} reports the measurements associated with \heii\ and contaminant lines. {Columns 2 to 7 list the $W$ and flux of \heii\ total profile, \heii\ BC and \heii\ blue shifted component respectively.} Col.\,8 lists  the \oiiiuv\ + \alii\ flux. The last column reports the \niv\ flux. The \niv\ flux is highly uncertain due to absorption lines and to the blending with the blue wing of \civonly, then in many of the cases only upper limits are reported.

On average, the \heii\ total flux is $\sim$20$\%$ that of \civ. Since the typical $W$ \civ\ $\approx 10 - 15$\,\AA, the typical $W$ \heii\ is around $\approx$3 \AA\ distributed over a broad profile with $4000\lesssim$\ FWHM\ $\lesssim11000$\kms. The weakness of the \heii\ emission with respect to \civ, with \civ/\heii\ $\approx  5 -  6  \gtrsim 3$\ is out of the question.  It is important however to stress that our approach tends to maximize the contribution of the \civbc\ to the total \civ\ emission. This may lead to an overestimation of the \civ/\heii\ BC ratio. Nevertheless, in several cases (e.g., SDSS J214009.1--064403.9) we observed a significant \civ\ emission at rest-frame, implying that the   \civ/\heii\ estimate $\approx 5 - 6$\ is a safe one. We mention the \civ/\heii\ ratio because its value is rather high  for the low ionization parameter expected  in the emitting regions, if metallicity is solar or slightly supersolar \citepalias[Sect. \S\ref{sec:feiii}, ][]{negreteetal12}. 

An important feature in our spectra involves the strong blueshifts and asymmetries associated with the HILs. In $\sim$50$\%$ of our sample, the \civonly\ blueshifted component contributes $\sim$50$\%$ of the total profile. \heii\ also shows a highly  blueshifted component, contributing $\sim$60$\%$\ of the total flux. The A.I. values also support the presence of outflows: the blueshifted component pumped by radiative forces is more prominent than the virial component located at rest-frame. This is a major  difference between High-Ionization Lines (e.g., \civ) on the one side, and low and intermediate--ionization lines on the other (e.g., \aliii). The last one are basically symmetric and  show no evidence of large shifts with respect to the rest-frame.  

An indication of the strength of the outflow is provided by the shift amplitude  of the blue shifted component or by the shift amplitude of the total \civonly\ profile measured by the centroid at half intensity, \cmp. The most negative values indicate stronger blueshifts suggesting stronger outflows. BAL quasars tend to show smaller values of  \cmp\ (closer to 0) in the \civonly\ profile, due to the presence of the absorption features in the blue side of the profile. The non-BAL quasars show values 1000 $\lesssim \mid$ \cmp $\mid \lesssim$ 5000 \kms. \heii\ shows even  stronger asymmetries than \civ, although they could be related to the faintness of \heii\ and the deblending uncertain of broad and blue components. The strong outflows could be important for feedback effects on the host galaxy (Section \S\ref{sec:feed}).

\begin{table*}
\setlength{\tabcolsep}{1.3pt} 
\begin{center}
\caption{Measurements on C{\sc iv}$\lambda$1549 region \label{tab:c4}   }
\scriptsize
\begin{tabular}{c c c c c c c c c c c c c c}\\ 
\hline\hline\noalign{\vskip 0.1cm}         
\multirow{2}{*}{SDSS Identification} & \multirow{2}{*}{$f_{\lambda}$(1350\AA)} &
 \multicolumn{5}{c}{C{\sc iv} 1549$_{\mathrm{TOTAL}}$} & \multicolumn{2}{c}{C{\sc iv} 1549$\mathrm{_{BC}}$} & \multicolumn{4}{c}{C{\sc iv} 1549$_{\mathrm{BLUE}}$}   \\
   \cmidrule(lr){3-7} \cmidrule(lr){8-9} \cmidrule(lr){10-13}  
   \noalign{\vskip 0.04cm}         
 &   & $W$ & $F$ & FWHM & \cmp\ & A.I. & $W$ & $F$ & $W$ & $F$ & FWHM & \cmp\ \\
 (1) & (2) & (3) & (4) & (5) & (6) & (7) & (8) & (9) & (10) & (11) & (12) & (13)  \\
\hline \noalign{\vskip 0.07cm}

SDSSJ000807.27-103942.7	&	2.9	&	37.7	$\pm$6.6$^\mathrm{a}$	&	71.9	$\pm$10.4$^\mathrm{a}$	&	6200$\pm$420	&	-1130$\pm$210	&	-0.044$\pm$	0.103	&	27.4	$\pm$6.0$^\mathrm{a}$	&	51.8	$\pm$10.1$^\mathrm{a}$	&	10.3	$\pm$5.3$^\mathrm{a}$	&	20.1	$\pm$10.2$^\mathrm{a}$	&	4770$\pm$300	&	-2740$\pm$150	\\		
SDSSJ004241.95+002213.9	&	3.7	&	19.5	$\pm$	4.3	&	61.8	$\pm$	12.2	&	10080	$\pm$	340	&	-4530	$\pm$	170	&	-0.438	$\pm$	0.087	&	5.2	$\pm$	4.4	&	16.4	$\pm$	13.6	&	14.3	$\pm$	2.7	&	45.4	$\pm$	7.4	&	9550	$\pm$	250	&	-5170	$\pm$	130	\\		
SDSSJ021606.41+011509.5	&	3.5	&	19.1	$\pm$	2.6$^\mathrm{b}$	&	62.4	$\pm$	5.5$^\mathrm{b}$	&	4130	$\pm$	230	&	-850	$\pm$	110	&	-0.074	$\pm$	0.086	&	13.4	$\pm$	2.5	&	43.8	$\pm$	6.8	&	5.7	$\pm$	1.1$^\mathrm{b}$	&	18.6	$\pm$	3$^\mathrm{b}$	&	3020	$\pm$	160	&	-1760	$\pm$	80 \\			
SDSSJ024154.42-004757.5	&	3.1	&	10.0	$\pm$	1.8$^\mathrm{a}$	&	25.4	$\pm$	3.7$^\mathrm{a}$	&	6830	$\pm$	450	&	-2420	$\pm$	220	&	-0.375	$\pm$	0.073	&	3.9	$\pm$	1.3$^\mathrm{a}$	&	9.8	$\pm$	3.2$^\mathrm{a}$	&	6.1	$\pm$	1.9$^\mathrm{a}$	&	15.6	$\pm$	4.7$^\mathrm{a}$	&	6570	$\pm$	380	&	-3560	$\pm$	1909	\\	
SDSSJ084036.16+235524.7	&	3.1	&	14.4	$\pm$	2.2	&	38.1	$\pm$	4.5	&	6630	$\pm$	370	&	-1670	$\pm$	180	&	-0.378	$\pm$	0.070	&	9.9	$\pm$	2.4	&	26.6	$\pm$	5.9	&	4.5	$\pm$	2	&	11.5	$\pm$	4.8	&	4590	$\pm$	240	&	-3850	$\pm$	120	\\		
SDSSJ101822.96+203558.6	&	4.4	&	12.1	$\pm$	2.7$^\mathrm{a}$	&	47.5	$\pm$	9.5$^\mathrm{a}$	&	9580	$\pm$	390	&	-3990	$\pm$	200	&	-0.533	$\pm$	0.069	&	2.6	$\pm$	1.7$^\mathrm{a}$	&	10.0	$\pm$	6.6$^\mathrm{a}$	&	9.5	$\pm$	2.1$^\mathrm{a}$	&	37.5	$\pm$	7.4$^\mathrm{a}$	&	9770	$\pm$	320	&	-4520	$\pm$	160	\\		
SDSSJ103527.40+445435.6	&	3.0	&	12.2	$\pm$	4.2$^\mathrm{a}$	&	33.4	$\pm$	11.1$^\mathrm{a}$	&	9390	$\pm$	450	&	-3460	$\pm$	230	&	0.193	$\pm$	0.073	&	5.9	$\pm$	3.3	&	16.2	$\pm$	8.8	&	6.3	$\pm$	2.5$^\mathrm{a}$	&	17.2	$\pm$	6.7$^\mathrm{a}$	&	6350	$\pm$	320	&	-5050	$\pm$	160	\\		
SDSSJ105806.16+600826.9	&	4.6	&	11.4	$\pm$	2.2	&	39.3	$\pm$	6.4	&	6880	$\pm$	400	&	-2890	$\pm$	200	&	-0.405	$\pm$	0.076	&	1.5	$\pm$	1.4	&	5.0	$\pm$	4.8	&	9.9	$\pm$	1.9	&	34.3	$\pm$	5.4	&	6840	$\pm$	380	&	-3160	$\pm$	190	\\		
SDSSJ110022.53+484012.6	&	3.7	&	24.3	$\pm$	2.8	&	75.9	$\pm$	4.1	&	8010	$\pm$	540	&	-4430	$\pm$	270	&	-0.203	$\pm$	0.071	&	7.1	$\pm$	1.7	&	22.0	$\pm$	4.9	&	17.2	$\pm$	2	&	53.9	$\pm$	3.4	&	7140	$\pm$	410	&	-5120	$\pm$	210	\\			
SDSSJ125659.79-033813.8	&	2.8	&	47.4	$\pm$	9.9$^\mathrm{a}$	&	111.2	$\pm$	20.4$^\mathrm{a}$	&	5650	$\pm$	460	&	-2020	$\pm$	230	&	-0.186	$\pm$	0.091	&	25.8	$\pm$	11.8	&	60.5	$\pm$	27.1$^\mathrm{a}$	&	21.6	$\pm$	8.2$^\mathrm{a}$	&	50.7	$\pm$	18.5	&	5000	$\pm$	400	&	-2950	$\pm$	200	\\			
SDSSJ131132.92+052751.2	&	2.0	&	6.6	$\pm$	0.9$^\mathrm{a,b}$	&	17.6	$\pm$	1.8$^\mathrm{a,b}$	&	5700	$\pm$	320	&	-1180	$\pm$	160	&	-0.220  $\pm$	0.088	&	4.7	$\pm$	0.9$^\mathrm{a}$	&	12.5	$\pm$	2.2$^\mathrm{a}$	&	1.9	$\pm$	0.7$^\mathrm{b}$	&	5.1	$\pm$	1.7$^\mathrm{b}$	&	4320	$\pm$	220	&	-2710	$\pm$	110	\\			
SDSSJ143525.31+400112.2	&	10.3	&	9.9	$\pm$	1.3$^\mathrm{a}$	&	85.2	$\pm$	7.2$^\mathrm{a}$	&	8210	$\pm$	680	&	-3090	$\pm$	340	&	-0.214	$\pm$	0.074	&	3.7	$\pm$	1.2	&	31.6	$\pm$	9.5	&	6.2	$\pm$	1.0$^\mathrm{a}$	&	53.6	$\pm$	6.8$^\mathrm{a}$	&	6990	$\pm$	430	&	-4330	$\pm$	220	\\		
SDSSJ144412.37+582636.9	&	1.0	&	13.9	$\pm$	1.8$^\mathrm{a}$	&	15.0	$\pm$	1.2$^\mathrm{a}$	&	8570	$\pm$	450	&	-1990	$\pm$	230	&	0.020	$\pm$	0.085	&	9.0	$\pm$	2.0$^\mathrm{a}$	&	9.7	$\pm$	2$^\mathrm{a}$	&	4.9	$\pm$	1.7$^\mathrm{a}$	&	5.3	$\pm$	1.7$^\mathrm{a}$	&	6310	$\pm$	320	&	-3480	$\pm$	160	\\
SDSSJ151258.36+352533.2	&	4.7	&	8.8	$\pm$	1.4$^\mathrm{a}$	&	32.6	$\pm$	4.0$^\mathrm{a}$	&	5300	$\pm$	220	&	-1880	$\pm$	110	&	-0.022	$\pm$	0.090	&	3.3	$\pm$	1.1$^\mathrm{a}$	&	12.1	$\pm$	3.7$^\mathrm{a}$	&	5.5	$\pm$	1.2$^\mathrm{a}$	&	20.5	$\pm$	4.2$^\mathrm{a}$	&	3990	$\pm$	200	&	-2630	$\pm$	100	\\			
SDSSJ214009.01-064403.9	&	1.9	&	16.1	$\pm$	2.4$^\mathrm{b}$	&	29.7	$\pm$	3.4$^\mathrm{b}$	&	5260	$\pm$	770	&	-1170	$\pm$	380	&	-0.424	$\pm$	0.069	&	10.8	$\pm$	2.6$^\mathrm{b}$	&	20.0	$\pm$	4.3$^\mathrm{b}$	&	5.3	$\pm$	1.8$^\mathrm{b}$	&	9.7	$\pm$	3.2$^\mathrm{b}$	&	5350	$\pm$	270	&	-3770	$\pm$	140	\\			
SDSSJ220119.62-083911.6	&	5.5	&	18.4	$\pm$	3.8$^\mathrm{b}$	&	103.4	$\pm$	19.0$^\mathrm{b}$	&	4430	$\pm$	510	&	-830	$\pm$	250	&	0.002	$\pm$	0.092	&	15.2	$\pm$	4.9$^\mathrm{b}$	&	85.7	$\pm$	26.4$^\mathrm{b}$	&	3.2	$\pm$	2.7$^\mathrm{b}$	&	17.7	$\pm$	15$^\mathrm{b}$	&	2920	$\pm$	170	&	-1160	$\pm$	90	\\			
SDSSJ222753.07-092951.7	&	2.8	&	31.9	$\pm$	5.5	&	77.9	$\pm$	11	&	5260	$\pm$	410	&	-1190	$\pm$	200	&	-0.277	$\pm$	0.051	&	16.3	$\pm$	4.8	&	39.5	$\pm$	11.2	&	15.7	$\pm$	6.8	&	38.4	$\pm$	16.2	&	5820	$\pm$	420	&	-2340	$\pm$	210	\\		
SDSSJ233132.83+010620.9	&	8.3	&	21.4	$\pm$	4.4	&	174.2	$\pm$	35.8	&	11230	$\pm$	700	&	-2430	$\pm$	350	&	-0.333	$\pm$	0.097	&	16.1	$\pm$	5	&	132.0	$\pm$	39.1	&	5.3	$\pm$	1.6	&	42.2	$\pm$	12.4	&	6990	$\pm$	380	&	-6960	$\pm$	190	\\		
SDSSJ234657.25+145736.0	&	4.4	&	11.1	$\pm$	1.7$^\mathrm{a}$	&	41.5	$\pm$	4.6$^\mathrm{a}$	&	6390	$\pm$	800	&	-1420	$\pm$	400	&	-0.392	$\pm$	0.071	&	7.2	$\pm$	1.5$^\mathrm{a}$	&	26.9	$\pm$	4.9$^\mathrm{a}$	&	3.9	$\pm$	1.5$^\mathrm{a}$	&	14.6	$\pm$	5.3$^\mathrm{a}$	&	6670	$\pm$	340	&	-4750	$\pm$	170	\\		
\hline																												
\end{tabular}
\end{center}
\small {\sc Notes.}  Columns are as follows: (1) SDSS name. (2) Continuum flux measured at 1350 \AA. (3), (8) and (10) report the equivalent width in unit of \AA. (4), (9) and (11) list line fluxes in units of \uflux.  (5) and (12) list the FWHM of the corresponding line in \kms. (6) and (13) report the centroid at half intensity in units of \kms. (7) lists the asymmetry index (A.I.).  
$^\mathrm{a}$Measurement contaminated by narrow absorption lines.  $^\mathrm{b}$Measurement contaminated by broad absorption lines.
\end{table*} 

\begin{table*}
\setlength{\tabcolsep}{4pt} 
\begin{center}
\caption{Measurements on He{\sc ii}$\lambda$1640 region \label{tab:he2}   
}
\scriptsize
\begin{tabular}{c c c c c c c c c c c c c}\\ 
\hline\hline\noalign{\vskip 0.1cm}      
\multirow{2}{*}{SDSS Identification} 
    &  \multicolumn{2}{c}{He{\sc ii} 1640 $_{\mathrm{TOTAL}}$} & \multicolumn{2}{c}{He{\sc ii} 1640$\mathrm{_{BC}}$} & \multicolumn{2}{c}{He{\sc ii} 1640$_{\mathrm{BLUE}}$} & \multicolumn{1}{c}{Al{\sc ii} + O{\sc iii}] 1667} & \multicolumn{1}{c}{N{\sc iv}] 1486} \\
   \cmidrule(lr){2-3} \cmidrule(lr){4-5} \cmidrule(lr){6-7} \cmidrule(lr){8-8}   \cmidrule(lr){9-9} \noalign{\vskip 0.04cm}         
 & $W$ &$F$ & $W$ & $F$ & $W$ & $F$ &  $F$ & $F$  \\
(1) & (2) & (3) & (4) & (5) & (6) & (7) & (8) & (9) \\

\hline
\noalign{\vskip 0.07cm}

SDSSJ000807.27-103942.7	&	7.9	u	&	13.2	u	&	2.5	u	&	4.0	u	&	5.4	u	&	9.2	u	&	8.5	$\pm$	4.1	&	2.4	u$^\mathrm{a}$ \\							
SDSSJ004241.95+002213.9	&	2.7	$\pm$	2.4	&	8.3	$\pm$	7.2	&	1.0	u	&	3.1	u	&	1.7	$\pm$	1.2	&	5.2	$\pm$	3.6	&	2.7	u	&	7.8	u\\				
SDSSJ021606.41+011509.5	&	2.1	$\pm$	1.9	&	6.6	$\pm$	5.8	&	1.3:	&	4.1:	&	0.8:	&	2.5:	&	2.2	$\pm$	1.4	&	$\ldots$\\					
SDSSJ024154.42-004757.5	&	3.3	$\pm$	0.9	&	7.7	$\pm$	2	&	0.5	&	1.1:	&	2.8	$\pm$	0.9	&	6.6	$\pm$	2.1	&	4.5	u	&	5.5	$\pm$	3.5$^\mathrm{a}$\\				
SDSSJ084036.16+235524.7	&	2.3	$\pm$	1.6	&	5.6	$\pm$	3.7	&	1.2	$\pm$	1	&	3.0	$\pm$	2.5	&	1.1	$\pm$	1	&	2.6	u	2.2	&	2.5	$\pm$	1.9	&	3.7	u\\	
SDSSJ101822.96+203558.6	&	2.2	u	&	8.2	u	&	0.7	u	&	2.5	u	&	1.5	$\pm$	0.7	&	5.7	$\pm$	2.5	&	5.5	u	&	8	u\\						
SDSSJ103527.40+445435.6	&	2.5	u	&	6.5	$\pm$	2.3	&	1.6	u	&	4.2	u	&	0.9	$\pm$	0.4	&	2.3	$\pm$	1	&	5.1	$\pm$	2.6	&	16.2	u$^\mathrm{a}$	\\				
SDSSJ105806.16+600826.9	&	1.1	u	&	3.2	u	&	0.5:	&	1.5:	&	0.6	u	&	1.7	u	&	3	u	&	10	u\\								
SDSSJ110022.53+484012.6	&	2.8	$\pm$	1.0$^\mathrm{a}$	&	8.2	$\pm$	2.8$^\mathrm{a}$	&	1.0:$^\mathrm{a}$	&	2.9:$^\mathrm{a}$	&	1.8	$\pm$	0.9	&	5.3	$\pm$	2.6	&	1.7	u	&	$\ldots$\\				
SDSSJ125659.79-033813.8	&	7.0	u$^\mathrm{a}$	&	15.5	u$^\mathrm{a}$	&	2.7:	&	6.0:	&	4.3	u$^\mathrm{a}$	&	9.5	u$^\mathrm{a}$	&	5.5	u	&	$\ldots$\\								
SDSSJ131132.92+052751.2	&	4.0	$\pm$	1.0$^\mathrm{a}$	&	10.4	$\pm$	2.3$^\mathrm{a}$	&	1.2	$\pm$	0.9	&	3.1	$\pm$	2.3	&	2.8	$\pm$	1.1$^\mathrm{a}$	&	7.3	$\pm$	2.7$^\mathrm{a}$	&	6.7	$\pm$	2.5	u$^\mathrm{a}$	&	$\ldots$\\
SDSSJ143525.31+400112.2	&	2.3	$\pm$	0.9$^\mathrm{a}$	&	18.2	$\pm$	6.5$^\mathrm{a}$	&	1.3	$\pm$	0.8$^\mathrm{a}$	&	10.3	$\pm$	5.9$^\mathrm{a}$	&	1	$\pm$	0.5	&	7.9	$\pm$	3.7	&	9.2	$\pm$	4.1$^\mathrm{a}$	&	5.8	u$^\mathrm{a}$\\	
SDSSJ144412.37+582636.9	&	1.3	$\pm$	0.7	&	4.0	$\pm$	2.5	&	1.1	$\pm$	0.8	&	2.9	$\pm$	2.2	&	0.2:	&	1.1:	&	4.9	u	&	5.7	$\pm$	4.2	u$^\mathrm{a}$\\				
SDSSJ151258.36+352533.2	&	2.0	u$^\mathrm{a}$	&	6.6	u$^\mathrm{a}$	&	1	$\pm$	0.7$^\mathrm{a}$	&	3.2	$\pm$	2.3$^\mathrm{a}$	&	1	u$^\mathrm{a}$	&	3.4	u$^\mathrm{a}$	&	5.3	$\pm$	4.1	&	$\ldots$\\					
SDSSJ214009.01-064403.9	&	3.8	$\pm$	1.4	&	7.0	$\pm$	2.5	&	1.9	$\pm$	0.8	&	3.5	$\pm$	1.4	&	1.9	$\pm$	1.2	&	3.5	$\pm$	2.2	&	7.6	u	&	$\ldots$\\		
SDSSJ220119.62-083911.6	&	$\ldots$	&	$\ldots$	&	$\ldots$	&	$\ldots$	&	$\ldots$	&	$\ldots$	&	$\ldots$	&	$\ldots$\\															
SDSSJ222753.07-092951.7	&	5.7	$\pm$	2.6	&	13.1	$\pm$	5.7	&	1.4	$\pm$	1.3	&	3.2	$\pm$	2.9	&	4.3	$\pm$	2.7	&	9.9	$\pm$	6	&	2.7	u	&	6.8	u$^\mathrm{a}$\\		
SDSSJ233132.83+010620.9	&	6.3	$\pm$	2.9	&	46.9	$\pm$	21.4	&	2.6	$\pm$	&	21.0	u	&	3.7	$\pm$	1.8	&	25.9	$\pm$	12	&	17.8	u	&	9.3	u\\				
SDSSJ234657.25+145736.0	&	1.7	$\pm$	1.1$^\mathrm{a}$	&	5.8	$\pm$	3.7$^\mathrm{a}$	&	1.7	$\pm$	1.1$^\mathrm{a}$	&	5.8	$\pm$	3.7$^\mathrm{a}$	&	$\ldots$	&	$\ldots$	&	7.5	$\pm$	3.6	&	6.7	u$^\mathrm{a}$\\					
\hline	

\end{tabular}
\end{center}
\small {\sc Notes.} Columns are as follows: (1) SDSS name. Equivalent width are listed in col. (2), (4) and (6) in units of \AA. Line fluxes are reported in cols. (3), (5), (7), (8) and (9) in units of \uflux.  $^\mathrm{a}$Measurement contaminated by narrow absorption lines. Two colons (:) indicate a high uncertainty in the measurement. Letter $u$ marks an upper limit to the measurement.
\end{table*} 

\subsection{\siiv+\oiv}
 
Table \ref{tab:si4} reports the values of the emission lines in the \siiv+\oiv\ spectral region (1400\AA\ blend). The blue side of the 1400\AA\ spectral region is seriously affected by absorption lines. The \siiv+\oiv\ blend is formed by high-ionization lines. A blueshift asymmetry is expected in the profile, but in $\sim$40$\%$ of the sample the blueshifted component can not be detected due to strong absorption lines. In the rest of the sample, the measurements of the \cmp\ and FWHM are also affected to the ever-present absorptions. \cii\ and O{\sc i}+S{\sc ii}$\lambda$1304 are also affected by the absorption. {In the majority of the sample, they cannot be detected or their flux uncertainty is large.}


\begin{table*}
\setlength{\tabcolsep}{2.5pt} 
\begin{center}
\caption{Measurements on 1400\AA\ blend region \label{tab:si4}   
}
\scriptsize
\begin{tabular}{c c c c c c c c c c c c c c c c c c c c}\\ 
\hline\hline\noalign{\vskip 0.1cm}      
\multirow{2}{*}{SDSS Identification} 
   &   \multicolumn{2}{c}{S{\sc iv} + O{\sc iv}] 1400 $_{\mathrm{TOTAL}}$} & \multicolumn{2}{c}{S{\sc iv} + O{\sc iv}] 1400 $\mathrm{_{BC}}$} & \multicolumn{2}{c}{S{\sc iv} + O{\sc iv}] 1400 $_{\mathrm{BLUE}}$} & \multicolumn{1}{c}{C{\sc ii} 1332} & \multicolumn{1}{c}{O{\sc i} + S{ \sc ii} 1304}\\
   \cmidrule(lr){2-3} \cmidrule(lr){4-5} \cmidrule(lr){6-7} \cmidrule(lr){8-8} \cmidrule(lr){9-9} \noalign{\vskip 0.04cm}         
 & $W$ & $F$ & $W$ & $F$ & $W$ & $F$ & $F$ & $F$ \\
 (1) & (2) & (3) & (4) & (5) & (6) & (7) & (8) & (9)  \\
\hline
\noalign{\vskip 0.07cm}

SDSSJ000807.27-103942.7	&	15.9	$\pm$	2.1$^\mathrm{a}$	&	40.0	$\pm$	3.3$^\mathrm{a}$	&	15.9	$\pm$	2.1$^\mathrm{a}$	&	40.0	$\pm$	3.3$^\mathrm{a}$	&	$\ldots$	&	$\ldots$	&	4.8	$\pm$	3.6$^\mathrm{a}$	&	1.9	$\pm$	1.1$^\mathrm{a}$	\\				
SDSSJ004241.95+002213.9	&	11.1	$\pm$	1.8$^\mathrm{a}$	&	38.0	$\pm$	5.0$^\mathrm{a}$	&	7.7	$\pm$	1.9$^\mathrm{a}$	&	26.3	$\pm$	5.9$^\mathrm{a}$	&	3.4	$\pm$	0.9$^\mathrm{a}$	&	11.7	$\pm$	2.8$^\mathrm{a}$	&	$\ldots$	&	$\ldots$	\\				
SDSSJ021606.41+011509.5	&	19.5	$\pm$	2.2$^\mathrm{b}$	&	67.6	$\pm$	3.8$^\mathrm{b}$	&	19.5	$\pm$	2.2$^\mathrm{b}$	&	67.6	$\pm$	3.8$^\mathrm{b}$	&	$\ldots$	&	$\ldots$	&	11.7	$\pm$	5.1	&	10	$\pm$	3.8	\\				
SDSSJ024154.42-004757.5	&	9.0	$\pm$	1.5$^\mathrm{a}$	&	25.9	$\pm$	3.4$^\mathrm{a}$	&	6.4	$\pm$	1.5$^\mathrm{a}$	&	18.4	$\pm$	4$^\mathrm{a}$	&	2.6	$\pm$	1.2	&	7.5	$\pm$	3.4	&	7.9	$\pm$	2.7$^\mathrm{a}$	&	8.3	$\pm$	2.1$^\mathrm{a}$	\\
SDSSJ084036.16+235524.7	&	7.1	$\pm$	1	&	21.0	$\pm$	2.1	&	7.1	$\pm$	1	&	21.0	$\pm$	2.1	&	$\ldots$	&	$\ldots$	&	5.4	$\pm$	2.8$^\mathrm{a}$	&	3.9	$\pm$	1.8	\\				
SDSSJ101822.96+203558.6	&	9.1	$\pm$	1.3$^\mathrm{a}$	&	39.3	$\pm$	3.8$^\mathrm{a}$	&	5.3	$\pm$	1.3$^\mathrm{a}$	&	23.0	$\pm$	5$^\mathrm{a}$	&	3.8	$\pm$	0.7$^\mathrm{a}$	&	16.3	$\pm$	2.7$^\mathrm{a}$	&	5.3	$\pm$	2.1$^\mathrm{a}$	&	11.4	$\pm$	3.5$^\mathrm{a}$	\\
SDSSJ103527.40+445435.6	&	9.1	$\pm$	1.7$^\mathrm{a}$	&	26.4	$\pm$	4.2$^\mathrm{a}$	&	6.8	$\pm$	2.1$^\mathrm{a}$	&	19.8	$\pm$	5.7$^\mathrm{a}$	&	2.3	$\pm$	0.8$^\mathrm{a}$	&	6.6	$\pm$	2.2$^\mathrm{a}$	&	{2.6	u}	&	4 u	\\		
SDSS105806.16+600826.9	&	12.1	$\pm$	1.8	&	49.7	$\pm$	5.4	&	8	$\pm$	1.7	&	33.0	$\pm$	6	&	4.1	$\pm$	1.2	&	16.7	$\pm$	4.7	&	14.3	$\pm$	4.7	&	8	$\pm$	2.1	\\
SDSSJ110022.53+484012.6	&	12.8	$\pm$	1.7	&	41.8	$\pm$	3.8	&	5.7	$\pm$	1.1	&	22.1	$\pm$	3.6	&	7.1	$\pm$	1.3	&	19.7	$\pm$	3	&	12.8	$\pm$	6.1	&	28	$\pm$	11.9	\\
SDSSJ125659.79-033813.8	&	21.0	$\pm$	3.3$^\mathrm{a}$	&	55.2	$\pm$	6.8$^\mathrm{a}$	&	15.8	$\pm$	10.0$^\mathrm{a}$	&	41.7	$\pm$	26.1$^\mathrm{a}$	&	5.2	$\pm$	2.9	&	13.5	$\pm$	7.3	&	14.9	$\pm$	10.8$^\mathrm{a}$	&	27.6	$\pm$	11.1$^\mathrm{a}$	\\
SDSSJ131132.92+052751.2	&  5.7:$^\mathrm{b}$ &	10.7:$^\mathrm{b}$	&	5.7:$^\mathrm{b}$	&	10.7:$^\mathrm{b}$	&	$\ldots$	&	$\ldots$	&	$\ldots$	&	$\ldots$	\\													
SDSSJ143525.31+400112.2	&	7.2	$\pm$	1.9$^\mathrm{a}$	&	70.2	$\pm$	17.4$^\mathrm{a}$	&	5.4	$\pm$	2.4$^\mathrm{a}$	&	52.2	$\pm$	22.9$^\mathrm{a}$	&	1.8	$\pm$	1.7$^\mathrm{a}$	&	18.0	$\pm$	16.7$^\mathrm{a}$	&	14.2	$\pm$	10.2	&	23.4	$\pm$	18.6$^\mathrm{a}$	\\
SDSSJ144412.37+582636.9	&	12.7	$\pm$	3.7$^\mathrm{b}$	&	14.7	$\pm$	4.0$^\mathrm{b}$	&	12.7	$\pm$	3.7$^\mathrm{b}$	&	14.7	$\pm$	4$^\mathrm{b}$	&	$\ldots$	&	$\ldots$	&	$\ldots$	&	$\ldots$	\\
SDSSJ151258.36+352533.2	&	6.4	$\pm$	1.2$^\mathrm{a}$	&	28.5	$\pm$	4.7$^\mathrm{a}$	&	4.9	$\pm$	1.5$^\mathrm{a}$	&	21.8	$\pm$	6.2$^\mathrm{a}$	&	1.5	$\pm$	0.7$^\mathrm{a}$	&	6.7	$\pm$	3.2$^\mathrm{a}$	&	$\ldots$	&	12.4	$\pm$	4.3$^\mathrm{a}$	\\		
SDSSJ214009.01-064403.9	&	20.5	$\pm$	2.9$^\mathrm{b}$	&	39.1	$\pm$	3.8$^\mathrm{b}$	&	18	$\pm$	3.2	&	34.3	$\pm$	5.1	&	2.5	$\pm$	1.8$^\mathrm{b}$	&	4.8	$\pm$	3.4$^\mathrm{b}$   &	3.7	$\pm$	2.1$^\mathrm{b}$	&	8.7	$\pm$	2.8	\\	
SDSSJ220119.62-083911.6	&	13.4	$\pm$	6.6$^\mathrm{b}$	&	71.7	$\pm$	34.4$^\mathrm{b}$	&	13.4	$\pm$	6.6$^\mathrm{b}$	&	71.7	$\pm$	34.4$^\mathrm{b}$	&	$\ldots$	&	$\ldots$	&	$\ldots$	&	$\ldots$	\\		
SDSSJ222753.07-092951.7	&	10.3	$\pm$	1.9$^\mathrm{a}$	&	28.2	$\pm$	4.4$^\mathrm{a}$	&	10.3	$\pm$	1.9$^\mathrm{a}$	&	28.2	$\pm$	4.4$^\mathrm{a}$	&	$\ldots$	&	$\ldots$	&	$\ldots$	&	$\ldots$	\\
SDSSJ233132.83+010620.9	&	$\ldots$	&	$\ldots$	&	$\ldots$	&	$\ldots$	&	$\ldots$	&	$\ldots$	&	$\ldots$	&	$\ldots$	\\
SDSSJ234657.25+145736.0	&	10.3	$\pm$	1.6$^\mathrm{a}$	&	41.7	$\pm$	4.8$^\mathrm{a}$	&	9.8	$\pm$	1.4$^\mathrm{a}$	&	39.7	$\pm$	3.9$^\mathrm{a}$	&	0.5	:$^\mathrm{a}$	&	2.0	:$^\mathrm{a}$	&	20.5	$\pm$	4.1$^\mathrm{a}$	&	$\ldots$	\\				
\hline	
\end{tabular}
\end{center}
\small {\sc Notes.}  Columns are as follows: (1) SDSS name. Cols. (3), (5), (7), (8) and (9)  list line fluxes in units of \uflux. Cols. (2), (4) and (6) list the equivalent width in \AA.  $^\mathrm{a}$Measurement contaminated by narrow absorption lines. $^\mathrm{b}$Measurement contaminated by broad absorption lines. Two colons (:) indicates a high uncertainty in the measurement. Letter $u$ marks an upper limit to the measurement. 
 
\end{table*}

\subsection{The \feii\ and \feiii\ emission} 
\label{sec:feii-iii}

{\feii\ emission is usually not very strong in the range 1700--2100 \AA, even in the case of the strongest optical \feii\ emitters. Only a few features are detected at $\sim$ 1715\,\AA, 1785\,\AA\ and 2020 \AA. We modeled these individual features with single Gaussians, without using a  model template emission. On the other hand, the \feiii\ emission has a most important role in xA sources. It was modeled using the scaled and broadened template of \citet{vestergaardwilkes01} based on I Zw 1. However, it is not enough to reproduce the observed \feiii\ emission. We add two extra {emission component at 1914 and 2080 \AA\ (modeled by two Gaussians), to account for the line of \feiii\ UV \#34 line that may be enhanced by \lya\ fluorescence, and for the broad hump at 2080 \AA, mainly ascribed to the blended features of \feiii\ UV \#Ê48.}   Most relevant \feii\ and \feiii\ features between 1700 and 2100\,\AA\ are individually discussed Appendix \ref{sec:feident}.}

{Table \ref{tab:fe2} reports the equivalent widths and the fluxes of the most important UV \feii\ multiplet \#191 (\feii\,1785\,\AA) and the ones of the total \feii\ emission, i.e., the sum of the \feii\ features described above: at 1715, 1785 and 2020\,\AA. Table \ref{tab:fe3} reports equivalent widths and fluxes of the most important individual  features ascribed to \feiii\ (1914 and 2080\,\AA) and the total \feiii\ emission: the sum of the individual features and the template contribution in the full  range. It is interesting to note that \feiii\ emission is almost always stronger than \feii, by a factor $\sim$2--3. The   strong \feiii\ emission will be discussed in Sect.\,\S \ref{sec:fluo} and Appendix \ref{sec:origin}.}  
	

\begin{table}
\setlength{\tabcolsep}{4pt} 
\begin{center}
\caption{Flux and Equivalent width of Fe{\sc ii} contribution \label{tab:fe2}   
}
\scriptsize
\begin{tabular}{c c c c c}\\ 
\hline\hline\noalign{\vskip 0.1cm}   
\multirow{2}{*}{SDSS Identification} 
 &  \multicolumn{2}{c}{Fe{\sc ii} $\lambda$1785} & \multicolumn{2}{c}{Fe{\sc ii} $_{\mathrm{TOTAL}}$} \\
   \cmidrule(lr){2-3} \cmidrule(lr){4-5} \noalign{\vskip 0.04cm}         
 & $W$& $F$ & $W$ & $F$ \\
 (1) & (2) & (3) & (4) & (5) \\
\hline
\\[-0.25cm]
\noalign{\vskip 0.07cm}
SDSSJ000807.27-103942.7	&	3.5	$\pm$	0.7	&	4.4	$\pm$	0.7	&	11.2	$\pm$	2.8	&	14.0	$\pm$	3.2	\\
SDSSJ004241.95+002213.9	&	2.1	$\pm$	0.6	&	5.7	$\pm$	1.4	&	6.2	$\pm$	2.0	&	16.7	$\pm$	5.2	\\
SDSSJ021606.41+011509.5	&	2.5	$\pm$	0.5	&	7.0	$\pm$	1.1	&	9.7	$\pm$	2.0	&	29.3	$\pm$	5.1	\\
SDSSJ024154.42-004757.5	&	1.5	$\pm$	0.2	&	3.1	$\pm$	0.4	&	6.5	$\pm$	1.1	&	12.4	$\pm$	1.8	\\
SDSSJ084036.16+235524.7	&	1.1	$\pm$	0.3	&	2.2	$\pm$	0.6	&	7.9	$\pm$	1.9	&	16.0	$\pm$	3.5	\\
SDSSJ101822.96+203558.6	&	1.7	$\pm$	0.6	&	5.5	$\pm$	1.8	&	6.8	$\pm$	1.9	&	21.0	$\pm$	5.5	\\
SDSSJ103527.40+445435.6	&	2.7	$\pm$	0.6	&	6.6	$\pm$	1.2	&	4.8	$\pm$	1.1	&	12.1	$\pm$	2.5	\\
SDSS105806.16+600826.9	&	0.4:	&	1.0:&	2.5	$\pm$	0.8	&	7.0	$\pm$	2.2 \\
SDSSJ110022.53+484012.6	&	1.3	$\pm$	0.4	&	3.2	$\pm$	1.0	&	3.4	$\pm$	1.2	&	8.7	$\pm$	3.1	\\
SDSSJ125659.79-033813.8	&	$\ldots$			&	$\ldots$			&	9.9	$\pm$	4.3	&	16.0	$\pm$	6.7	\\
SDSSJ131132.92+052751.2	&	1.0	$\pm$	0.4	&	2.9	$\pm$	1.0	&	2.3	$\pm$	0.7	&	6.4	$\pm$	1.9	\\
SDSSJ143525.31+400112.2	&	1.1	$\pm$	0.3	&	7.4	$\pm$	1.6	&	3.8	$\pm$	1.0	&	23.6	$\pm$	5.4	\\
SDSSJ144412.37+582636.9	&	$\ldots$			&	$\ldots$			&	5.5	$\pm$	2.2	&	5.0	$\pm$	1.9	\\
SDSSJ151258.36+352533.2	&	1.9	$\pm$	0.4	&	5.3	$\pm$	1.0	&	3.8	$\pm$	0.8	&	11.0	$\pm$	1.9	\\
SDSSJ214009.01-064403.9	&	0.7	$\pm$	0.3	&	1.2	$\pm$	0.5	&	6.5	$\pm$	1.5	&	11.7	$\pm$	2.4	\\
SDSSJ220119.62-083911.6	&	$\ldots$			&	$\ldots$			&	1.1	$\pm$	0.5	&	3.8	$\pm$	1.8	\\
SDSSJ222753.07-092951.7	&	0.2	$\pm$	0.1	&	0.4	$\pm$	0.2	&	5.3	$\pm$	1.3	&	10.5	$\pm$	2.3	\\
SDSSJ233132.83+010620.9	&	1.3	$\pm$	0.9	&	9.6	$\pm$	6.7	&	5.5	$\pm$	2.9	&	35.5	$\pm$	18.0	\\
SDSSJ234657.25+145736.0	&	$\ldots$			&	$\ldots$			&	1.3	$\pm$	0.5	&	4.5	$\pm$	1.5	\\
\hline

\end{tabular}
\end{center}
\small {\sc Notes.}  Columns are as follows: (1) SDSS name.  (2) and (4) report the equivalent width in units of \AA. (3) and (5) correspond to the flux in units of \uflux.\\
\end{table}

\begin{table*}
\setlength{\tabcolsep}{3pt} 
\begin{center}
\caption{Flux and Equivalent width of Fe{\sc iii} contribution \label{tab:fe3}   
}
\scriptsize
\begin{tabular}{c c c c c c c c  }\\ 
\hline\hline\noalign{\vskip 0.1cm}   
\multirow{2}{*}{SDSS Identification} 
 & \multicolumn{2}{c}{Fe {\sc iii} $\lambda$1914} & \multicolumn{2}{c}{Fe {\sc iii} $\lambda$2080} & \multicolumn{2}{c}{Fe {\sc iii} $_{\mathrm{TOTAL}}$} \\
   \cmidrule(lr){2-3} \cmidrule(lr){4-5} \cmidrule(lr){6-7} \noalign{\vskip 0.04cm}         
  & {$W$} & $F$ & $W$ & $F$ & $W$ & $F$ \\
 (1) & (2) & (3) & (4) & (5) & (6) & (7) \\
\hline
\noalign{\vskip 0.07cm}

SDSSJ000807.27-103942.7	&	1.9	$\pm$	0.7	&	1.9	$\pm$	0.7	&	7.5	$\pm$	0.8	&	5.9	$\pm$	0.6	&	9.9	$\pm$	2.0	&	17.1	$\pm$	3.0	\\
SDSSJ004241.95+002213.9	&	2.9	$\pm$	0.8	&	7.1	$\pm$	2.0	&	6.4	$\pm$	1.1	&	13.8	$\pm$	2.5	&	4.5	$\pm$	1.3	&	31.6	$\pm$	8.8	\\
SDSSJ021606.41+011509.5	&	5.8	$\pm$	0.9	&	16.2	$\pm$	2.4	&	6.6	$\pm$	0.7	&	17.7	$\pm$	1.9	&	25.1	$\pm$	3.4	&	103.9	$\pm$	9.8	\\
SDSSJ024154.42-004757.5	&	2.3	$\pm$	1.0	&	3.8	$\pm$	1.7	&	0.1	u	&	0.2	u	&	12.6	$\pm$	3.9	&	24.6	$\pm$	7.2	\\
SDSSJ084036.16+235524.7	&	3.4	$\pm$	1.8	&	6.1	$\pm$	3.1	&	$\ldots$		&	$\ldots$		&	23.0	$\pm$	3.2	&	44.5	$\pm$	4.3	\\
SDSSJ101822.96+203558.6	&	4.7	$\pm$	0.9	&	13.7	$\pm$	2.7	&	6.8	$\pm$	1.3	&	17.2	$\pm$	3.2	&	11.8	$\pm$	2.7	&	63.9	$\pm$	13.1	\\
SDSSJ103527.40+445435.6	&	5.8	$\pm$	0.7	&	13.1	$\pm$	1.7	&	9.0	$\pm$	0.8	&	18.3	$\pm$	1.6	&	3.1	$\pm$	0.6	&	38.2	$\pm$	6.7	\\
SDSS105806.16+600826.9	&	3.5	$\pm$	0.5	&	8.0	$\pm$	1.1	&	1.8	$\pm$	0.9	&	3.8	$\pm$	1.8	&	12.9	$\pm$	2.7	&	40.6	$\pm$	7.3	\\
SDSSJ110022.53+484012.6	&	0.9	$\pm$	0.6	&	2.0	$\pm$	1.4	&	3.2	$\pm$	1.0	&	6.1	$\pm$	1.9	&	5.6	$\pm$	2.5	&	20.0	$\pm$	8.6	\\
SDSSJ125659.79-033813.8	&	2.6	$\pm$	2.3	&	4.3	$\pm$	3.8	&	6.5	$\pm$	2.7	&	9.0	$\pm$	3.8	&	7.8	$\pm$	6.0	&	25.7	$\pm$	19.5	\\
SDSSJ131132.92+052751.2	&	3.1	$\pm$	0.4	&	8.6	$\pm$	1.0	&	3.1	$\pm$	0.5	&	8.1	$\pm$	1.2	&	4.4	$\pm$	0.9	&	28.7	$\pm$	5.0	\\
SDSSJ143525.31+400112.2	&	3.2	$\pm$	0.7	&	19	$\pm$	4.4	&	2.1	$\pm$	0.6	&	11.0	$\pm$	3.2	&	8.7	$\pm$	2.0	&	79.4	$\pm$	16.1	\\
SDSSJ144412.37+582636.9	&	3.6	$\pm$	1.3	&	3.2	$\pm$	1.1	&	4.0	$\pm$	2.1	&	3.2	$\pm$	1.7	&	4.3	$\pm$	3.3	&	10.2	$\pm$	7.9	\\
SDSSJ151258.36+352533.2	&	3.0	$\pm$	0.6	&	7.3	$\pm$	1.5	&	2.1	$\pm$	0.6	&	4.5	$\pm$	1.3	&	6.7	$\pm$	1.7	&	27.2	$\pm$	6.3	\\
SDSSJ214009.01-064403.9	&	3.7	$\pm$	0.8	&	6.5	$\pm$	1.4	&	5.3	$\pm$	0.8	&	9.0	$\pm$	1.4	&	4.3	$\pm$	1.2	&	21.9	$\pm$	5.9	\\
SDSSJ220119.62-083911.6	&	2.0	$\pm$	0.6	&	7.6	$\pm$	2.5	&	1.1	$\pm$	0.7	&	3.6	$\pm$	2.5	&	8.0	$\pm$	2.6	&	41.2	$\pm$	13.0	\\
SDSSJ222753.07-092951.7	&	1.0	$\pm$	0.5	&	1.8	$\pm$	0.9	&	7.1	$\pm$	1.1	&	10.4	$\pm$	1.6	&	8.0	$\pm$	2.1	&	25.4	$\pm$	6.1	\\
SDSSJ233132.83+010620.9	&	1.3	$\pm$	0.3	&	8.7	$\pm$	2.3	&	3.3	$\pm$	0.8	&	20.5	$\pm$	4.9	&	4.4	$\pm$	2.2	&	57.8	$\pm$	28.8	\\
SDSSJ234657.25+145736.0	&	2.1	$\pm$	0.3	&	6.0	$\pm$	1.0	&	1.1	$\pm$	0.8	&	2.7	$\pm$	2.1	&	8.6	$\pm$	2.4	&	32.9	$\pm$	8.7	\\

\hline	
\end{tabular}
\end{center}
\small {\sc Notes.}  Columns are as follows: (1) SDSS name.  Cols. (2), (4) and (6) list the equivalent width in units of \AA. Cols. (3), (5) and (7) correspond to the flux in units of \uflux.\\
\end{table*}

\begin{figure}[!ht]			
\begin{center}
\includegraphics[scale=0.35]{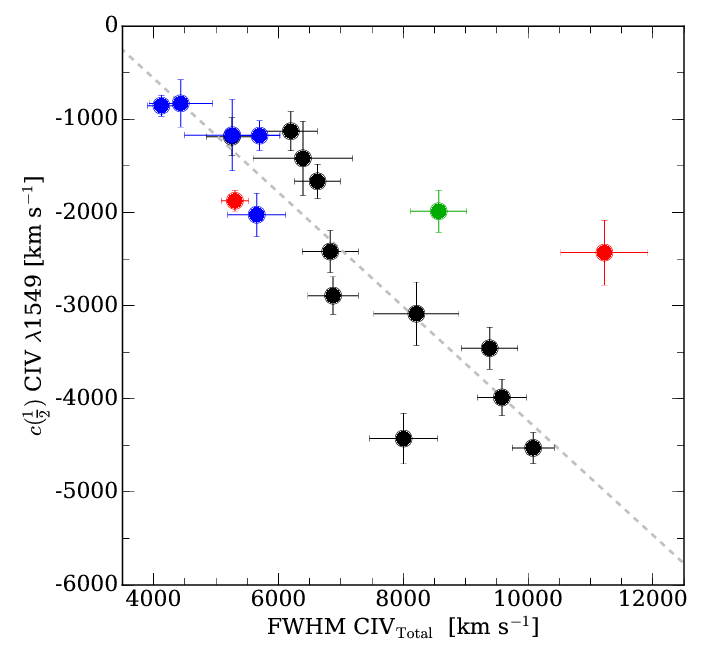}
\end{center}
\caption{Relation between the \cmp\ and the FWHM of the total \civ\ emission. Dashed gray line corresponds to the best fit using an orthogonal method. Black, blue, green and red filled circles  correspond respectively to normal, BAL, mini--BAL and RL quasars for the GTC-xA sample.
\label{fig:c12fwhm}
}
\end{figure}

\subsection{The origin of the \civ\ line broadening}
\label{sec:c12fwhm}

The GTC-xA sample shows a very tight correlation between FWHM and \cmp\ of the total \civ\ profile (Fig. \ref{fig:c12fwhm}).  The best fitting to our new sample is {\cmp $\approx (-0.613\pm0.081) \cdot$ FWHM $+(1893\pm545)$ \kms}. This relation is consistent within uncertainties with the one found  by \citetalias{sulenticetal17} involving Pop. A sources. The range of blueshifts is from $\approx$ -1000 to -5000 \kms\ in both cases; however, the distribution of the GTC-xA is weighted in favor of larger shifts and broader FWHM with respect to the one of \citetalias{sulenticetal17}. As noted by \citetalias{sulenticetal17}, the correlation of Fig. \ref{fig:c12fwhm} justify the decomposition of \civ\ profile into the BC and a blueshifted excess, which becomes the dominant source of broadening in the case of the largest FWHM.

\subsection{\aliii\ as a High Redshift Virial Estimator}
\label{sec:cival}
  
Identification of a UV virial estimator for high redshift quasars ($z\gtrsim$ 2) has been an important quest for several years \citep[e.g., review by][]{marzianietal17e}. Once \hb\ is lost to ground based spectroscopy, \mgii\ can be exploited up to z$\sim$2.0 \citep{trakhtenbrotnetzer12, shenliu12, marzianietal13}.  \civ\ is the obvious candidate for quasars beyond $z\sim \ $2.0, but the \civonly\ line profiles often show large blue shifts/asymmetries whose amplitude is a function of source location along the 4DE1 sequence and also of luminosity \citepalias{sulenticetal17}. This implies that \mbh\ estimation using FWHM  \civonly\ can be uncertain by a large factor  \citep{sulenticetal07,netzeretal07,mejia-restrepoetal18}. There have been several recent attempts to account for the bias associated with the non-virial broadening of \civ\  \citep{parketal13,matsuokaetal13,brothertonetal15,denneyetal16,coatmanetal17}, however no correction yields safe \mbh\ estimates especially if Population A and B sources are not considered separately.  We focus here on extreme Pop. A (xA) sources and attempt to use  \aliii\ as an \hb\ virial surrogate. The first composites of low-$z$ FOS sources suggested that \aliii\ becomes more prominent at the extreme Pop. A end of the 4DE1 sequence \citep{bachevetal04}.  The \aliii\ line shows a FWHM similar to the one of \hb, which suggests that both lines are emitted in similar dynamical condition \citep{negreteetal13}. Analysis of VLT ISAAC +  FORS  spectra of high-$z$\ Pop. A sources \citep{marzianietal17e} show also that FWHM \aliii\ agrees well with FWHM \hb. Previously, the emission lines associated with the 1900\AA\ blend such as \aliii\ or \ciii\ were avoided due to severe blending  \citep{shenliu12}, and  the 1900\AA\ blend remains poorly studied to date. Our earlier work motivated us to identify xA source candidates and to obtain high S/N spectra of \civ\ and the 1900\AA\ blend.  In the majority of the xA candidates observed so far the \aliii\ profile is well defined and is not strongly affected by absorption features (See Section \S\ref{sec:blend}). 

\begin{figure*}[!ht]
\begin{center}
\includegraphics[width=8.5cm]{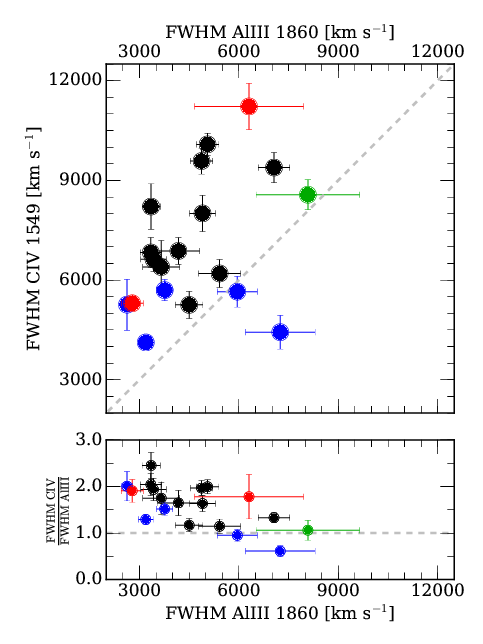}
\includegraphics[width=8.5cm]{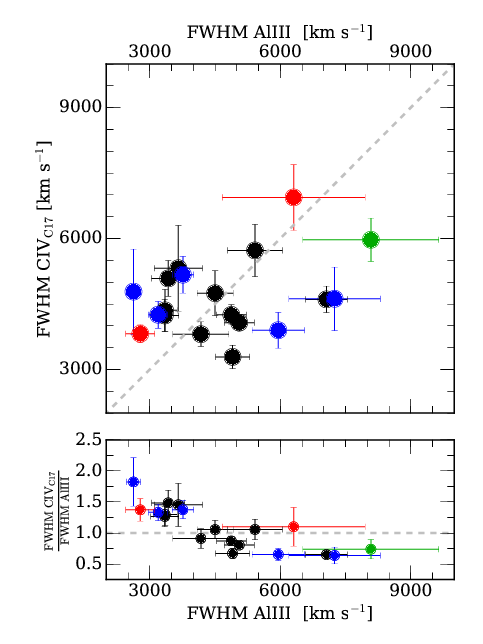}
\end{center}
\caption{Relation between the FWHM \aliii\ and FWHM \civ\ before (top) and after (bottom) \citet{coatmanetal17} correction. Dashed gray line corresponds to the 1:1 line. The bottom panels show the ratio between the FWHM \civ\ and \aliii. \label{fig:fwhm}. The colors are the same as the Figure \ref{fig:critsel} and \ref{fig:c12fwhm}.}
\end{figure*}

The left panel of  Figure \ref{fig:fwhm} shows a comparison between FWHM \civ\ and FWHM \aliii\ for the xA sample. The ratio between both FWHMs as a function of the FWHM \aliii\ is shown in the bottom part of the panel. \citet{coatmanetal17} proposed an empirical correction to the \civonly\ profile to obtain the contribution associated with the virial component. Its factor is based on a measure of the \civonly\ blueshift analogs to the \cmp\ we measure. We then applied a correction to the FWHM \civ\ proportional to \cmp\  \civ\ to correct for the excess broadening due to the blueshifted component.  The result  is shown in the right panel of Figure \ref{fig:fwhm}.  All \civ\ measurements have FWHM larger than FWHM \aliii\ before correction. Sources with FWHM \aliii\ $\approx 3000$ \kms\, show a bias factor of $\approx$ 2.5-3 in \civonly\ FWHM, implying almost an order-of-magnitude excess in \mbh\ estimates. After correction, even if the bias is almost zeroed, we find no significant improvement over the uncorrected panel. This is especially true when we examine the spectra of the 4-6 sources with largest differences with FWHM \aliii. \civonly, in most of them, is affected by absorptions.  Most of the highest confidence measures involve the cluster of sources in the lower left corner of the figure, which can be argued to be the highest confidence xA sources. However the sample is too small to allow us to go further. In a forthcoming paper comparing FWHM \hb\ and FWHM \aliii\ measures for our low and higher $z$\ samples (Marziani et al. 2018, in preparation),  we find a clear correlation for Pop. A sources. FWHM \aliii\ provides the safest equivalent of FWHM \hb\ for virial estimation in Pop. A sources.

\subsection{Prevalence and properties of absorbed systems}
\label{sec:abs}

The \civ\ emission line profile of our GTC-xA sources frequently shows  absorptions: only in 9 cases the profile is not contaminated by deep broad absorptions and in only 5 cases the profile is not affected by any narrow or broad absorption. In order to ascertain whether our sample suffers from unusually strong  line absorption compared to a general quasar population at $2.0 \lesssim z \lesssim 2.9$, we considered estimates of the absorption index (AI) and Balnicity index (BI) as defined in \citet{scaringietal09}. The AI \citep{halletal02} includes semibroad absorptions continuously detected over a range of at least 1000 \kms, from 0 to --29000 \kms; the BI \citep{weymannetal91} measures contiguous absorption features over 2000 \kms\ in the range of --25000 to --3000 \kms, and is  intended to identify quasars with broad through features almost certainly not associated with external absorbers. Table \ref{tab:abs} reports the AI and BI measures for our sample.
Large \civ\ BI absolute values ($\lesssim$ -1000 \kms) are associated with Pop.  A and with ``strong'' BALs with large equivalent widths and high up to terminal velocity, while mini-BALs are also found at low-$z$ among Pop. B and RL objects  \citep{sulenticetal06a}.  \citet{trumpetal06} provide both AI and BI measurements obtained from a SDSS DR3 catalog containing 5719 quasars in redshift range $z\sim$2 -- 2.6 \citep{schneideretal05}. There are 542 sources with \civ\ BI $<$0 suggesting  $\approx$ 9.5 \%\ for ``strong'' BAL sources, consistent with  conventionally quoted prevalence values for BAL QSOs.  In the GTC-xA sample, 5 over 19 sources show \civ\ BI$<0$. If we also include one source with \civonly\ BI=0  (\object{SDSSJ144412.37+582636.9}), but with strong absorption in the blue side of \siiv+\oiv\ blend at 1400 \AA\ and BI$\ll$0 (See Fig. \ref{fig:specSDSSJ144412.37+582636.9}), we obtain a prevalence of BAL QSOs around 25 -- 30\%. AI is less than 0 in 10 out of 19 sources of the GTC-xA sample; about 30\%\ of \citet{trumpetal06} satisfy the same condition. If one considers that about 1/2 of SDSS quasars are Pop. A quasars \citep{zamfiretal10}, the prevalence of strong BAL sources should be doubled to $\sim$ 20\%. We can therefore conclude that the GTC-xA sample does not show an unusually large fraction of intrinsically absorbed systems, possibly at a prevalence somewhat higher than the general quasar population.

\begin{table}
\setlength{\tabcolsep}{4pt} 
\begin{center}
\caption{Balnicity and Absorption indices \label{tab:abs}}%
\scriptsize
\begin{tabular}{l c c c }\\ 
\hline\hline\noalign{\vskip 0.1cm}       
  \multicolumn{1}{c}{SDSS Identification} & \multicolumn{1}{c}{BI}  & AI \\
   \multicolumn{1}{c}{(1)} & \multicolumn{1}{c}{(2)} & \multicolumn{1}{c}{(3)} \\ \hline \\[-0.2cm]
      SDSSJ000807.27-103942.7   &           0  &         0 \\ 
   SDSSJ004241.95+002213.9   &           0  &   0 \\ 
   SDSSJ021606.41+011509.5   &      -675  &  -2575 \\ 
   SDSSJ024154.42-004757.5   &           0  &    0 \\ 
   SDSSJ084036.16+235524.7   &           0  &   -210 \\ 
   SDSSJ101822.96+203558.6   &           0  &     0 \\ 
   SDSSJ103527.40+445435.6   &           0  &    -160 \\ 
   SDSS105806.16+600826.9   &           0  &      0 \\ 
   SDSSJ110022.53+484012.6   &           0  &    0 \\ 
   SDSSJ125659.79-033813.8   &      -720  &     -970 \\ 
   SDSSJ131132.92+052751.2   &       -2440  &   -5430 \\ 
   SDSSJ143525.31+400112.2   &           0  &     0 \\ 
   SDSSJ144412.37+582636.9   &           0  &    -770 \\ 
   SDSSJ151258.36+352533.2   &           0  &    -270 \\ 
   SDSSJ214009.01-064403.9   &       -2530  &    -3620 \\ 
   SDSSJ220119.62-083911.6   &       -5800 &    -7460 \\ 
   SDSSJ222753.07-092951.7   &           0  &     0 \\ 
   SDSSJ233132.83+010620.9   &           0  &     0\\
   SDSSJ234657.25+145736.0   &           0  &    -240  \\
\hline	
\end{tabular}
\end{center}
\small {\sc Notes.}  Columns are as follows: (1) SDSS name. (2) Balnicity index in units of \kms. (3) Absorption index in units of \kms. \\
\end{table}



\section{Discussion}

\begin{figure*}[!ht]
\begin{center}
\includegraphics[width=7cm]{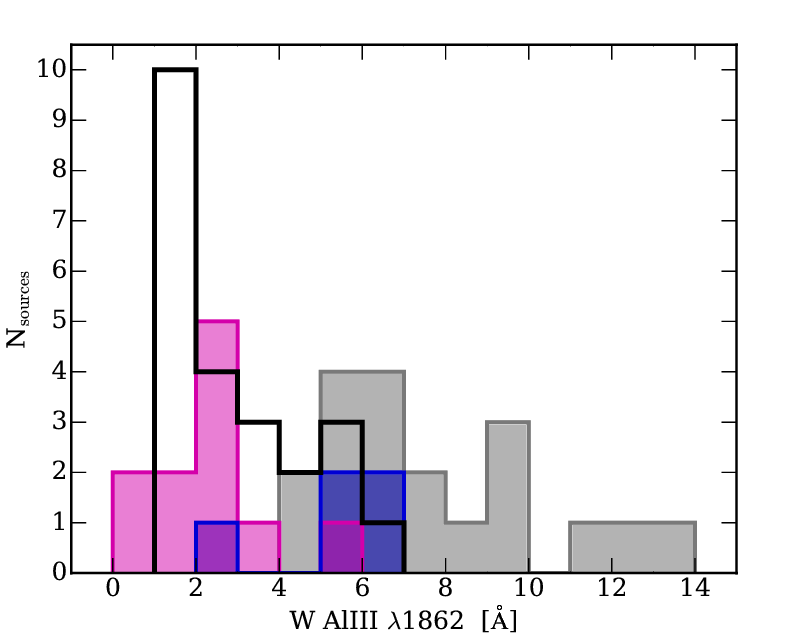}
\includegraphics[width=7cm]{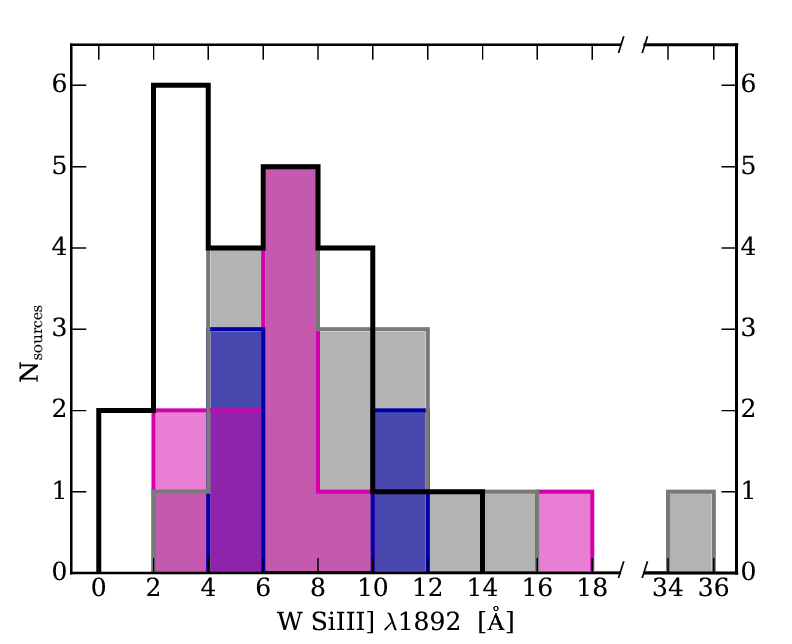}\\
\includegraphics[width=7cm]{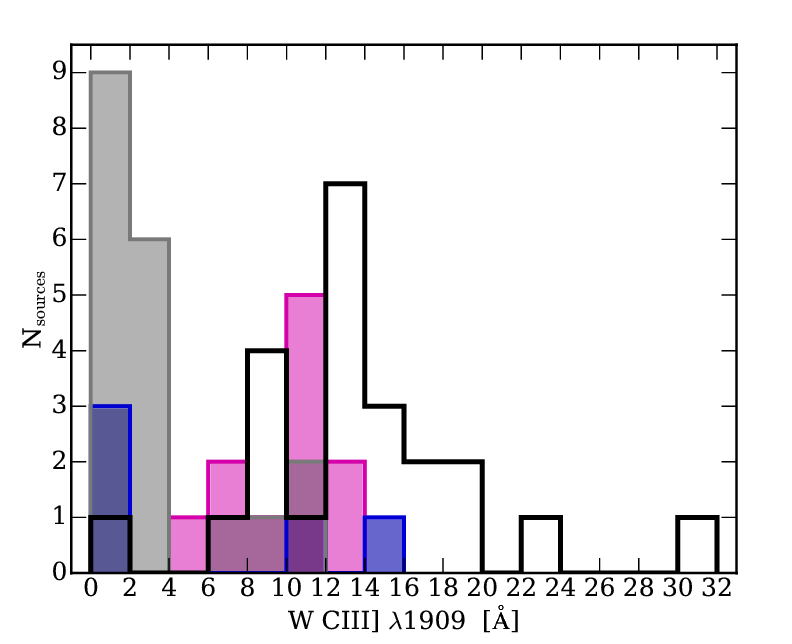}
\includegraphics[width=7cm]{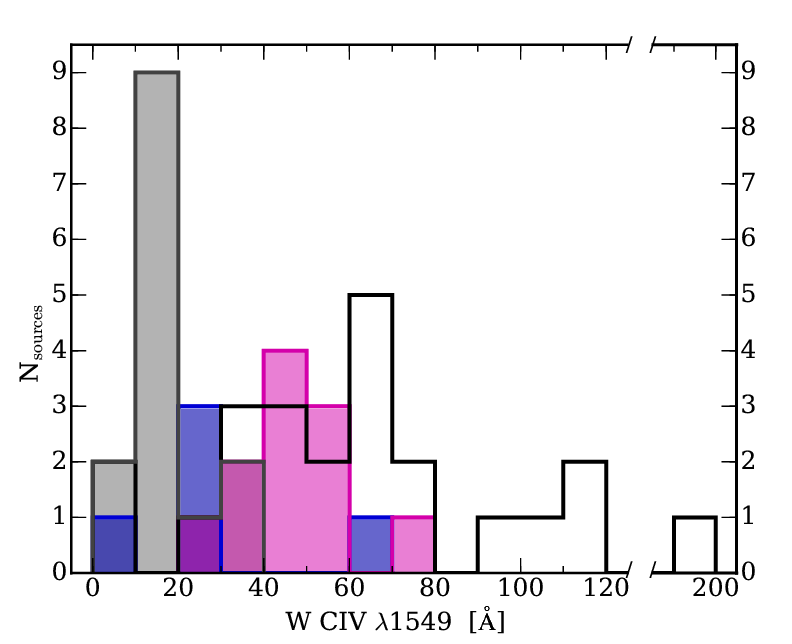}

\end{center}
\caption{Rest-frame equivalent width distributions of the \aliii, \siiii, \ciii\ and \civ\ for the GTC-xA (gray),  FOS-xA (blue), FOS-A (not filled) and S14-A (magenta) samples. \label{fig:ew}}
\end{figure*}

\subsection{Equivalent width distributions}
\label{sec:ewdis}

Fig. \ref{fig:ew} shows a comparison between the equivalent width distributions of \aliii, \siiii, \ciii\ and \civ\ for the GTC-xA sources and the three control samples: the low-$z$ (FOS-A and FOS-xA) and the high-$z$ (S14-A) samples. The \aliii\ line of the GTC-xA shows an enhancement relative to the other samples (upper left panel). The \aliiionly\ values distribution for the GTC-xA sources is significantly different more than 6$\sigma$ respect to the ones observed in FOS-A and S14-A, according to a T-test (6.7$\sigma$ and 6.1$\sigma$ respectively). In fact, a Kolmogorov-Smirnov (KS) test gives a very low probability of being equal population between  GTC-xA, and FOS-A and S14-A ($8\cdot10^{-6}$ and $5\cdot10^{-6}$ respectively). {Meanwhile, there is no statistically significant difference between the two Pop. A samples (FOS-A and S14-A) or the two Pop. xA (GTC-xA and FOS-xA) samples.}

The \siiii\ equivalent width (upper right panel) covers a wide range of values with no statistical significant difference between the samples. Only at a level of 2$\sigma$, the GTC-xA shows slightly higher values of $W$ \siiii\ than FOS-A sample. But neither the T-test nor the KS test yield a significant probability of being different populations.

Instead, in the GTC-xA sample the $W$ \ciiionly\ distribution (bottom left panel) appears to be markedly weaker than a Pop. A. Both T-test and KS-test give significant differences at 6.5$\sigma$ and 5.2$\sigma$ level for the GTC-xA sample to be a different population respect to FOS-A and S14-A respectively. The probability according to the KS test is very low ($3.2\cdot10^{-6}$ and $1\cdot10^{-4}$, respectively). In xA sources \ciii\ is weak and  blended with the \feiii\ contribution, to the point of not being clearly detectable in several cases. On the other hand, in the Pop. A samples (FOS-A and S14-A) this ion is perfectly appreciable by naked eye and specially prominent in Pop. B sources \citep[e.g., ][]{bachevetal04,kuraszkiewiczetal04}

Bottom right panel of Fig. \ref{fig:ew} shows a comparison between the equivalent width distributions of \civ\ for xA and Pop. A sources. xA values are the ones with the lowest $W$\ \civ\  among Pop. A sources, with a wide majority of sources distributed below $\sim 40$ \AA. On the converse, the FOS-A and S14-A samples show a much broader distribution, with no evidence of a systematic difference between them, within the limitations of the relatively small sample sizes. The statistical tests gives a significant difference between GTC-xA and FOS-A and S14-A at 5$\sigma$ and 7$\sigma$, and a KS-test a low probability ($\sim 10^{-5}$) to be the same population. 

These histograms give a first hint to the fact that xA quasars show a different spectroscopic behavior compared with the one shown by samples of the full Population A. Due to there are not any significant differences in the equivalent width distributions of two xA samples (GTC-xA and FOS-xA) and neither between the two Pop. A samples (FOS-A and S14-A), we can consider a sample with 24 xA sources (19+5) and 34 Pop. A sources (23+11). A comparison of the \aliii\ equivalent width of the 24 xA respect to 34 Pop. A sources shows a significant difference between them at a level of 7.8$\sigma$ with a low probability to be the same distribution ($7\cdot10^{-8}$). There is also a statistically significant difference in the distributions of \ciiionly\ and \civ, but in the opposite direction: values of the xA sources are significantly lower than Pop. A (6.6$\sigma$ and 5$\sigma$ level respectively with probabilities lower than $2\cdot10^{-7}$).

\subsection{Composite spectra at low and high-$z$}
\label{sec:composites}

\begin{figure}[!ht]
	\begin{center}
		\includegraphics[width=8.5cm]{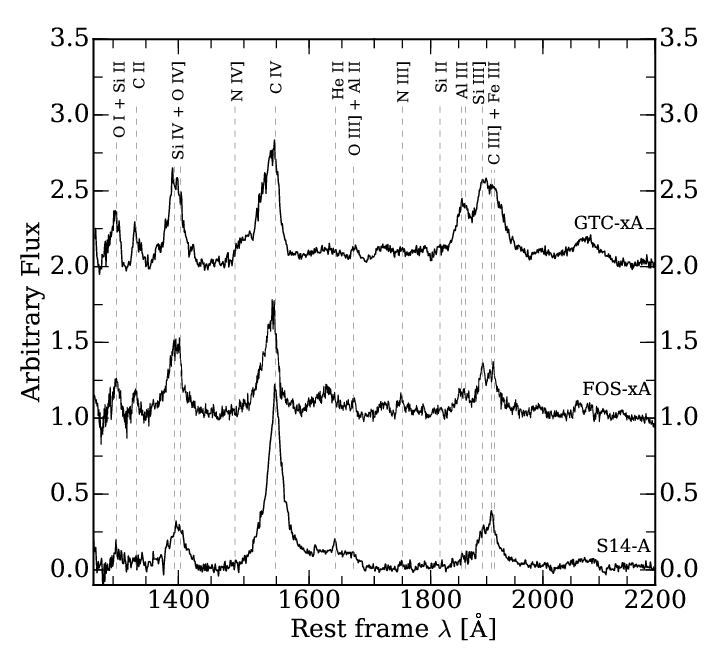}
	\end{center}
	\caption{From top to bottom, rest-frame composite spectra of the GTC-xA, FOS-xA, and S14-A. Intensity units are arbitrary flux. The  GTC-xA and FOS-xA have been displaced by a constant value  for clarity.
		\label{fig:compo} }
\end{figure}

With the objective of highlight main features observed in the xA spectrum and make a comparison with control samples, we built a composite spectrum using the full GTC-xA sample. Multicomponent fits and its analysis are shown in Section\,\S{\ref{sec:composite1}}. At the top level of Figure \ref{fig:compo} is represented the composite GTC-xA spectrum. It clearly shows an enhancement of \aliiionly\ and \siiiionly, in contrast with the weakness of \ciiionly, which has mean $W$ \ciiionly\ $\sim$ 4 \AA. Also, \feiii\ becomes the largest contribution in the red side of the 1900\AA\ blend, which is also prominent in the \feiii\ feature detected at $\sim$ 2080\,\AA. Another noticeable feature is the almost fully blueshifted \civ\ profile, which shows a low equivalent width. A blueshift is also observed in \siiv+\oiv, although in this case the profile is more symmetric due to the presence of strong absorptions in its blue side. As we point out in Section \S\ref{sec:civheii}, \heiiuv\ is very weak with a flat topped appearance profile. Low-ionization features, \cii\ and \oi+\siiia, are also clearly appreciated.

In order to compare the general behavior of xA sources, we also built composite spectra for two comparison samples: FOS-xA and S14-A samples, middle and bottom level of Figure \ref{fig:compo}, respectively. A multicomponent analysis was also applied to composite control samples. The width line of \aliii\ in FOS-xA ($\sim$2740 \kms) is narrower than GTC-xA ($\sim$4500 \kms). The average bolometric luminosity of the FOS-xA sample is $\sim 2.4 \cdot 10^{45}$ erg s$^{-1}$ and $\sim 10^{47}$ \ergs\ for the GTC-xA, therefore a FWHM change can be associated with the higher luminosity in the GTC-xA sample.

The S14-A composite spectrum shows a comparable luminosity ($\log L \approx 46$) than the FOS-xA, and a similar redshift that GTC-xA sample. S14-A shows a strong and symmetric \civ\ profile with a median equivalent width of 47 \AA, more than two times large than the median values obtained in xA samples: 14 and 23 \AA\, for GTC-xA and FOS-xA, respectively. Conversely, the 1900\AA\, blend of S14-A composite spectrum is dominated by \ciiionly\, and \siiiionly\ profiles. The xA samples show a weaker \ciiionly, with median values of equivalent width 2.6 \AA\ for GTC-xA, 1.8 \AA\ for FOS-xA and 11 \AA\ for S14-A. Also they shows a stronger \aliii\ that Pop. A samples, with the next median values: 7 \AA\ for the GTC-xA, 6 \AA\ for FOS-xA and 2 \AA\ for S14-A. Other notable features of xA sources include a smaller \heii, as well as a higher (\oiv+ \siiv)/\civ$\mathrm{_{BC}}$ ratio. Features like \oi+\siiia\ and \cii\ are also clearly observed, {supporting the inference of a low-ionization degree for xA objects} (See Section \S\ref{sec:feiii}).  There is a fundamental caveat between GTC-xA and S14-A samples: among the faint sources of S14-A there is no xA sources. A systematic difference in \lledd\ of $\approx$ 2 can be accounted, if the Eddington ratio of GTC-xA sources is \lledd $\approx 1$ and if the S14-A are instead radiating at \lledd $\approx 0.3$ \citep{sulenticetal14}, without invoking any effect of luminosity.

\subsection{A low--ionization spectrum with prominent \feiii}
\label{sec:feiii}

\begin{figure}[!ht]
\begin{center}
\includegraphics[width=10cm]{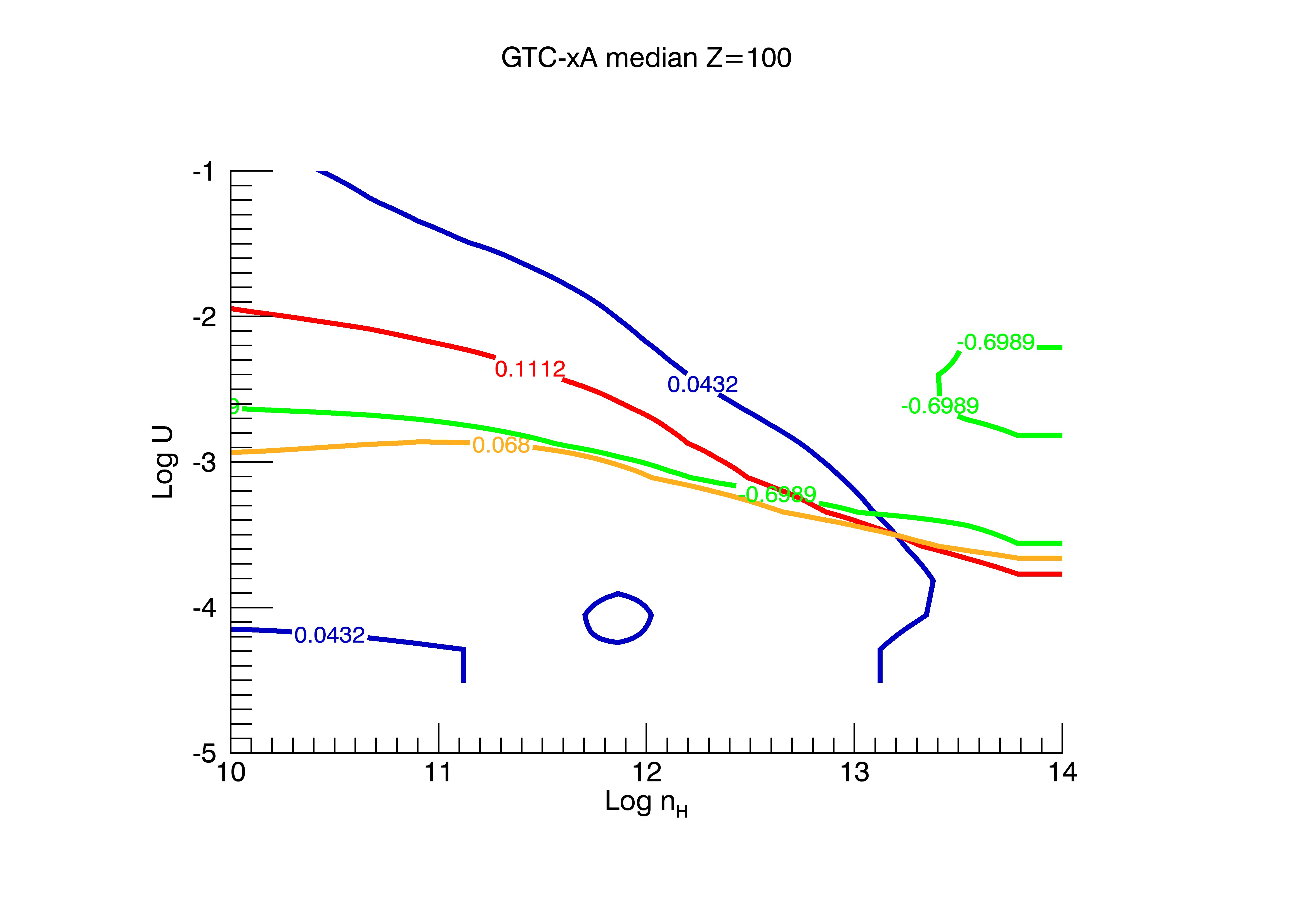}
\end{center}
\vspace{-1cm}
\caption{{Isocontours in the plane log of ionization parameter ($U$) versus log of hydrogen density (\nh)  from {\tt CLOUDY} simulations for the line intensity ratios \aliii/\siiii\ (blue), \heii/\civ\ (green),  \civ/\aliii\ (yellow), \civ/\siiii\ (red) using $Z = 100 Z_{\odot}$. Labels report the median values of the intensity ratios for the objects of the GTC-xA sample.}}
\label{fig:z50gtcxa}
\end{figure}

The UV spectral range provides several diagnostic ratios that can constrain physical parameters. For instance the \ciii/\siiii\ and \aliii/\siiii\ ratios are sensitive to hydrogen density, \civ/\aliii\, and \civ/\siiii\ to ionization, and \heii/\civ\ and (\oiv+\siiv)/\civ\ to metallicity  \citep[\citetalias{negreteetal12}, ][]{negreteetal13,negreteetal14}. {The diagram of Fig. \ref{fig:z50gtcxa} has been computed from an array of {\tt CLOUDY 13.05} \citep{ferlandetal13} simulations covering the range in ionization parameter $-4.5 \le \log U \le 0$ and density $7 \le n_\mathrm{H} \le 14$\ \cmc, with steps of 0.25 dex. A total of 551 simulations were carried out assuming a standard AGN continuum appropriate for Pop. A, a column density $N_\mathrm{c} = 10^{23}$ cm$^{-2}$ and a metallicity of 100$Z_\odot$. The median flux ratios given by \aliii/\siiii, \civ/\aliii, and \civ/\siiii\ converge toward $\log U\approx-3$\ and $\log$ \nh $\approx 13$ [cm$^{-3}$].} The suggestion of low-ionization and high density is in agreement with the previous study of two xA sources by \citetalias{negreteetal12}. Consistently with the inference of high density, \ciii\ is usually weak although not entirely negligible with a $\log$ \ciii/\siiii\ $\approx-0.42$.  Due to the critical electronic density of the \ciii\ is n$_\mathrm{H}$=10$^{10}$ \cmc, this line is expected to  be collisionally suppressed with respect to \aliii\ and \siiii, as it is shown by the GTC-xA composite spectrum. {The strong \feiii\ contribution observed in the GCT-xA sources also supports the existence of a very low--ionization spectrum, $\log U\leq-2$. A more detailed analysis of \feiii\ emission is presented in Appendix \ref{sec:feappen}.}

The intensity ratios considered above are not strongly sensitive to metallicity except for the \heii/\civ\ and the (\oiv+\siiv)/\civ\ ratios. According to {\tt CLOUDY} simulations, at solar metallicity we expect \civ/\heii\ $\approx 1$\, for the low-$U$ and high density solution, which is very far from the value that we derive for the xA sample  ($\sim 0.19$).  A possible solution is offered by a large increase in metallicity above solar values. {The \heii/\civ\ is the ratio between a recombination and a collisionally excited line: at high-$Z$, the metals absorb a larger fraction of the ionizing flux, and the He recombination line become weaker. This effect leads to a decrease of the \heii/\civ\ over the value expected at solar $Z$\ \citep{hamannferland93,hamannferland99}. }  In this case, the $Z$-sensitive \heii/\civ\ would be small ($\sim 0.2$) even at low ionization. An increment in the metallicity to $Z \approx 100 Z_{\odot}$ gives such low \heii/\civ\ ratios {(Fig. \ref{fig:z50gtcxa})}. The value $Z \approx 100 Z_{\odot}$ may appear unrealistically high. However, \citetalias{negreteetal12}   found evidence of high-$Z$, as well as of possible abundance anomalies. It appears reasonable to assume very high values of $Z$\ for xA sources: Pop. A sources at low-$z$ and high-$z$ have \nv/\civ\ and (\oiv+\siiv)/\civ\ implying $Z \gtrsim 10 Z_{\odot}$ \citep[][\citetalias{sulenticetal14}]{shinetal13}. The median ratio  \siiv/\civ\ $\sim$\, 1 may suggest a metallicity significantly higher than 10 times solar for the GTC-xA sample. The high \siiv+\oiv/\civ\ ratio also supports this suggestion. {In agreement with the interpretation of \citet{hamannferland99}, the equivalent width of \heiiuv\ is extremely low, making the line almost undetectable above noise}.

We should also consider that our approach may overestimate the \civbc\ intensity, as it attempts to maximize its contribution attempting to account for the red side of the profile, even when the profile if almost fully blue shifted. There is the possibility that the \civbc\ is significantly weaker.  A slightly lower value of  $Z$=50$Z_\odot$  is again in agreement with the observations assuming an overestimation of the \civbc\ intensity by factor $2$. Therefore, a $Z$ value of several tens $Z_{\odot}$  might be considered as a lower limit if we consider a systematic overestimation of the \civbc\footnote{{It is unlikely that the overestimate might be much larger than a factor 2, as the \civbc\ is clearly detectable while \heiiuv\ equivalent width is extremely low.}}.  More precise inferences require dedicated photoionization computations beyond the scope of this paper. 

\subsection{Evidence of  2080 \AA\ enhancement}
\label{sec:fluo}

{Several sources show a significant excess with respect to the template at around $\sim$2080 \AA, indicating that UV \#48 is selectively enhanced with respect to the ``normal'' emission in xA sources as represented by \citet{vestergaardwilkes01} template. In eight sources the equivalent width of the feature in excess respect to the template is W(2080)$\gtrsim 5$\,\AA. This excess is evident from the spectra, and is even more outstanding because the HILs in the adjacent spectral regions are of low equivalent width. A mock profile built from the transition of multiplet \feiii\ UV \#\,48  between the terms a$^{5} S  -  z^{5} P^{\rm o}$, assuming relative intensity for optically thin transitions \citep{kovacevicetal10} accounts for the 2080\AA\ feature: the mock profile resulting from the lines broadened to $4000$\,\kms\ is a feature at $\approx$ 2073\,\AA\ with total FWHM $\approx$ 5000\,\AA, consistent with the observations.}

\begin{figure}[!ht]
\begin{center}
\includegraphics[width=9.0cm]{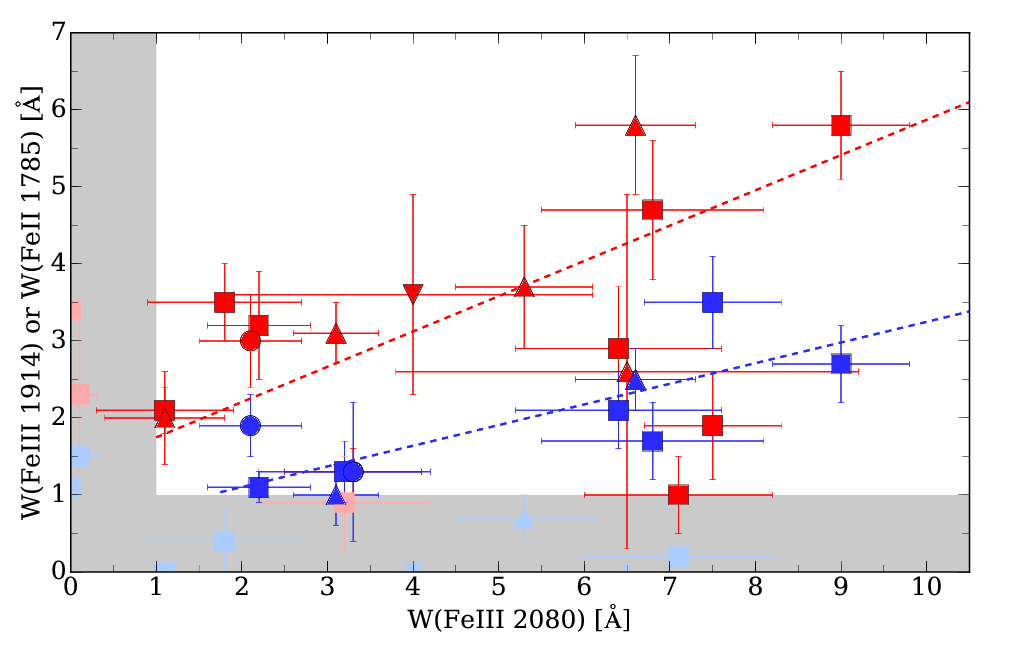}
\end{center}
\caption{Relation between intensity of the \feiii\ feature at 2080 \AA\ with Fe UV 1785 \AA\ (blue) and \feiii$\lambda$1914 (red). Squares, up and down triangles, and circles mark the normal, BAL, mini--BAL and RL quasars, respectively. The gray shaded area indicates an approximate detection limit. Blue and red lines mark the orthogonal fit for each case. 
\label{fig:fe3} }
\end{figure}

{Figure \ref{fig:fe3} shows the equivalent width of the \feiii$\lambda$1914 and of the \feii\,UV\,\#191 multiplet versus the equivalent width of the 2080\,\AA\ feature. Only 10 sources satisfy the detection limit we imposed at $W \approx $1\,\AA. Strong \feiii$\lambda$2080 is usually associated with strong \feiii$\lambda$1914. According to an orthogonal fit taking into account errors in both axis, these two emission lines follow the relation \feiii$_{1914}$=(0.46 $\pm$ 0.14)$\cdot$\feiii$_{2080}$+(1.29 $\pm$ 0.55), which verified the null hypothesis of a slope different from 0 at 3.3$\sigma$ significant level. There is a clear relation, although the errors associated with \feiii\,$\lambda$1914 are high because of the blend with \ciii\ and the contribution of the \feiii\ template. On the another hand, the \feii$\mathrm{_{UV}}$ \#191\ is generally rather weak, but the 5 cases with a clear detection ($W \gtrsim$ 2\,\AA) are associated with strong 2080\,\AA\ features. An orthogonal fit gives a relation { \feii$\mathrm{_{UV}}$=(0.27$\pm$0.06)$\cdot$ \feiii$\lambda{2080}$+(0.57$\pm$0.30)} with a slope different from 0 at a 4$\sigma$ significance level. This relation considers measurements with $W >$ 1.0\,\AA\ of both \feii\,UV ${\#191}$\ and \feiii$\lambda{1914}$ (Figure \ref{fig:fe3}). {A fluorescence mechanism (described in Appendix \ref{sec:origin}) may account for the selective enhancement of the 2080 \AA\ feature. }}

\subsection{Estimates of the accretion parameters}
\label{sec:acc}

\begin{table}
	\setlength{\tabcolsep}{4.5pt} 
	\begin{center}
		\caption{Accretion parameters \label{tab:accretion}}
			\scriptsize
		\begin{tabular}{c c c c c}\\ 
			\hline\hline\noalign{\vskip 0.1cm}       
			\multicolumn{1}{c}{SDSS Identification}  &\multicolumn{1}{c}{log $L \mathrm{_{Bol}}$} & \multicolumn{1}{c}{log \mbh$\mathrm{_{, AlIII}}$} & \lledd \\
			\multicolumn{1}{c}{(1)} & \multicolumn{1}{c}{(2)} & \multicolumn{1}{c}{(3)} & (4)  \\ \hline
			\noalign{\vskip 0.12cm}
SDSSJ000807.27-103942.7	&	46.9		&	9.29	&	0.29	\\
SDSSJ004241.95+002213.9	&	47.0		&	9.24	&	0.34	\\
SDSSJ021606.41+011509.5	&	47.5		&	9.13	&	1.54	\\
SDSSJ024154.42-004757.5	&	46.9		&	8.88	&	0.77	\\
SDSSJ084036.16+235524.7	&	46.9		&	8.88	&	0.70	\\
SDSSJ101822.96+203558.6	&	47.1		&	9.28	&	0.42	\\
SDSSJ103527.40+445435.6	&	47.1		&	9.63	&	0.21	\\
SDSS105806.16+600826.9	&	47.2		&	9.21	&	0.65	\\
SDSSJ110022.53+484012.6	&	47.0		&	9.22	&	0.37	\\
SDSSJ125659.79-033813.8	&	47.0		&	9.41	&	0.26	\\
SDSSJ131132.92+052751.2	&	47.3		&	9.19	&	0.93	\\
SDSSJ143525.31+400112.2	&	47.4		&	9.14	&	1.31	\\
SDSSJ144412.37+582636.9	&	46.7		&	9.49	&	0.10	\\
SDSSJ151258.36+352533.2	&	47.1		&	8.80	&	1.31	\\
SDSSJ214009.01-064403.9	&	47.2		&	8.79	&	1.61	\\
SDSSJ220119.62-083911.6	&	47.3		&	9.72	&	0.23	\\
SDSSJ222753.07-092951.7	&	46.9		&	9.09	&	0.39	\\
SDSSJ233132.83+010620.9	&	47.7		&	9.84	&	0.50	\\
SDSSJ234657.25+145736.0	&	47.1		&	9.02	&	0.73	\\
			\hline	
		\end{tabular}
	\end{center}
	\small {{\sc Notes.} Columns: (1) SDSS identification. (2) Log of Bolometric luminosity in unit of erg s$^{-1}$. {The associated uncertainty is 10\% the luminosity value.} (3) Black hole mass in unit of \mo. (4) Eddington ratio.   }  
\end{table} 

\subsubsection{Bolometric correction} 
\label{sec:bolcorr}

{There has been no accurate determination of the bolometric correction (hereafter B.C.) specific to xA sources. An attempt has been made to compute the bolometric correction from the model SEDs of \citetalias{marzianisulentic14}.  \citetalias{marzianisulentic14} suggested that the difference in the optical/UV/X range between a typical NLSy1 and an extreme NLSy1 is about 10\%\ (see their  section 6.1.2.3). It is unlikely that the uncertainty in the B.C. is much larger than $\approx 20$\%\ (at 1$\sigma$\ confidence level).  SDSS quasars have  with B.C.$_{1350} = 3.5$ and a scatter $\sim20\%$ \citep{richardsetal06}. SDSS quasars are however an heterogenous sample composed by quasar of different spectral types. On the contrary, the GTC-xA sample is composed by similar sources. If we consider 27 spectra of sample 3 of \citetalias{marzianisulentic14} i.e., at redshift in the range 2.0 -- 2.6, the   rms averaged between $\approx$ 1100 and 2900 \AA\ is $\approx $10\%.   }

Considering that the bulk of the ionizing energy is coming from the big blue bump in correspondence  of the Lyman limit, the SED is not significantly different from the one of \citet{mathewsferland87}. The \citetalias{marzianisulentic14} SED differs from the reference \citet{mathewsferland87} SED only for the slope in the X-ray domain,  and yields a B.C.$_{1350} \approx 5.4$. { The B.C. value  is consistent with the one of \citet{elvisetal94}. The value of \citet{elvisetal94} may perhaps provide a more appropriate comparison, as the sources in that paper implied a detection from the IPC instrument on board the {\tt Einstein} observatory \citep{giacconietal79}, i.e., in the soft X-ray domain. Bolometric luminosities ${ L\mathrm{_{Bol}}}$ of the GTC-xA sources are presented in Table \ref{tab:accretion}. They have been computed through the monochromatic luminosities at 1350 \AA\ ($\lambda$\ $L_{\lambda}$(1350\AA)) and a   {B.C.=5.4 $\pm$ 0.6.} } {A more careful estimate would require a detailed analysis of the mid- and far-IR domain, where part of the accretion luminosity is reradiated, and is deferred to an eventual work.}


\subsubsection{Black hole mass}
\label{sec:mbh}

We estimate black hole mass (\mbh) using the UV \civonly\ scaling relation of \citet{vestergaardpeterson06}[Eq. 7] applied to {FWHM(\aliiionly)\footnote{{log\mbh(\aliiionly)=log $\left\lbrace \left[\frac{\mathrm{FWHM(Al\,{\sc III})}}{1000\,\mathrm{kms^{-1}}}\right]^2 \left[\frac{\lambda L_\lambda(1350\,\AA)}{10^{44}\,\mathrm{erg\,s^{-1}}}\right]^{0.53} \right\rbrace$ + (6.73$\pm$0.01).}}}. The \aliiionly\ profile is usually not affected by the strong outflows often observed in the \civ\ profile, thus providing a good approximation to the virial motion of the BLR \citep[e.g., ][see also Section \S\ref{sec:cival}]{marzianietal17e}. We adopt  \mbh$\mathrm{_{,AlIII}}$ as the virial mass estimator. Second column of Table \ref{tab:accretion} show the \mbh\ estimations. {Typical formal uncertainties in the estimation of the masses, through error propagation,  are about 25\% taking into account an uncertainty of 10\% in the flux at 1350\AA, and including the estimated errors for the FWHM \aliii.} 

The GTC-xA sample covers a black hole mass range of 8.8$ \ \leq$\mbh$\leq \ $9.8 \mo, with a median value of $\langle$\mbh$\rangle \sim \ $9.2 \mo. The GTC-xA sample is covering the high-\mbh\ counterparts of the FOS-xA sample. Also these values confirm the well--know tendency of larger black hole mass ($\sim$10$^9$ \mo) at high redshift ($z\ \approx$ 2) \citep{shemmeretal04, netzeretal07, trakhtenbrotnetzer12}, a redshift-dependent selection effect common to all flux-limited samples \citep{sulenticetal14}.

\subsubsection{Eddington ratio}
\label{sec:eddratio}

Eddington ratios (\lledd) have been calculated based on  masses obtained from the FWHM \aliiionly, explained in previous section and following the relation: \ledd $\approx 1.5\times10^{38}\cdot$ (\mbh/\mo) [\ergs], as recommended by \cite{netzermarziani10}. \lledd\ values for our sample are reported in   Table \ref{tab:accretion}. {Applying  the usual method of propagation of errors and using previous formulae, the typical mean uncertainty in the Eddington ratio is about 30-35\% (only in one case the estimated error is larger than the 50\%, in \object{SDSSJ233132.83+010620.9}   \lledd$\approx$ 0.5$\pm$0.3)}. 

The average \lledd\ value for the GTC-xA sample is $\langle$ \lledd$ \rangle \approx 0.66$, comparable to the one obtained for I Zw 1, a prototype highly accreting quasar at low-$z$. \  An appropriate sample for a comparison is also the sample $\#3$ of \citetalias{marzianisulentic14} that involves xA quasars in the redshift range $2 \le z \le 2.6$\  with the \aliii\ used as a virial broadening estimator. The obtained average \lledd\ of the GTC-xA sample is consistent with the value  obtained for sample $\#$3 of \citetalias{marzianisulentic14} (\lledd$\approx 0.7$), once the estimates are computed with the same bolometric correction (5.4 in this work versus 6.3 in \citetalias{marzianisulentic14}), and the same \lledd\ constant (\citetalias{marzianisulentic14} used $1.3\,10^{38}$\ erg s$^{-1}$).  A Welch-test on the average \lledd\ confirms that two averages are not significantly different, and therefore the xA nature of the sources in the present sample.

The scatter in \lledd\ in the GTC-xA sample after extinction correction is $\sigma_{ L/L_\mathrm{Edd}}\sim$ 0.47\ which is somewhat above the one of sample-$\#$3 of \citetalias[][]{marzianisulentic14} ($\sigma_\mathrm{L/L_\mathrm{Edd}} \approx 0.16$), although this may be in part due to the small size of the GTC-xA, with only 19 objects. Also the larger scatter in FWHM may contribute to the \lledd\ scatter: average FWHM \aliiionly\ $\sim 4220 \pm 900$ \kms\ for \citetalias{marzianisulentic14} versus an average FWHM \aliiionly\ $\sim 4720 \pm 1600$ \kms\ for the GTC-xA sample. As an example, \object{SDSSJ144412.37+582636.9} has a FWHM \aliiionly\ $\sim 8080 $ \kms\ which yields the minimum \lledd\ value (0.1) in the sample. Additionally this source is a mini-BAL with clear absorptions in its spectrum (See Fig. \ref{fig:specSDSSJ144412.37+582636.9}) introducing additional uncertainties in the extinction correction.  In the absence of an orientation correction, a large scatter in the observed FWHM translates into a large scatter in \lledd.

\begin{figure}[!ht]
\begin{center}
\includegraphics[scale=0.37]{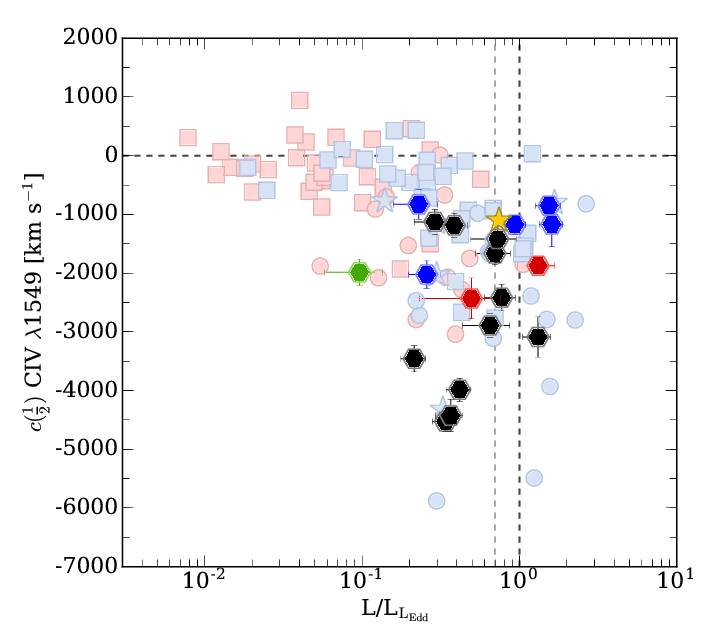}
\end{center}
\caption{Behavior of the centroid at 1/2 of total \civ\ profile as a function of  Eddington ratio. Hexagons represent the GTC-xA sample. The colors of hexagons are the same as Figures \ref{fig:critsel} and \ref{fig:c12fwhm}. Blue and red pale symbols correspond to the Population A and B, respectively. Squares and stars correspond to the FOS-A and xA samples \citep{sulenticetal07}. Circles are from HE sample \cite{sulenticetal17} sample. Yellow star marks the position of I Zw 1. The black dashed horizontal line indicates $c(\frac{1}{2})$=0 \kms. Gray and black dashed vertical lines indicate \lledd=0.2 and 1.0, respectively. \label{fig:c12-lledd}} 
\end{figure}

A correlation between the \cmp\ shift and  \lledd\ \citep{sulenticetal17} implies that highest \cmp\ values are associated with the highest \lledd. Figure \ref{fig:c12-lledd} shows the plane \cmp\ vs. \lledd, where GTC-xA sources occupy a fairly extreme position, where also I Zw 1 is located. Two-third of the 11 sample (unabsorbed sources) have \cmp\ $\lesssim -2000$ \kms, and the blueshifts distribution is more skewed toward high \cmp\ values than in the sample of \citetalias{sulenticetal17}. Comparing with the HE sample \citep{sulenticetal17} and FOS-xA and FOS-A sample \citep{sulenticetal07}, we observe that FOS-xA sources (the pale blue squares around \lledd\ $\sim 1$)\ also cover a similar domain of blueshift amplitudes, while Pop. A sources show lower amplitude shifts,  $-900$ \kms\ on average. This diagram provides further support to the idea that the Eddington ratio is the physical driver of the outflows, {and that xA sources are associated with high Eddington ratio values}.

\subsection{Weak-lined and xA quasars}
\label{sec:wlq}

\begin{figure}[!ht]
\begin{center}
\includegraphics[width=8.5cm]{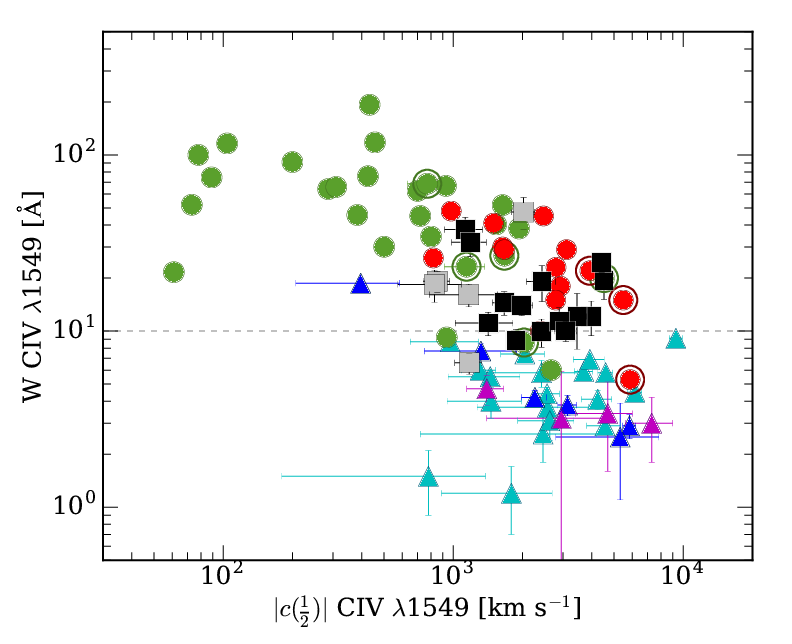}
\end{center}
\caption{\label{fig:wshift}{ Relation between rest-frame equivalent width of \civ\ and \cmp\ in \kms. Black squares: GTC-xA. Gray squares: BAL GTC-xA. Blue triangles: \citet{plotkinetal15}, magenta: \citet{wuetal12} and cyan: \citet{luoetal15}. Red and green circles correspond to the Pop. A samples from \citetalias{sulenticetal17} and FOS-A sample, respectively. Circled symbols correspond to FOS-xA and \citetalias{sulenticetal17}-xA. }}
\end{figure}

The Weak-Line Quasars (WLQs) show very low equivalent width in the \civ\ and other UV emission lines \citep{diamond-stanicetal09,wuetal12,shemmeretal10}. A standard definition of WLQ involves sources with $W$(\civonly)\,$\le 10$\,\AA. WLQs also show strong \feii\ emission and \civonly\ blueshift. \citet{luoetal15} provides one of the largest samples of WLQ candidates, with detailed measures on shifts and $W$ of \civonly\ for 21 of them. It is an almost exclusively RQ sample as are 4DE1 Population A, and especially xA quasars. Most WLQ candidates are high-$z$\ making estimates of source rest-frame uncertain in many cases. When \hb\ measures are available \citep{plotkinetal15,sulenticetal07,sulenticetal17}, we confirm that WLQs belong to the Population A domain of 4DE1 with \rfe$\gtrsim 1.0$, independently of $z$ and $L$  \citep[Fig. 9 of][]{marzianietal16a}. At least 80\% of WLQs are xA sources and, if doubtful cases (e.g., affected by absorptions, so most likely not real WLQs) are excluded {all}  WLQs might belong to the xA population. The majority of WLQ show FWHM \hb\ $\gtrsim$ 2000\,\kms. WLQ are simply Pop. A sources with strong \rfe, high amplitude \civonly\ blueshift and the lowest values for $W$ \civonly.

According to the standard WLQ definition our GTC-xA sample includes three WLQs, excluding BALs. If we consider the prevalence of WLQs among xAs in different samples, we found: 4 out of 42 in \citetalias{marzianisulentic14};  1 out of 3  (\citetalias{sulenticetal17}); 3 out of 13 in the present sample (excluding BAL and mini-BAL QSOs), and 2 out of 10 (\citetalias{sulenticetal07}), for a total $f_{\rm WLQ} = n_{WLQ}/n_{\rm xA} = 10/68\approx 15$\%. 
 
As we see in Section \S\ref{sec:ewdis}, the peak of the GTC-xA $W$ \civ\ distribution is slightly above 10 \AA\  (See also bottom right panel of Fig. \ref{fig:ew}). There is no solution of continuity at $W$ \civ\ $\approx$ 10 \AA. WLQs are located at the low end of the $W$ \civ\ distribution of the GTC-xA sources and there is no suggestion that they constitute a special class. Moreover, as in the case xA sources, WLQs are characterized by extreme outflow velocities. {Figure \ref{fig:wshift} shows the $W$ \civ\ vs. \civ\ shift domain for the 19 GTC-xA sources of our sample and several WLQs, Pop. xA and Pop. A samples: 19 WLQs from \citet{luoetal15}, 6 WLQs from \citet{plotkinetal15}, 4 WLQ from \citet[][which includes PHL 1811 from \citealt{wuetal11}]{wuetal12}, 3 Pop. xA and 11 Pop. A from \citet{sulenticetal17}, and 5 xA and 23 Pop. A from \citetalias{sulenticetal07}.} The measurements in our samples are the centroid at half maximum, while \citet{luoetal15} consider interactive measurement, presumably of the peak shift. In the context of the present analysis, we expect that the effect may not be relevant because, in the case of the largest blueshift, the total profile is shifted so that peak and centroid at half maximum yield consistent values. 

{The distributions of xA and WLQ blueshifts are largely overlapping, with an excess of WLQs showing blueshifts in the range 5000--10000\,\kms. A trend between $W$ \civ\ and \civ\ shift amplitude is suggested, in the sense that the blueshift amplitude appears to increase among weaker \civ\ emitters. It is important to consider that this is part of a general trend among quasars, with a much higher statistical significance if Pop. A and B at high luminosity are included \citep{marzianietal16a}. }

Restricting attention to xA and WLQs means to consider only the extreme end of the equivalent width-shift distribution. Both xAs and WLQs show a coupling between physical and dynamical conditions of the outflowing gas which seems a well-defined, common phenomenon affecting outflow-dominated sources \citepalias{sulenticetal17}.  WLQs can therefore interpreted as extreme xA sources in the context of 4DE1. The previous discussion also validates the assumption that at least the wide majority (if not all) of RQ WLQs are xA quasars. In addition, it suggests that WLQs are an ill-defined class, and that their observational properties should be analyzed within the context of the wider class of xA sources. 

At the present time we suggest that observational biases may be playing a serious role. If one considers the data for the most extreme WLQs, extremely low $W$ and extremely high \civonly\ blueshifts, we find that some of the lowest $W$ values in Fig. \ref{fig:wshift} involve sources with \civonly\ absorptions, that reduce $W$ \civonly, and UV redshift uncertainties, that affect estimates of \civonly\ shifts \citep{luoetal15}. Only a few WLQ candidates have data available for the \hb\ region \citep{plotkinetal15}. Two of these sources report doubtful detections of an extremely broad and very low $W$ \civonly\ (SDSSJ094533.98+100950.1 and SDSSJ141730.92+073320.7). Sources with \cmp\,$<$\,-5000\,\kms\ and/or $W$ \civonly\,$<$\,3-4\,\AA\ require high S/N confirmatory spectra. Even if confirmed, we see no evidence for a break in the distribution at $W$ \civonly\,=\,10\AA\  suggesting that WLQ are the most extreme Pop. A quasars (xA), with a lowest values for $W$ \civonly.

\subsection{Outflows and feedback}
\label{sec:feed}

Hereafter we will consider only median luminosity estimates after extinction correction excluding BAL and mini-BAL sources. The  median value of the blueshifted component (excluding the BAL cases)  is {$\log L_{\rm BLUE} \approx 44.0 \, \pm\,$0.3  [erg s$^{-1}$].}  It is relatively straightforward to compute the mass of ionized gas needed to sustain the \civ\ blue shifted component luminosity: $M^\mathrm{ion} \sim   95 \, Z_{10}^{-1} n_{9}^{-1}  M_{\odot}$,   where the density is assumed $10^{9}$ cm$^{-3}$, and the relation  is normalized to  a metal content ten times solar \citep[e.g.,][]{canodiazetal12,marzianietal16a,marzianietal17c}.  Computation of the mass outflow rate, of its thrust, and of its kinetic power  requires several assumptions (discussed in the recent  review of \citealt{marzianietal17c}). In the present context, we assume a BLR radius estimated from the radius luminosity relation derived by \citet{kaspietal07}. At the average  continuum luminosity after extinction correction $\lambda L_{\lambda}$(1350) $\approx  2.4 \cdot 10^{46}$ \ergs, the $r_{\rm BLR} \approx 2.6  \cdot 10^{17}$ cm $\approx$ 0.086 pc.  The mass outflow rate can then be written as: $ \dot{M}^\mathrm{ion}  \sim   150  L_{45} v_\mathrm{o,5000} r^{-1}_{\rm BLR, 0.1pc} Z_{10}^{-1} n_{9}^{-1} M_{\odot}$ [yr$^{-1}$], where the outflow radial velocity is in units of 5000 \kms. For the median centroid displacement of the blue shifted component $\approx -3560$ \kms, we obtain a { $ \dot{M}^\mathrm{ion} \approx 12.5$ $M_{\odot}$ yr$^{-1}$. } The final thrust can be expressed as: $ \dot{M}^\mathrm{ion}   k v_\mathrm{o} \sim 5 \cdot 10^{36}  L_{45} k v^{2}_\mathrm{o,5000} r^{-1}_{\rm BLR, 0.1pc}  Z_{10}^{-1} n_{9}^{-1}  $   [g cm s$^{-2}$], where we have now introduced a factor $k$ that takes into account that the gas within the BLR is still being accelerated by radiation forces. In the  scenario of \citet{netzermarziani10}, for Eddington ratio $\rightarrow$ 1, as it is likely the case of xA sources,  $k$ can be as high as 10. We prudentially assume $k = 5$,  {and that the full blue shifted component is accelerated to $k v$}.  For the average values of the xA sources we obtain $ \dot{M}^\mathrm{ion}   k v_\mathrm{o} \sim 1.5 \cdot 10^{36}$ g cm s$^{-2}$.  The kinetic power of the outflow is $ \dot{\epsilon}  \sim 1.2 \cdot 10^{45} L_{45} k^{2} v^{3}_\mathrm{o,5000}   r^{-1}_{\rm 0.1pc}  Z_{10}^{-1}     n_{9}^{-1}  $\   erg s$^{-1}$, and becomes   $\dot{\epsilon}  \sim 1.3 \cdot 10^{45}$\ erg s$^{-1}$\ for the average parameter values of the xA sample.

The large outflow radial velocity (the average \cmp\ of the blue shifted component is close to $\sim -4000$ \kms), as well as the high fraction of the blue shifted  component to the total \civ\ luminosity ($\sim 0.45$, Sect. \S\ref{sec:civheii}) ensure  the maximization of the quasar mechanical feedback even if  $W$ \civ\ is in general relatively low \citep[c.f., ][]{kingmuldrew16}.  The ratio $\dot{\epsilon}/L \sim 0.01$\ comes close to  the threshold value for exerting  significant mechanical feedback on the entire host galaxies especially if the quasar is radiating at high Eddington ratio and at high luminosity \citep[e.g.,][and references therein]{kingpounds15}. High luminosity outflows may be even more powerful if one considers that the \civ\ emission traces only  a phase of multiphase outflowing medium \citep{tombesietal10,harrisonetal14,ciconeetal14,carnianietal15,feruglioetal15}. On the other hand, if $Z$ is as high as 100 $Z_{\odot}$, then all estimates should be reduced by a factor of 10. This emphasizes the need of accurate metallicity estimates not only to study the enrichment of the ISM by nuclear outflows, but also to ascertain their dynamical and evolutionary relevance on the host galaxy.

\section{Conclusion}
\label{sec:concl}

We present a sample of 19 quasar satisfying the 4DE1 selection criteria  for finding high redshift highly accretors xA sources. We  provide a description of the rest-frame UV spectra of our candidate sources.  

\begin{itemize}

\item Our GTC-xA sample shows the same characteristics of xA quasars at low-$z$, of which a prototype is  I Zw 1. Figure \ref{fig:critsel}  shows that GTC-xA sample occupies a very restricted \aliii/\siiii\  vs. \ciii/\siiii\ domain space, compared to more general quasar populations.  GTC-xA sources show different spectroscopic properties than other Pop. A (Bins A1, A2) quasars. We show that our sample is likely radiating close to the Eddington limit (\lledd$\geq$0.2) with an average consistent with the one computed on the \citetalias{marzianisulentic14} xA sample. 

\item \aliii\ and \siiii\ lines are well-modeled with a symmetric profile that is narrower than that of \civ\ total profile. These results support our previous suggestions that \aliii\ is likely a reliable virial estimator in analogue to \hb\ for lower $z$\ quasars. On the converse, the FWHM \civonly\ remains unreliable as a virial estimator showing no significant correlation with FWHM \aliii.

\item \civ\ profiles in GTC-xA sources show large blueshifts, $|$\cmp$|$=1000--5000 \kms.  Figure \ref{fig:c12fwhm} shows evidence for a  correlation between FWHM and \cmp\ for \civ.  {{The uniformity of the source distribution along the full blueshift range can be used to argue for a quasi isotropic wind in xA quasars. Another possibility sees a slight source excess around \cmp$\sim$--2000 \kms. Only a larger sample can permit a more robust speculation.}}

\item The emission line spectrum of the GTC-xA sources is characterized by low equivalent width of the most prominent emission lines.  WLQs appear as extreme xA sources and not as an independent class. All WLQ might be xA sources, if we consider observations with the \hb\ spectral range covered.

\item The appearance of the xA spectrum at high-$L$\  is of very low ionization. \feii\ and \feiii\ contributions appear to be more prominent in the spectra, although this is in part due to the low equivalent width of all the emission features. The \feiii\ strength is only in apparent contradiction with very low values of the ionization parameter: Fe$^{+2}$ becomes the dominant ionization stage of iron at the expense of Fe$^{+3}$, therefore increasing  Fe$^{+2}$ with respect to  Fe$^{+1}$.

\end{itemize}

Numerous results merit discussion and further study.  {Among them, we list two that are relevant to our global understanding of the black hole-host galaxy connection. }

\begin{itemize}
\item  The \civonly\ shift and the luminosity suggest  high values of mass outflow rate, thrust, and kinetic power of the outflow. This is expected for sources of high-$L$\ and maximized radiation forces per unit mass. {However, more refined computations are needed to assess the relevance of the outflow dynamical parameters for feedback effects on the host galaxy. }

\item The intensity ratio \heii/\civ\ requires high $Z$ ($\gtrsim 50 Z_{\odot}$) to be consistent with a low-ionization, high density ($\log U \sim -3$, $\log $ \nh $\sim 13$ \cmc) that we found from our data and that is consistent with the analysis of \citepalias{negreteetal12}. {This result claims for accurate estimates of the line emitting region chemical composition that are still missing today. } \end{itemize}

This preliminary paper provides a description of several key properties of xA at high-$L$. One of the most interesting result is that \aliii\ is providing a tracer of the dynamics of a virialized LIL-emitting region. If this is true at least approximately \aliii\ may offers a virial broadening estimator. The \aliii\ is a resonant doublet which implies resonant absorption and therefore efficient acceleration by continuum scattering. Then a non virial contribution due to outflows may remain unresolved in the \aliii\ profile. {Simultaneous observations of the \aliii\ and of the \hb\ spectral ranges are needed to make an exhaustive assessment of \aliii\ as a virial broadening estimator.} {The conceptual validity of luminosity estimates based on the \hb\ line width measurements	has been confirmed by the Hubble diagrams shown by \citetalias{marzianisulentic14} and \citet{negreteetal17}. The large scatter displayed by quasars with respect to supernov\ae\ demands that random and systematic sources of error should be understood and, if possible, corrected or reduced  before the xA sources can be exploited as useful distance indicators for cosmology. }

A reliable virial broadening estimator would make possible an application of the virial luminosity estimates proposed by \citetalias{marzianisulentic14} to a large number of quasars in the redshift range $1 \lesssim z \lesssim 3.5$. This redshift range is especially well suited for obtaining constraints on the cosmic density of matter $\Omega_\mathrm{M}$, since the effect of the cosmological constant in the dynamics of the Universe is felt only at $z \lesssim 1$.

\begin{acknowledgements}
MLMA acknowledges a CONACyT postdoctoral fellowship. AdO,  MLMA and JWS acknowledge financial support from the Spanish Ministry for Economy and Competitiveness through grants AYA2013-42227-P and AYA2016-76682-C3-1-P. MLMA, PM and MDO acknowledge funding from the INAF PRIN-SKA 2017 program 1.05.01.88.04. JP acknowledge financial support from the Spanish Ministry for Economy and Competitiveness through grants AYA2013-40609-P and AYA2016-76682-C3-3-P. MLMA is thankful for the kind hospitality at the Padova Astronomical Observatory,  PM for the hospitality at IAA-CSIC. DD and AN  acknowledge support from CONACyT through  grant CB221398. DD and AN thank also for  support from grant IN108716 PAPIIT, UNAM. We would like to thank Drs. J. Masegosa, I. Marquez, M. Povic and S. Cazzoli for all the fruitful discussions on the GANG meetings. This work is based on observations made with the Gran Telescopio Canarias (GTC), installed in the Spanish Observatorio del Roque de los Muchachos of the Instituto de Astrof\'{\i}sica de Canarias, in the island of La Palma. We thank all the GTC Staff for their support with the observations. This research has made use of the NASA/IPAC Extragalactic Database (NED) which is operated by the Jet Propulsion Laboratory, California Institute of Technology, under contract with the National Aeronautics and Space Administration.

\end{acknowledgements}

\bibliographystyle{aa}

\clearpage

\setcounter{figure}{0} 
\setcounter{section}{0}
\appendix
\section{\\ Rest-frame spectra and multicomponent fits}

\begin{figure*}
\begin{center}
\includegraphics[width=17.cm]{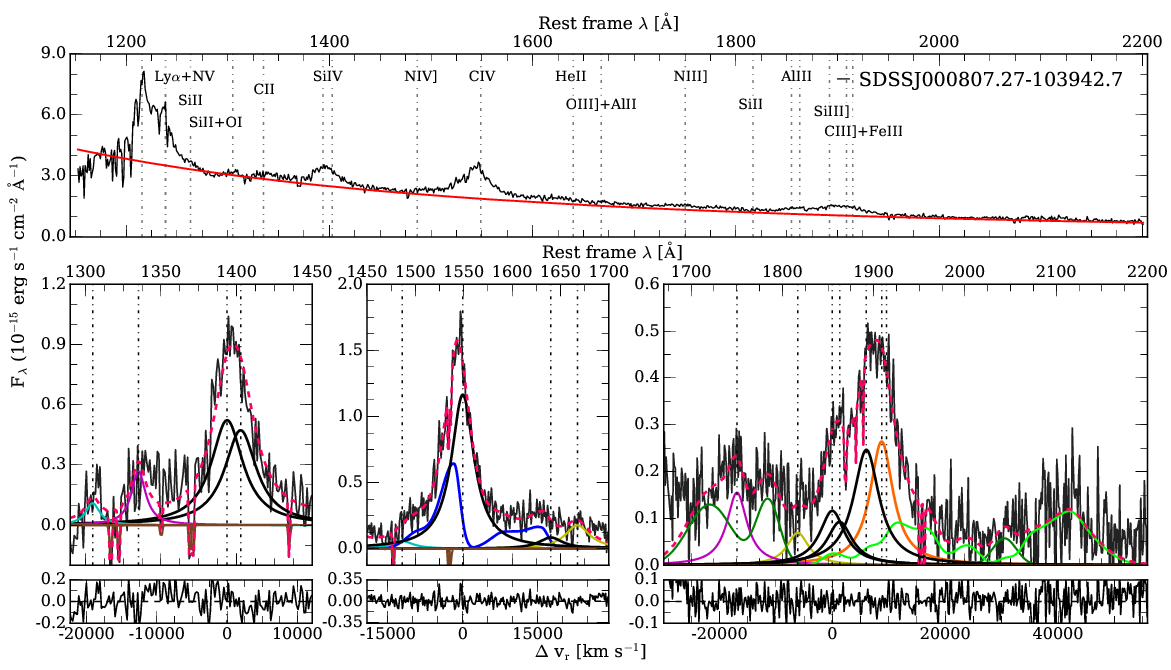}
\end{center}
\caption{Top panel: calibrated rest-frame spectrum of \object{SDSS000807.27-103942.7} before continuum substraction. Abscissa is rest-frame wavelength in \AA, while ordinate is specific flux in units 10$^{-15}$ erg s$^{-1}$ cm$^{-2}$ Hz$^{-1}$. Global or local continuum are specified by a continuous line. Dot-dashed vertical lines identify the position at rest-frame of the strongest emission lines. Bottom: multicomponent fits after continuum subtraction for the \siiv, \civ\ and 1900\AA\ blend spectral ranges. In all the panels the continuous black line marks the broad component at rest-frame associated to \siiv, \civ, \aliii\ and \siiii\ respectively, while the blue one corresponds to the blueshifted component associated to each emission. Dashed pink line marks the fit to the whole spectrum. Absorption lines are indicated by a brown line.  Dot-dashed vertical lines correspond to the rest-frame of each emission line. In the \siiv\ spectral range, the cyan line marks the contribution of  O{\sc i} + S{ \sc ii} 1304 blend, while the magenta line corresponds to the \cii\ emission line. In the \civ\ region, \niv\ is represented by a cyan line, while the yellow one corresponds to the \oiiiuv\ + \alii\ blend. In the 1900\AA\ blend range, \feiii\ and \feii\ contributions are traced by dark and pale green lines respectively, magenta lines marks the \niii\ and the yellow one corresponds to the \siii. Lower panels correspond to the residuals, abscissa is in radial velocity units \kms. 
\label{fig:specSDSSJ000807.27-103942.7} }
\end{figure*}

\begin{figure*}
\begin{center}
\includegraphics[width=17.cm]{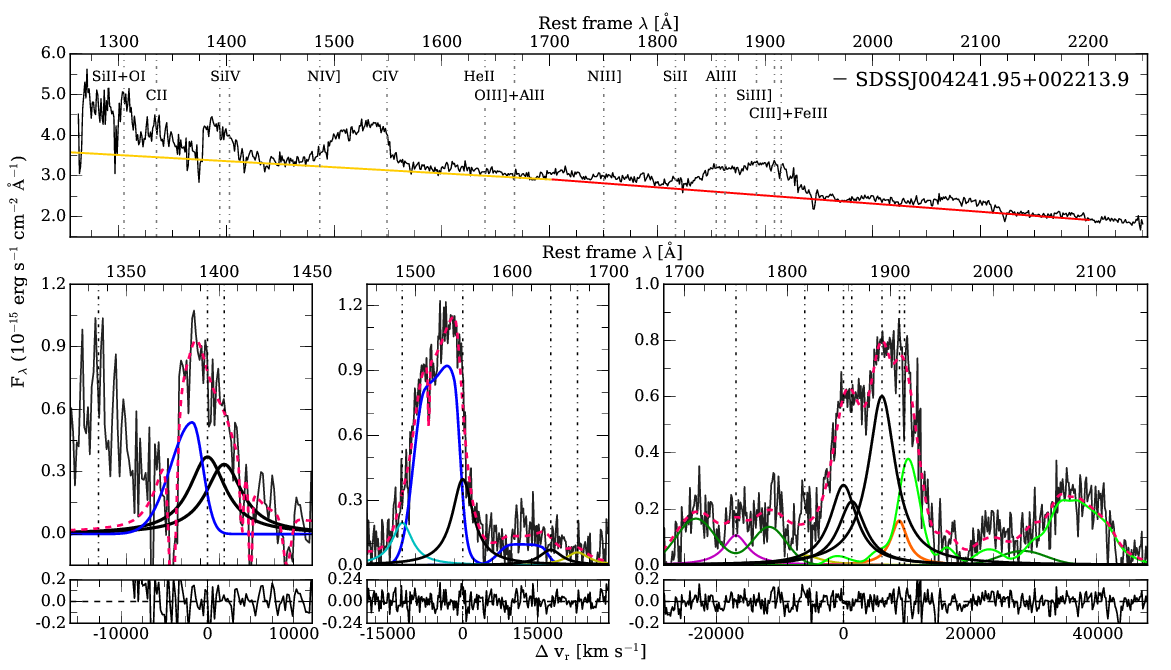}
\end{center}
\caption{(cont.) Same of the previous panel, for \object{SDSSJ004241.95+002213.9}.
\label{fig:specSDSSJ004241.95+002213.9}
}
\end{figure*}

\begin{figure*}
\begin{center}
\includegraphics[width=17cm]{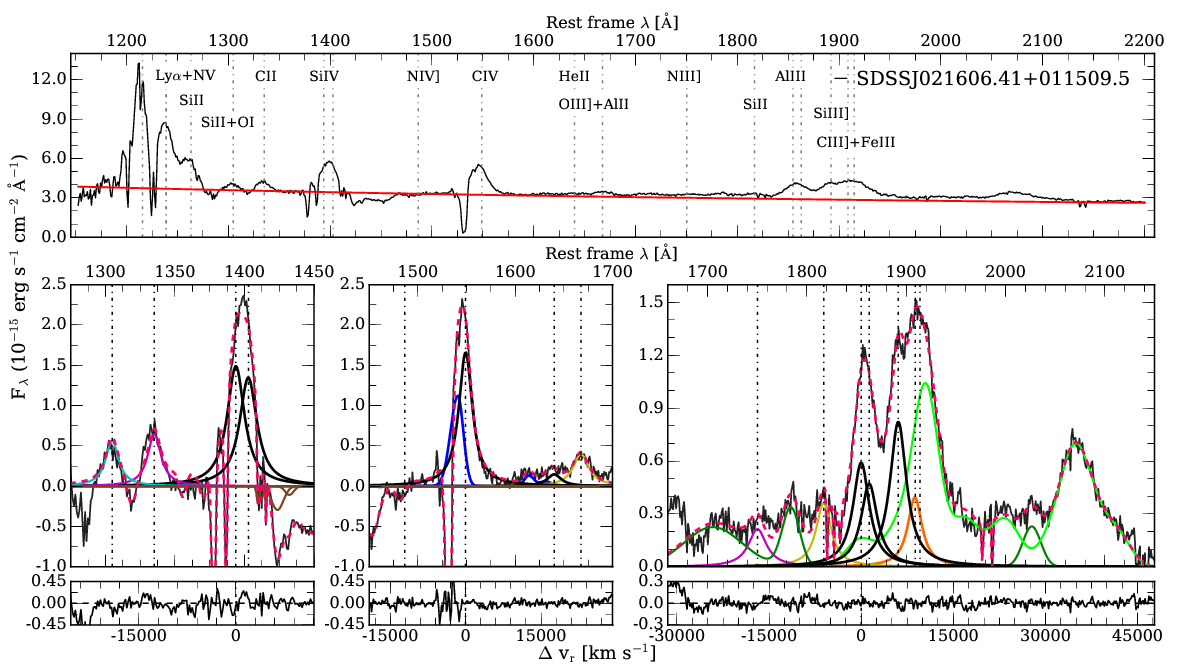}
\end{center}
\caption{(cont.) Same of the previous panel, for \object{SDSSJ021606.41+011509.5}.
\label{fig:specSDSSJ021606.41+011509.5}
}
\end{figure*}

\begin{figure*}
\begin{center}
\includegraphics[width=17.cm]{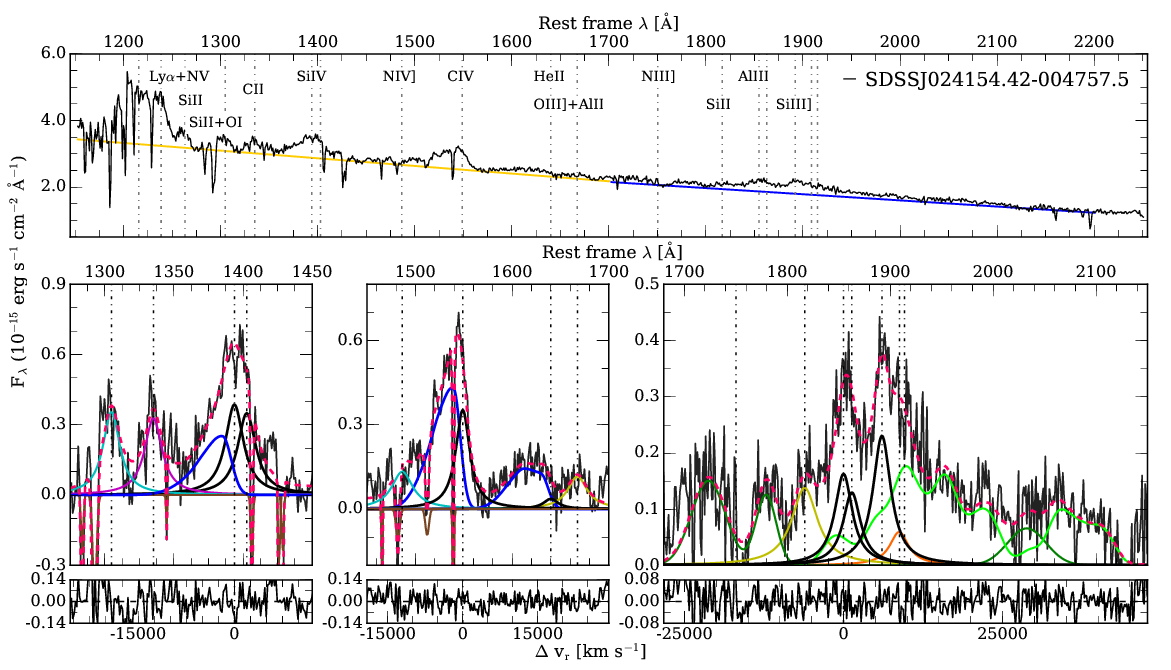}
\end{center}
\caption{(cont.) Same of the previous panel, for \object{SDSSJ024154.42-004757.5}.
\label{fig:specSDSSJ024154.42-004757.5}
}
\end{figure*}


\begin{figure*}
\begin{center}
\includegraphics[width=17.cm]{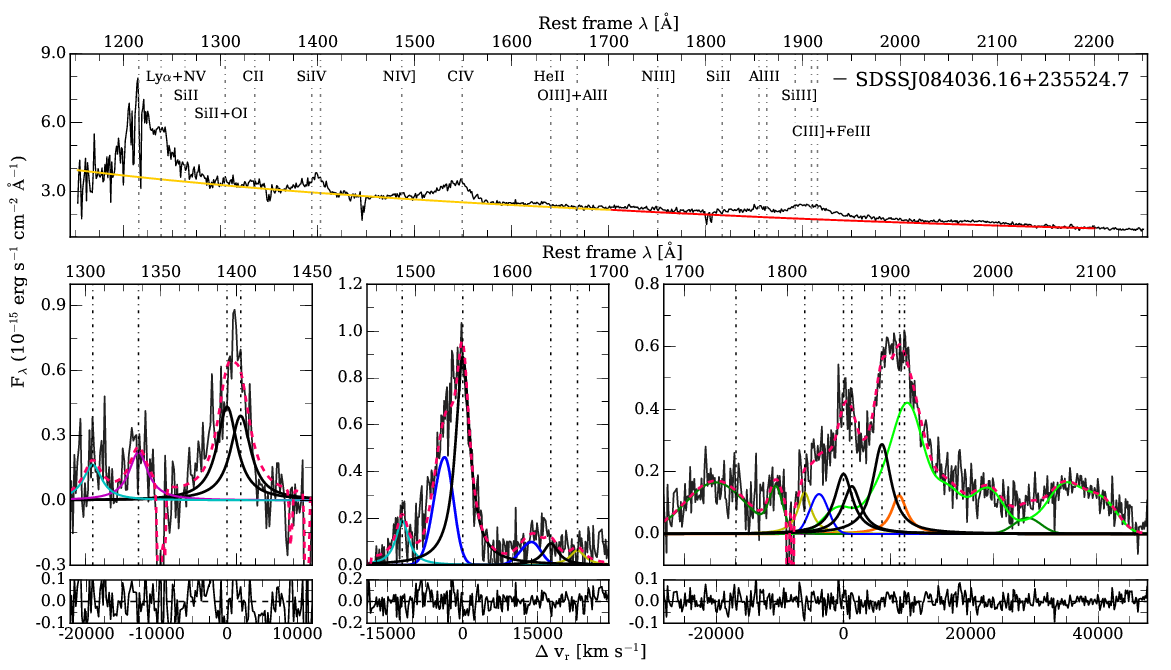}
\end{center}
\caption{(cont.) Same of the previous panel, for \object{SDSSJ084036.16+235524.7}.
\label{fig:specSDSSJ084036.16+235524.7}
}
\end{figure*}

\begin{figure*}
\begin{center}
\includegraphics[width=17.cm]{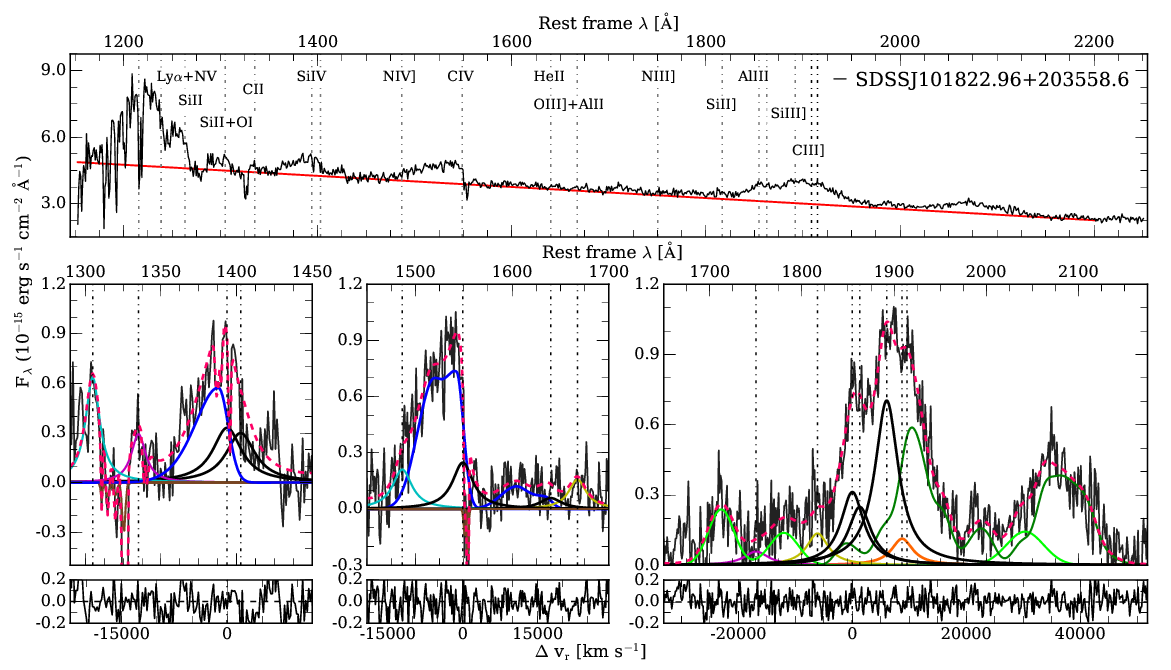}
\end{center}
\caption{(cont.) Same of the previous panel, for \object{SDSSJ101822.96+203558.6}.
\label{fig:specSDSSJ101822.96+203558.6}
}
\end{figure*}

\begin{figure*}
\begin{center}
\includegraphics[width=17.cm]{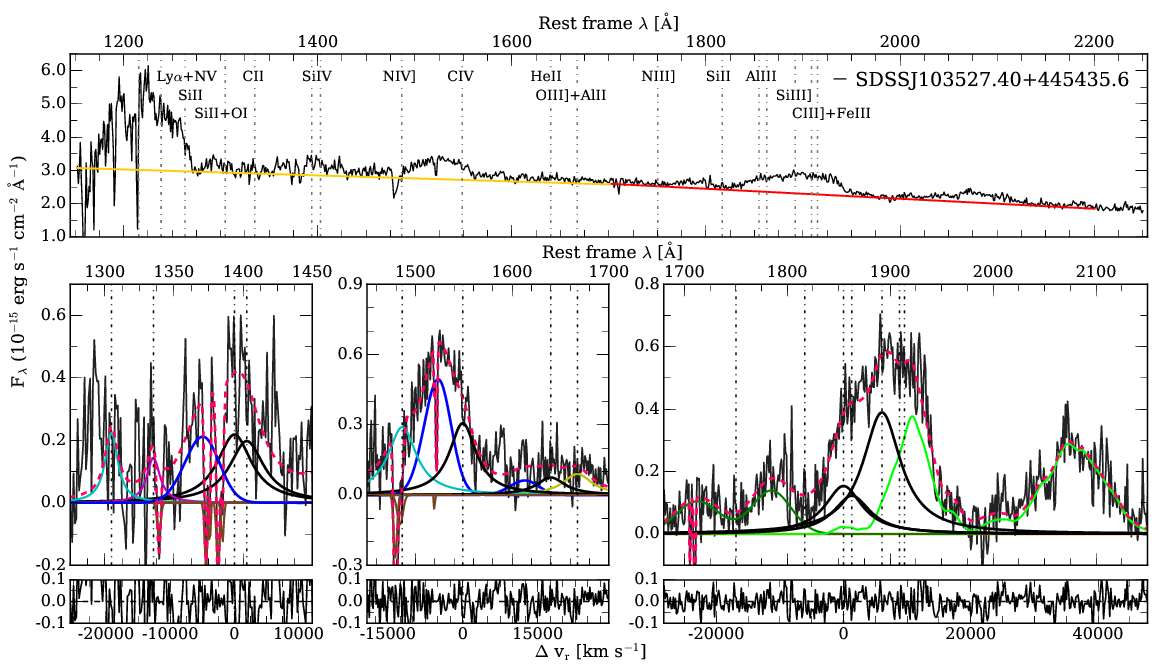}
\end{center}
\caption{(cont.) Same of the previous panel, for \object{SDSSJ103527.40+445435.6}.
\label{fig:specSDSSJ103527.40+445435.6}
}
\end{figure*}

\begin{figure*}
\begin{center}
\includegraphics[width=17.cm]{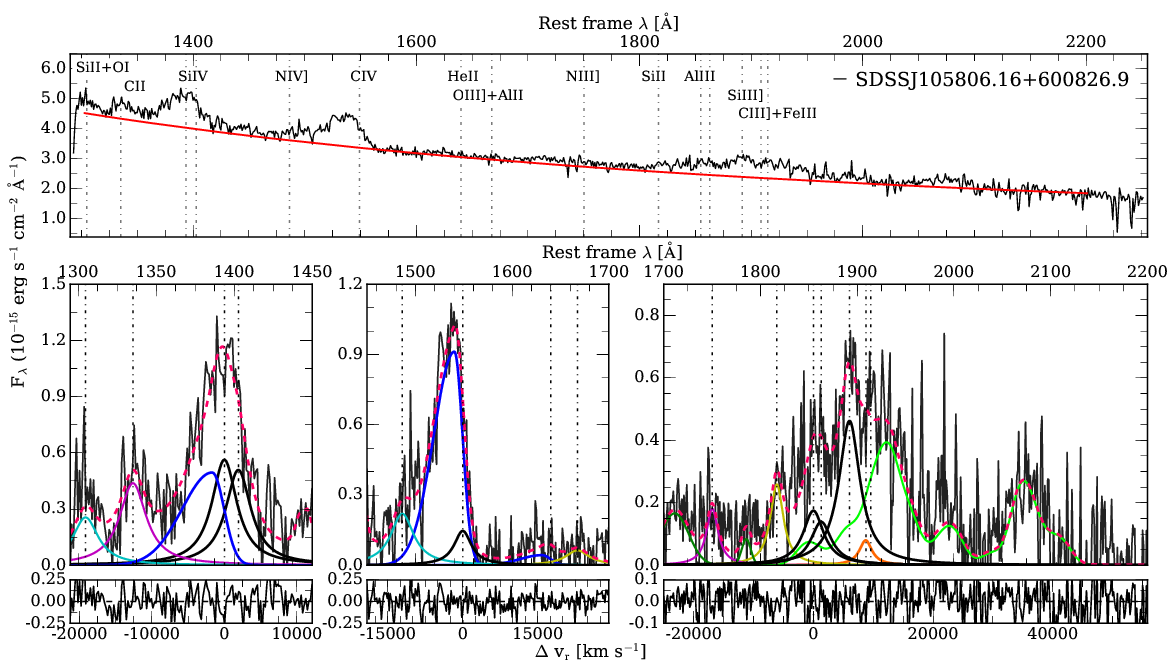}
\end{center}
\caption{(cont.) Same of the previous panel, for \object{SDSSJ105806.16+600826.9}.
\label{fig:specSDSSJ105806.16+600826.9}
}
\end{figure*}

\begin{figure*}
\begin{center}
\includegraphics[width=17.cm]{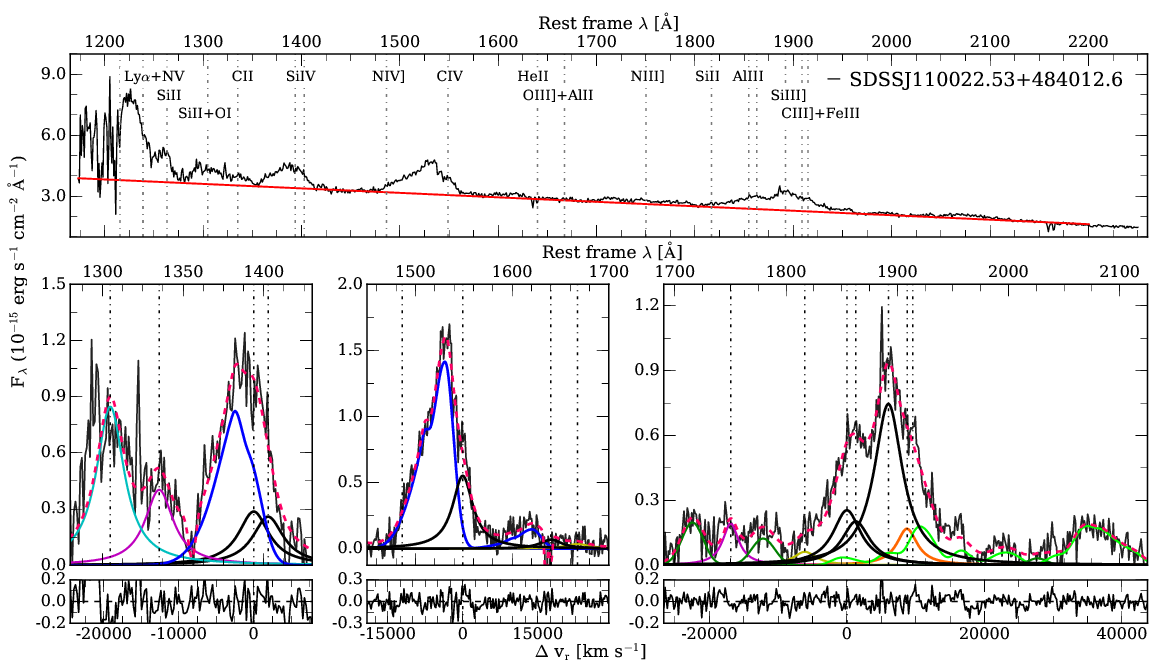}
\end{center}
\caption{(cont.) Same of the previous panel, for \object{SDSSJ110022.53+484012.6}.
\label{fig:specSDSSJ110022.53+484012.6}
}
\end{figure*}

 \begin{figure*}
\begin{center}
\includegraphics[width=17.cm]{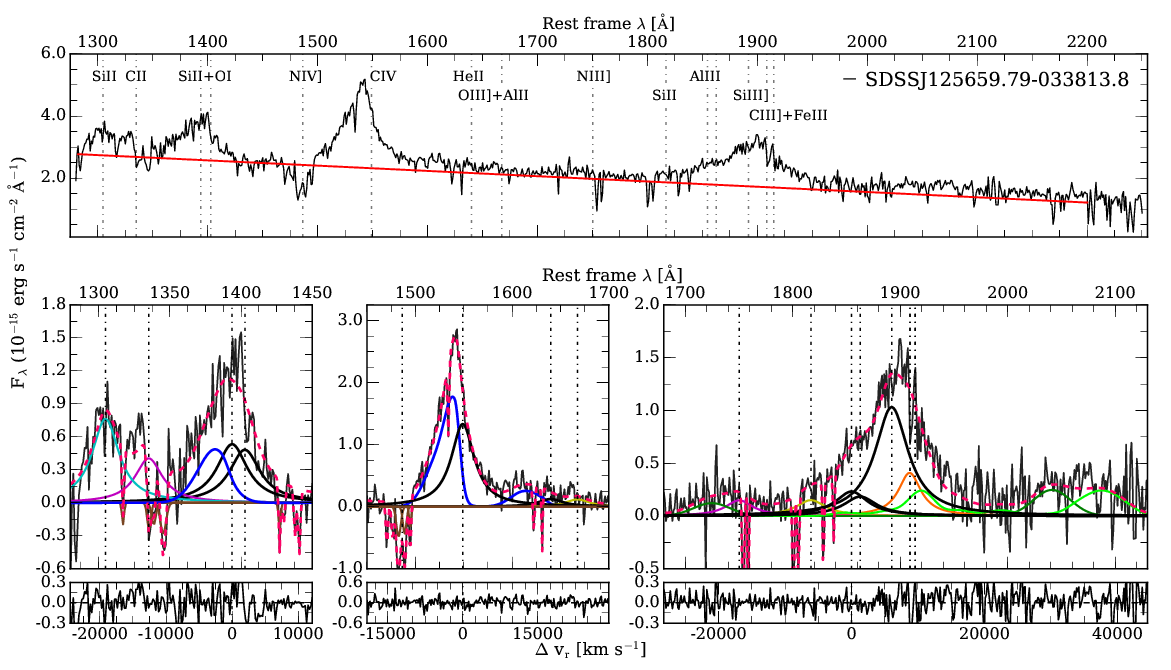}
\end{center}
\caption{(cont.) Same of the previous panel, for \object{SDSSJ125659.79-033813.8}.
\label{fig:specSDSSJ125659.79-033813.8}
}
\end{figure*}

\begin{figure*}
\begin{center}
\includegraphics[width=17.cm]{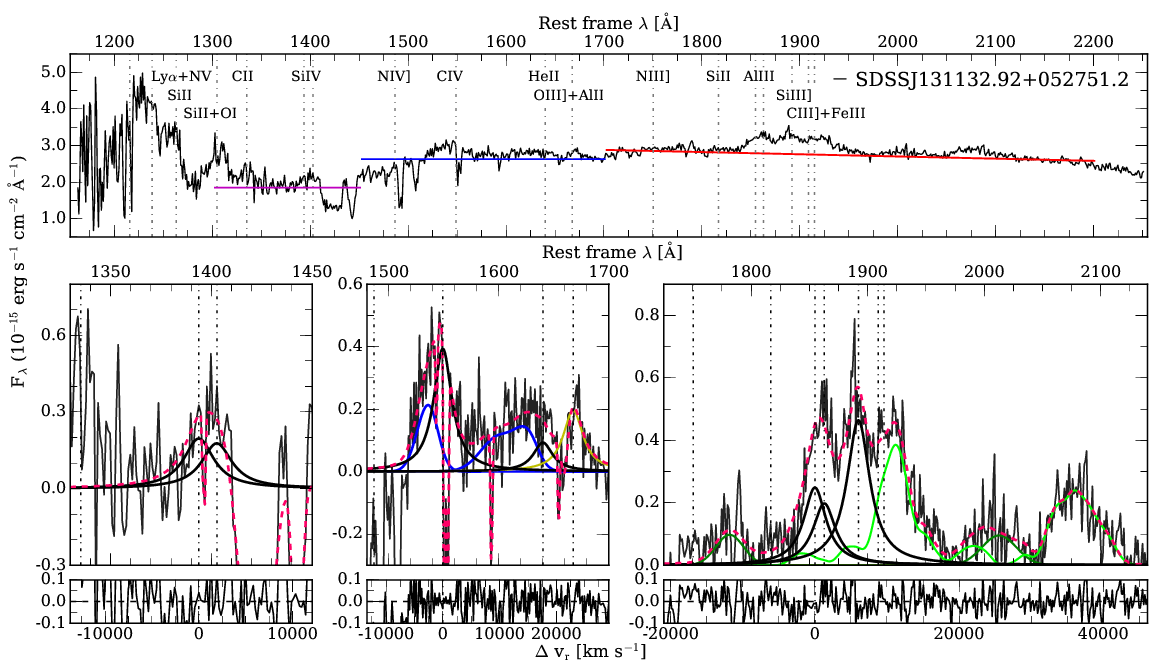}
\end{center}
\caption{(cont.) Same of the previous panel, for \object{SDSSJ131132.92+052751.2}.
\label{fig:specSDSSJ131132.92+052751.2}
}
\end{figure*}

\begin{figure*}
\begin{center}
\includegraphics[width=17.cm]{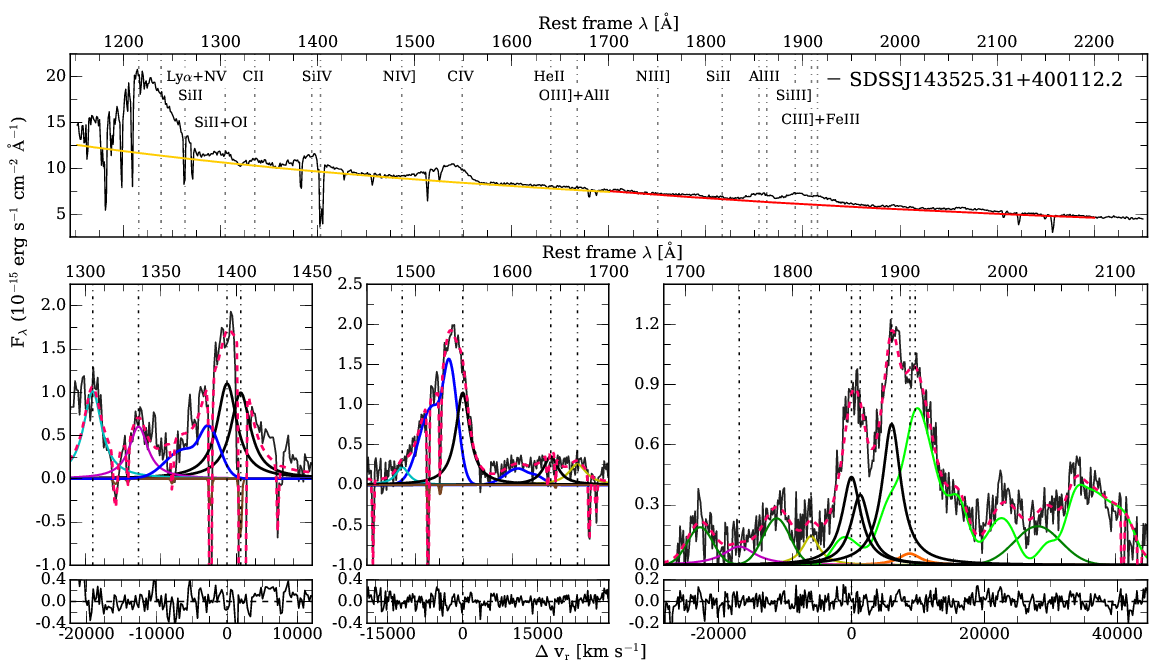}
\end{center}
\caption{(cont.) Same of the previous panel, for \object{SDSSJ143525.31+400112.2}.
\label{fig:specSDSSJ143525.31+400112.}
}
\end{figure*}

\begin{figure*}
\begin{center}
\includegraphics[width=17.cm]{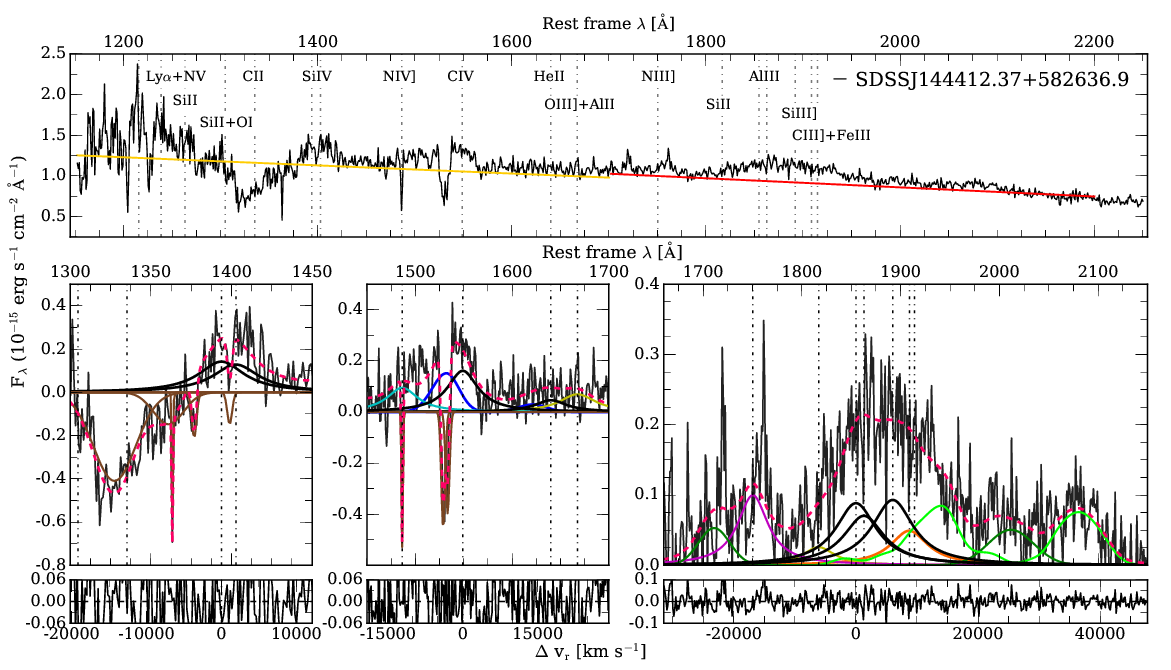}
\end{center}
\caption{(cont.) Same of the previous panel, for \object{SDSSJ144412.37+582636.9}.
\label{fig:specSDSSJ144412.37+582636.9}
}
\end{figure*}

\begin{figure*}
\begin{center}
\includegraphics[width=17.cm]{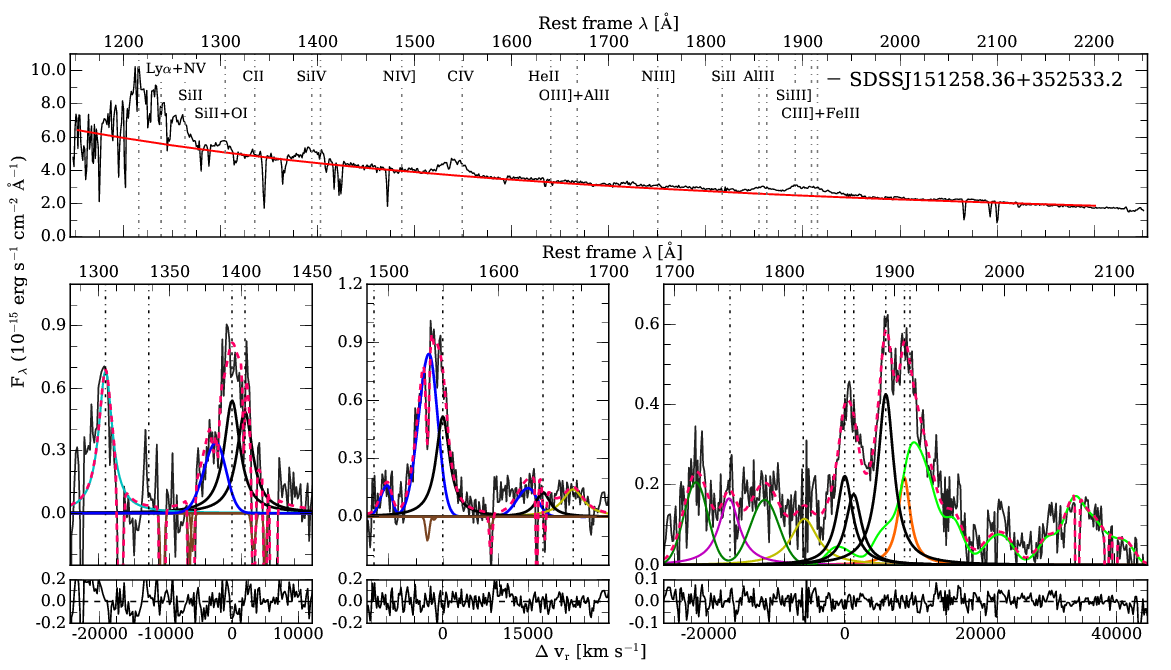}
\end{center}
\caption{(cont.) Same of the previous panel, for \object{SDSSJ151258.36+352533.2}.
\label{fig:specSDSSJ151258.36+352533.2}
}
\end{figure*}

\begin{figure*}
\begin{center}
\includegraphics[width=17.cm]{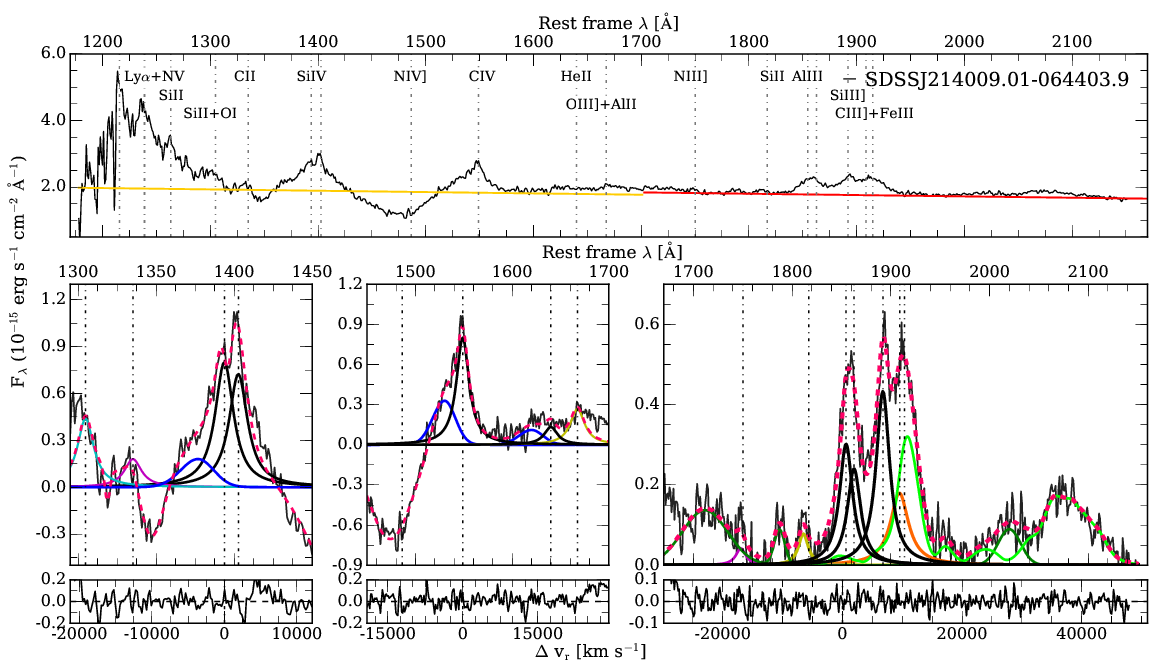}
\end{center}
\caption{(cont.) Same of the previous panel, for \object{SDSSJ214009.01-064403.9}.
\label{fig:specSDSSJ214009.01-064403.9}
}
\end{figure*}

\begin{figure*}
\begin{center}
\includegraphics[width=17.cm]{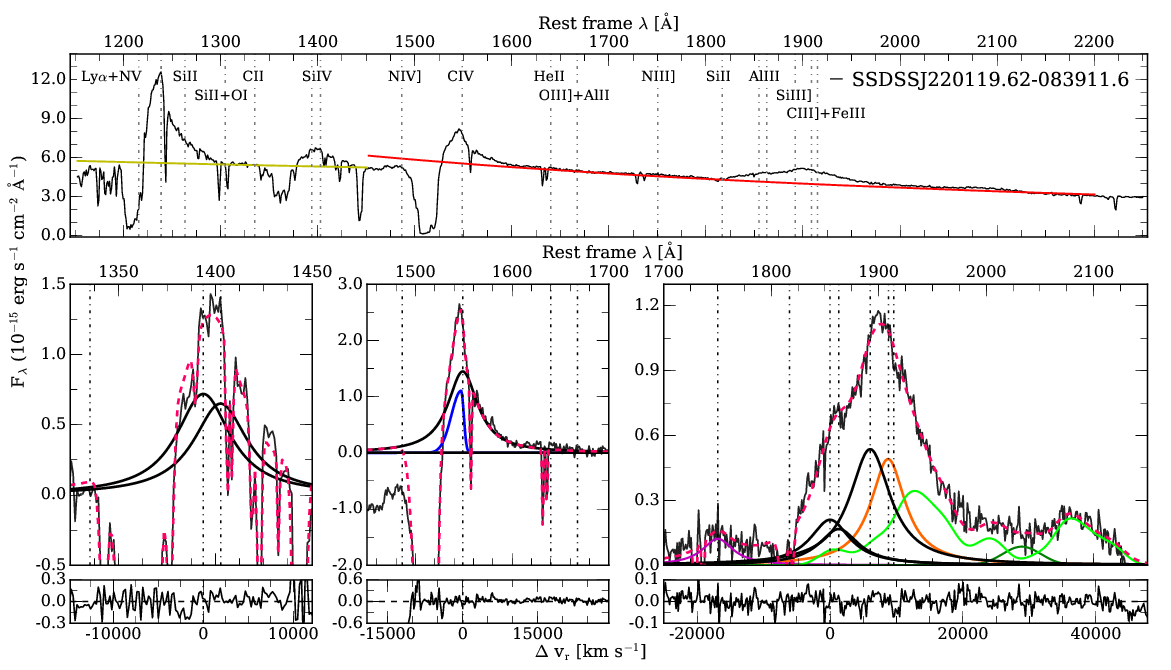}
\end{center}
\caption{(cont.) Same of the previous panel, for \object{SDSSJ220119.62-083911.6}.
\label{fig:specSDSSJ220119.62-083911.6}
}
\end{figure*}

\begin{figure*}
\begin{center}
\includegraphics[width=17.cm]{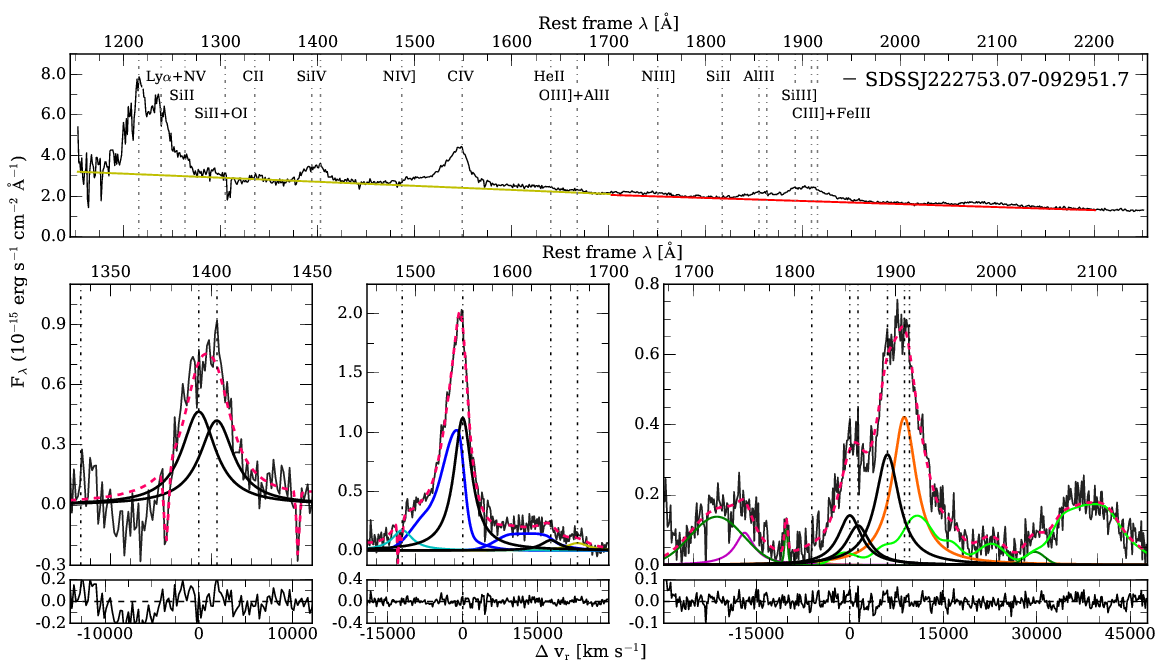}
\end{center}
\caption{(cont.) Same of the previous panel, for \object{SDSSJ222753.07-092951.7}.
\label{fig:specSDSSJ222753.07-092951.7}
}
\end{figure*}

\vfill\eject\clearpage

\begin{figure*}
\begin{center}
\includegraphics[width=17.cm]{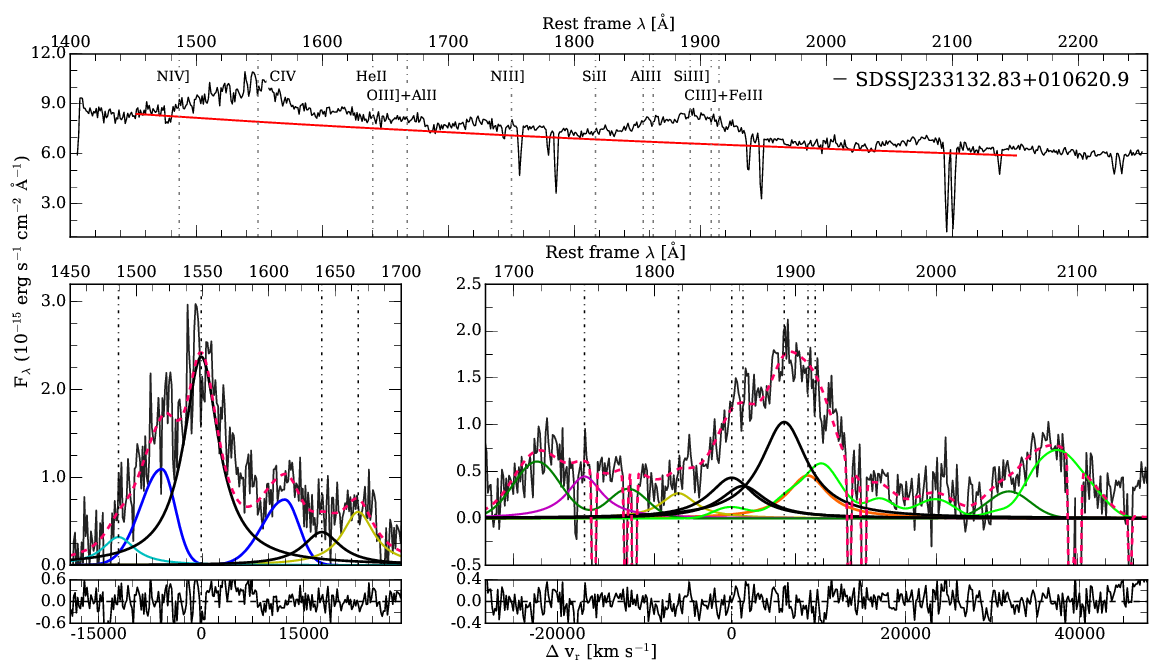}
\end{center}
\caption{(cont.) Same of the previous panel, for \object{SDSSJ233132.83+010620.9}.
\label{fig:specSDSSJ233132.83+010620.9}
}
\end{figure*}

\begin{figure*}[!ht]
\begin{center}
\includegraphics[width=17.cm]{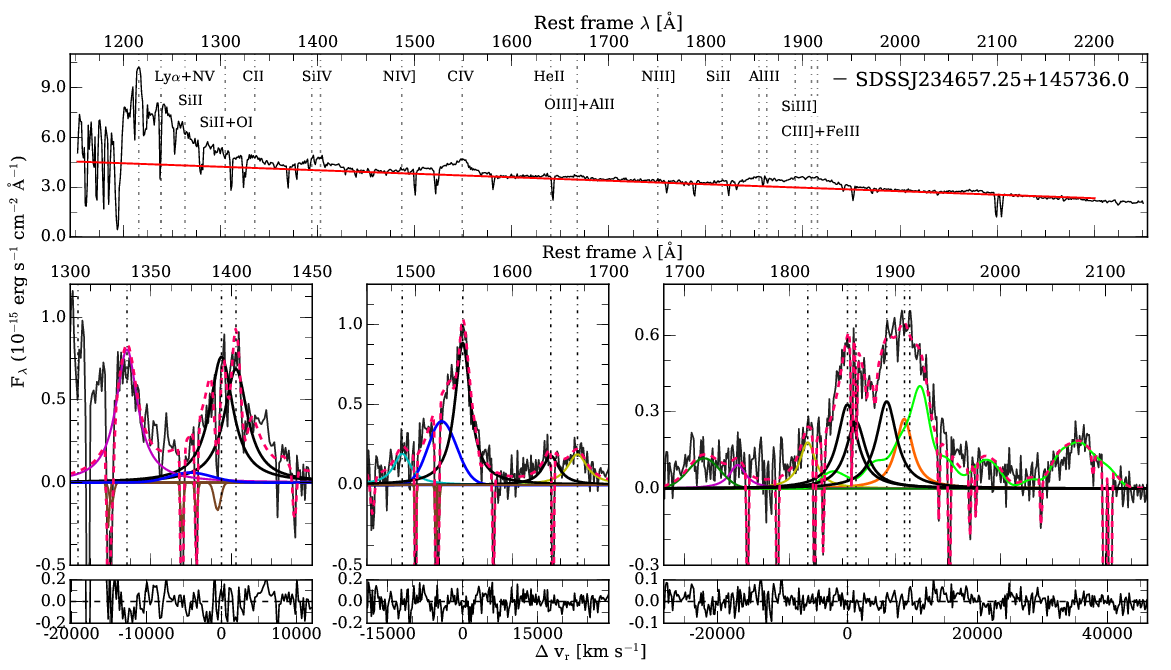}
\end{center}
\caption{(cont.) Same of the previous panel, for \object{SDSSJ234657.25+145736.0}.
\label{fig:specSDSSJ234657.25+145736.0}
}
\end{figure*}

\vfill\eject\clearpage


\section{{Error estimates} }
\label{sec:errors}

{Emission lines of the xA spectra are blended in almost all cases. For example, in the 1900\AA\ blend the emission of \ciii\ and \feiii\ $\lambda$1914 cannot be deblended; or in the case of \civonly\ the separation between broad and blue components is not obvious.We built Monte Carlo (MC) simulations to determine the error estimates, in order to consider the effects of blending. The MC method allows us to consider all emission components and include a variation of the flux and FWHM simultaneously. The method is based on minimizing the $\chi^2$ \citep{barlow89}. A variation in parameters of the line properties causes changes in the $\chi^2$. Then, the parameters at 1$\sigma$ confidence level are those that satisfy the constraint: $\chi^2_{\mathrm{1\sigma}} \le \chi^2_{\mathrm{min}}$+1 \citep{andrae10}.}

\begin{figure*}[ht!]
	\begin{center}
		\includegraphics[width=6cm]{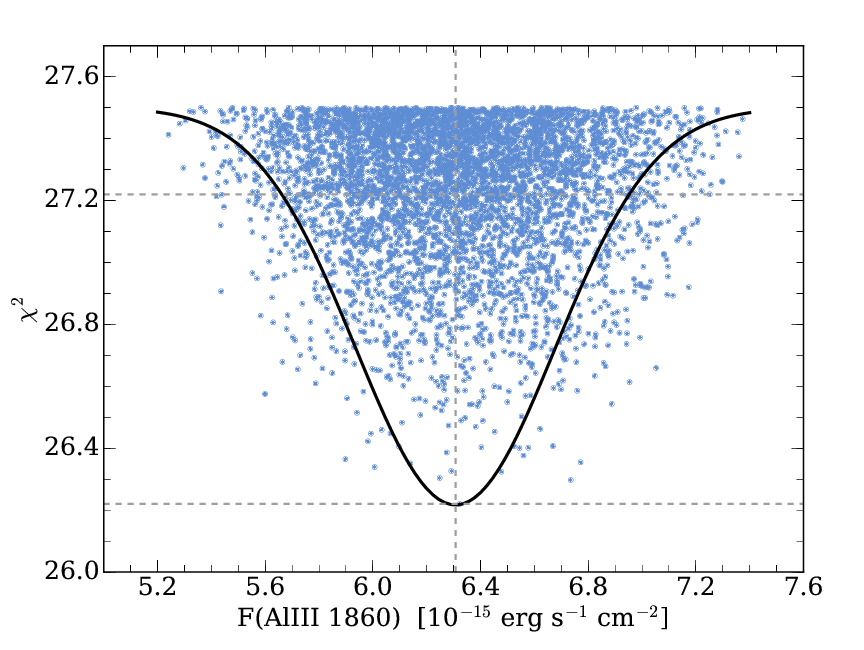}
		\includegraphics[width=6cm]{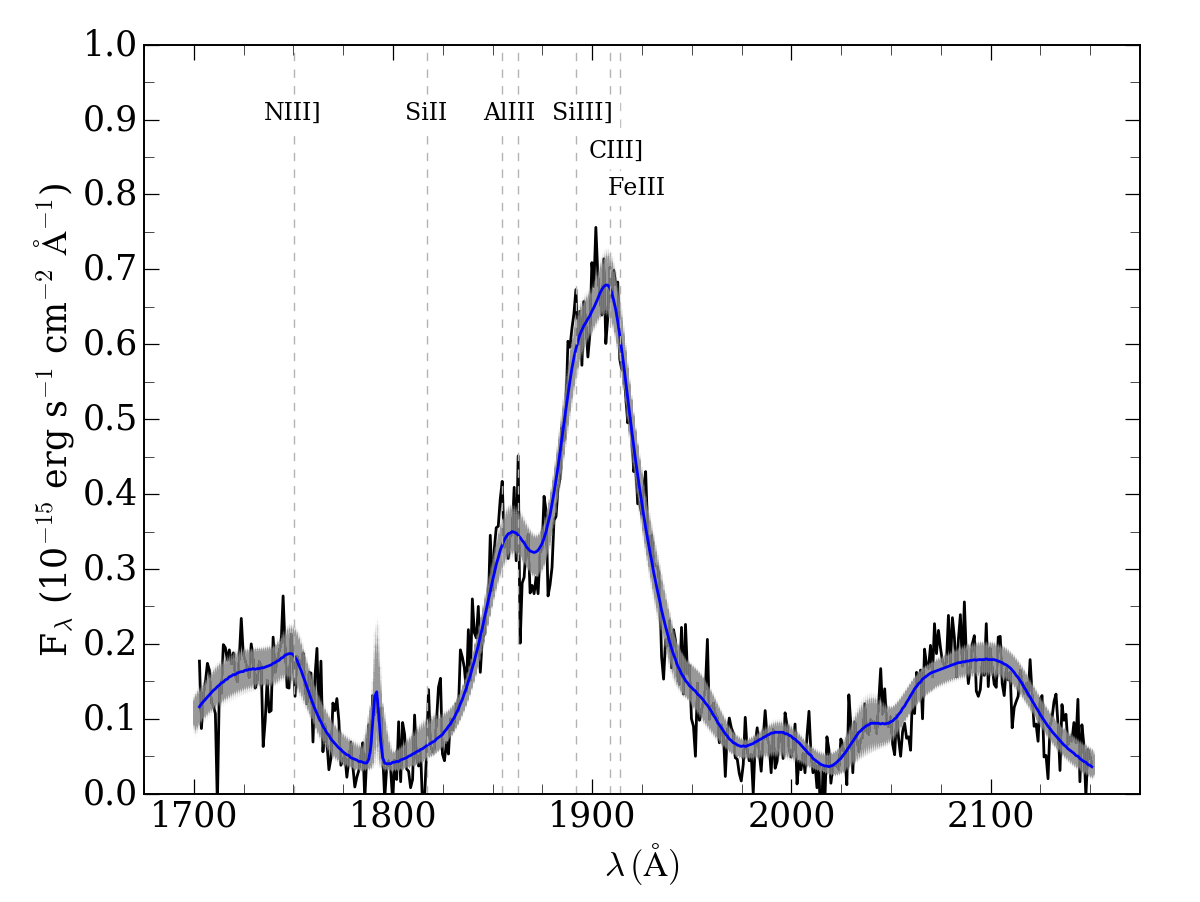}
		\includegraphics[width=6cm]{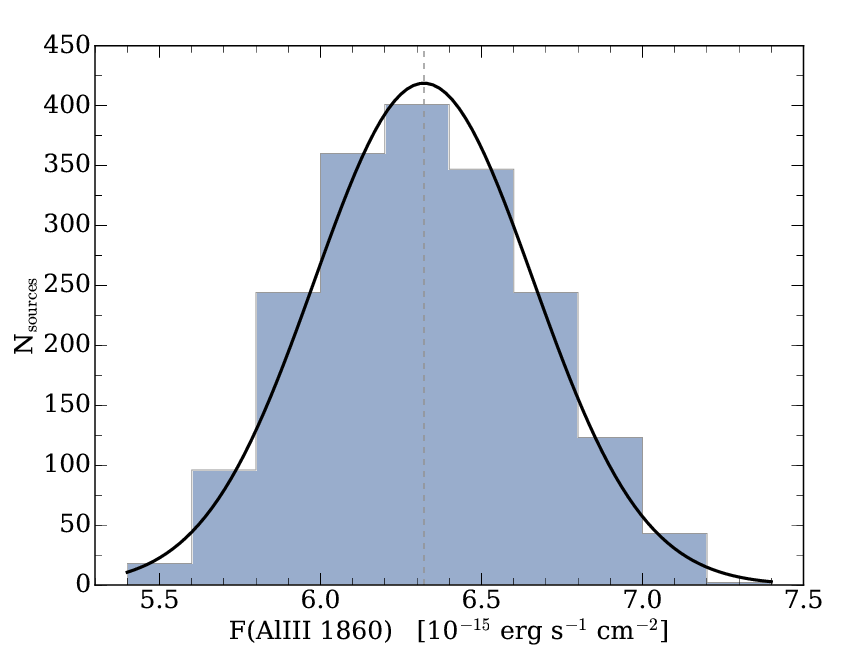} \\
		\includegraphics[width=6cm]{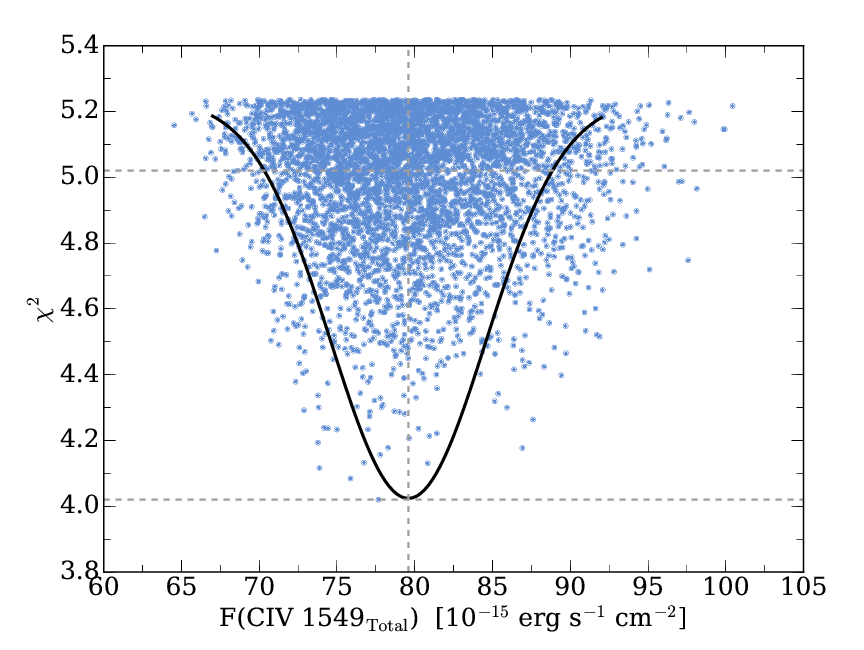}
		\includegraphics[width=6cm]{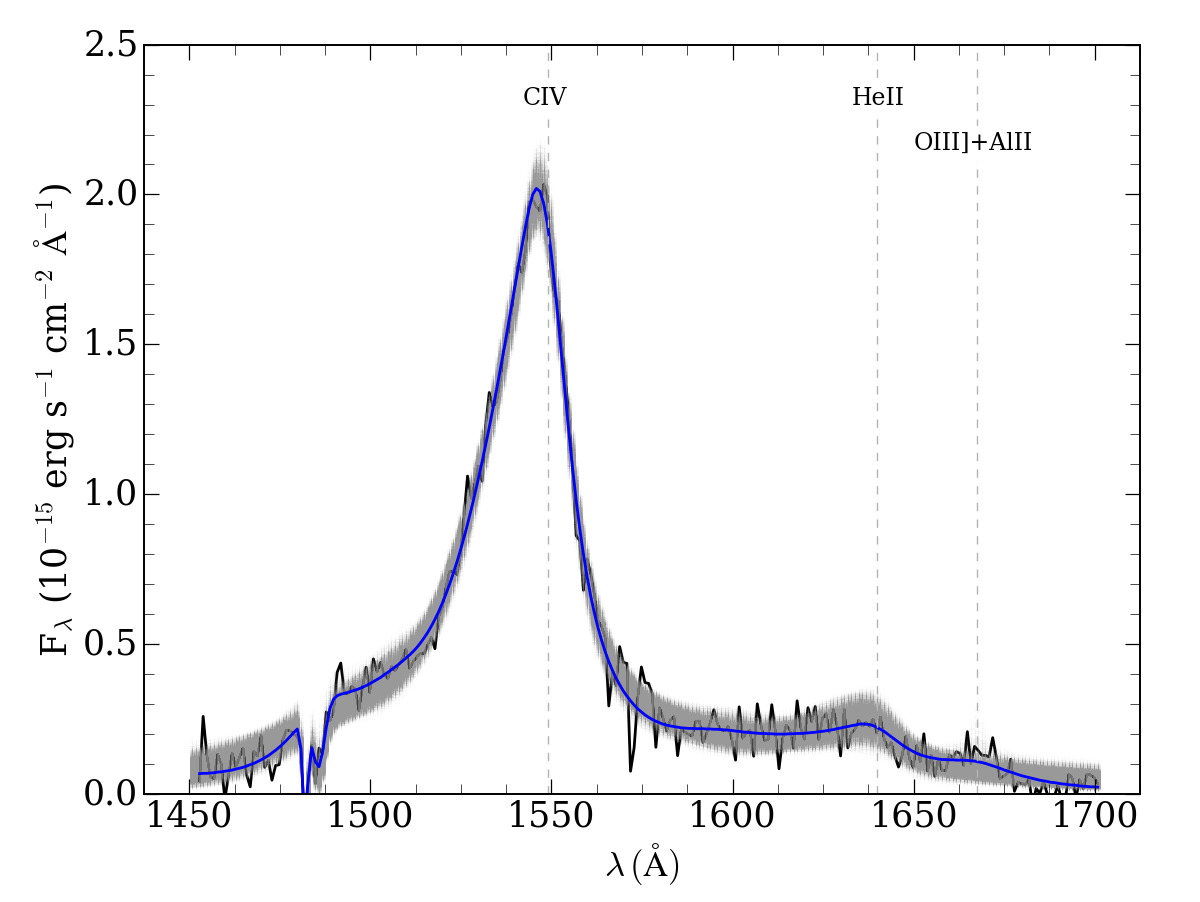}
		\includegraphics[width=6cm]{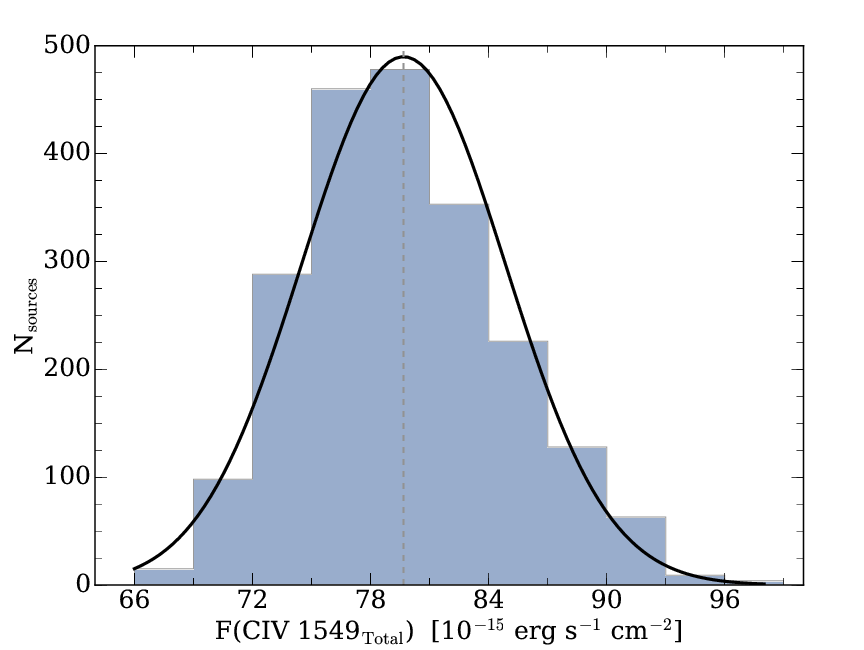}
	\end{center}
\caption{Left panels: distribution of $\chi^2$ as a function of the flux of \aliiionly\ (top) and total \civonly\ profile (bottom). Black continuous lines mark the normal distribution associated with simulations. Gray horizontal lines indicate the value at $\chi^2\mathrm{_{min}}$ and at $\chi^2_{\mathrm{min}}$+1. The gray vertical line indicates the mean of the normal distribution. Middle panels: spectra of the quasar SDSSJ222753.07-092951.7 in the 1900\AA\,blend (top) and total \civ\,(bottom) spectral ranges. Gray shadow regions correspond to the models at the $\chi^2_{\mathrm{min}}$+1 confidence level obtained from the Monte Carlo simulations. The blue line corresponds to the {\tt SPECFIT} result. Right panels show the distributions of the fluxes of \aliiionly\,(top) and total \civonly\,(bottom) for all the models under the 1$\sigma$ confidence level plotted in the left panels. Vertical lines has the same meaning as in left panels.
\label{fig:errors} }
\end{figure*}

{The MC method is applied to the same spectral regions identified for the multicomponent fits with {\tt SPECFIT} (Section \S\ref{sec:fitting}). All line components considered in modeling the observed quasar emission spectrum were taken into account. For example, in the 1900\AA\ blend region variations in flux and FWHM of \aliiionly, \siiiionly, \ciiionly, \siii, \niii, \feii\ and \feiii\ are considered for each simulation. Parameters vary around the values given by the {\tt SPECFIT} model without any constraint and the simulations are therefore totally independent of the {\tt SPECFIT} model.}

{We ran 5000 simulations for each quasar in order to obtain good sampling. Left panels of Figure \ref{fig:errors} show the distributions of $\chi^2$ for the flux of \aliiionly\ (top) and for the total \civonly\ profile (bottom), for the quasar SDSSJ222753.07-092951.7. Usually $\sim$ 1000 -- 3000 simulations satisfy the criterion $\chi^2_{\mathrm{1\sigma}} \le \chi^2_{\mathrm{min}}$+1. The total width of the distribution at  $\chi^2_{\mathrm{min}}$+1 is the 1$\sigma$ error confidence level we consider for each parameter. The central panels of Figure \ref{fig:errors} show the 1900\AA\ blend (top) and \civonly\ (bottom) spectra, and the quasar emission associated with the 1$\sigma$ confidence level derived from the MC simulations (gray shadow regions). }  

{Strongest emission lines like \aliiionly\ or the total emission of \civonly\ show Gaussian distributions for the flux and FWHM MC replications (See right panels of Figure \ref{fig:errors}). However, for weak lines like \ciii, \niii\ or \heii$_\mathrm{BC}$, the distribution has an asymmetric or flat behavior. This is because faint lines are strongly affected by the presence of the stronger ones (for example \siii\ affected by \aliiionly\,). In these cases, we consider the flux found by {\tt SPECFIT} as an upper limit. We report the FWHM values followed by a double colon to signify a high uncertain value (Tables \ref{tab:blend}, \ref{tab:c4}, \ref{tab:he2} and \ref{tab:si4}).}

\clearpage

\section{{\feii\ and \feiii}}
\label{sec:feappen}

\subsection{{Identification of \feii\ and \feiii\ transitions }}
\label{sec:feident}

{In next lines are described the most important \feii\ and \feiii\ transitions in the spectral range 1700 -- 2100\,\AA. Table \ref{tab:feid} reports the approximate central wavelength of the \feii\ and \feiii\ features, the multiplet identification, and the sources where they are prominent.}

\paragraph{{\feii}}
\begin{description}
\item {[$\lambda$1715\,\AA] -- This feature has not been identified previously as \feii\ emission. \citet{grahametal96} suggest Al{\sc ii} associated with the $3p ^{3}P^{o}  -   3d ^{3}D$\ transitions. However, the wavelength consistency is poor, the Al{\sc ii} transition is fairly high level, and its intensity is predicted negligible even for very low ionization parameter ($\sim 10^{-3}$). We therefore ascribe the feature to \feii\ UV multiplet \#38 (associated with the transition between terms  $a ^{4}F  -  ^{4}D^{o}$), and specifically to the transition $^{4}D^{o}_{\frac{9}{2}} \rightarrow ^{4}F^{o}_{\frac{7}{2}} $\ that has the higher oscillator strength among all \feii\ lines listed by \citet[][$\log g f \approx -0.395$]{kuruczbell95}. }

\item {[$\lambda$1785\,\AA] -- The \ UV multiplet \#191 is strong in several quasars, most notably I Zw 1 \citep{marzianietal96,laoretal97a,vestergaardwilkes01}. The emission is believed to be enhanced by a \lya\ fluorescence mechanism \citep{baldwinetal04} also operating in symbiotic stars and in stars with chromospheric activity \citep{johanssonhansen88,johanssonetal95}. }

\item {[$\lambda$2020\,\AA] -- This feature is never strong ($\lesssim 2$ \AA), but is apparently detected in a few objects (Tab. \ref{tab:feid}). The main feature could be associated with the line at 2020.739 \AA\ and other two lines of \feii\ multiplet UV \#83.}   
\end{description}

\paragraph{{\feiii}}

\begin{description}
\item {[$\lambda$1914\,\AA] -- The line of the \feiii\ UV multiplet \#34\  is included in the template. We add a  component to take into account the possibility of extraordinary enhancement due to \lya\ fluorescence, as outlined in \citet{marzianietal10}.}

\item {[$\lambda$2005\,\AA\ \& $\lambda$2045\,\AA] -- These features have been identified as due to \feiii\ UV \#55 \citep{grahametal96,vestergaardwilkes01}.}

\item {[$\lambda$2080\,\AA] -- This feature is perhaps the most prominent one among the \feiii\ features. Its identification  \feiii\ UV \#48 is supported by the consistency in wavelength of the three multiplet components. The peculiarity of the GTC-xA sources is that the feature is not reproduced well by the template: an additional component is needed. The analysis  of the 2080 \AA\ feature is reported in Section \S \ref{sec:fluo}, and its origin is discussed in Appendix \ref{sec:origin} below.}

\item {[$\lambda$2093\,\AA] -- We identify this feature as due to \feiii\ UV \#77\ \citep{vestergaardwilkes01}. A previous identification as Fe{\sc i} seems less likely, considering that Fe{\sc i} features are predicted  to be exceedingly weak \citep{sigutetal04} even at the lowest ionization degrees. }

\item {[$\lambda$2115\,\AA] -- \citet{vestergaardwilkes01} associates this feature with \feiii\ UV \#58.}

\end{description}

\subsection{ {On the origin of \feiii\ emission}}
\label{sec:origin}

{The strength of \feiii\ emission is surprising, given that the overall appearance of the spectrum is suggestive of extremely low ionization. The \feiii\ features have also been observed  in Pop. A sources or associated with extreme Pop. A  \citep[e.g., ][]{baldwinetal96,grahametal96,richardsetal11}. The \feiii\ lines are mainly emitted in a region at the boundary between the fully/partially ionized zone. At ionization parameter $\log U \gtrsim -2$, the region of \feiii\ line formation is usually a region where the opacity in the Lyman continuum grows to values $\tau_{912} \gtrsim 1$\ (i.e., a small fraction of the geometric depth of the emitting gas of slab). Deeper into the slab from its illuminated face, the dominant ionization stage is Fe$^{+3}$. A  significant fraction of \feiii\ is present at the illuminated face of the cloud within the H{\sc ii} zone, but only if the ionization parameter is very low  ($\log U \lesssim -3$, Fig. 3 of \citealt{sigutetal04}).  Fe$^{+2}$ is $\lesssim 2$\% of Fe$^{+1}$ in the partially-ionized zone (PIZ): Fe$^{+1}$ still remains the dominant ionization stage of iron in the PIZ, as required by the strong optical \feii\ emission required for these sources. In the spectral range between 1800 and 2100\,\AA, however, the dominant contribution may be due to \feiii, as predicted by the models of \citet{sigutetal04}, which assumes $\log U \sim -3$.}

{The role of \lya\ fluorescence in explaining the UV and IR \feii\ emission has been known since the late 1980s from photoionization simulations \citep[e.g., ][]{sigutpradhan98,verneretal04,sigutetal04} as well as from observational evidence \citep[e.g.,][and references therein]{penston87,marinelloetal16,clowesetal16}. As mentioned, a specific feature that is believed to be enhanced by \lya\ fluorescence is the \feii$\mathrm{_{UV}}$ \#191\ multiplet at 1785 \AA. Also the \feiii\ line at $\lambda$1914, ascribed to  the transition of the \feii$\mathrm{_{UV}}$ \#34\ multiplet z$^{7}P^{o}_{3} \rightarrow$ a$^{7}S_{3}$, is expected to be enhanced by \lya\ fluorescence. Although this is not a resonant line (the lower level is $\sim$3 eV above ground), the line appears  often stronger than in the \citet{vestergaardwilkes01} template, presumably because the upper level is populated by \lya\ fluorescence \citep{marzianietal10}.  Figure \ref{fig:fe3} suggest a close association between the 2080 feature and the \feii\ and \feiii\ features enhanced by \lya\ fluorescence. }

\begin{figure}[!ht]
\begin{center}
\includegraphics[width=8.5cm]{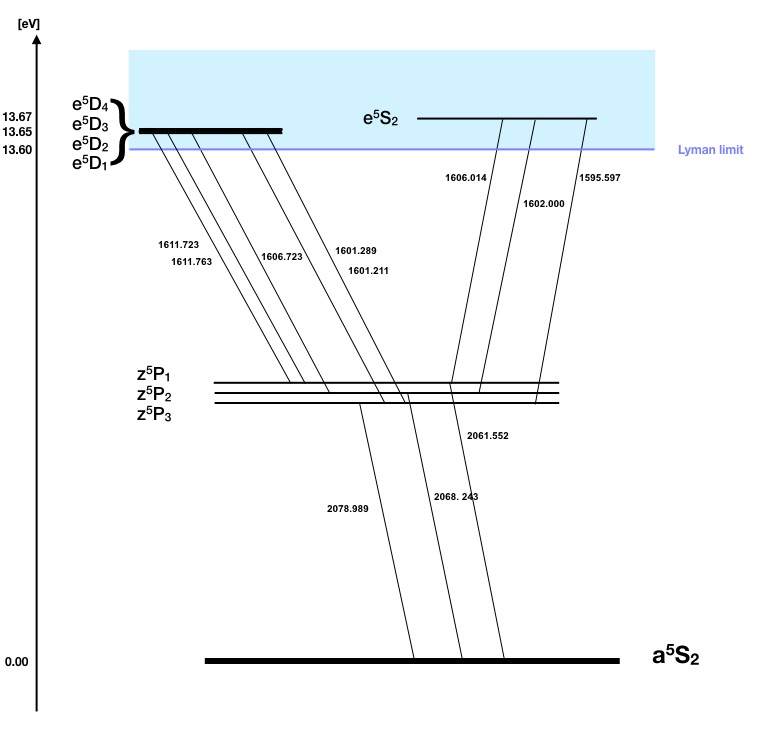}
\end{center}
\caption{Partial Grotrian diagram showing the energy levels and the spectroscopic terms associated with the emission of \feiii\ multiplet \#48, \#118 (top left) and \#119 (top right). Energy level separation is not drawn to scale for clarity. \label{fig:grotrian}}
\end{figure}

{Generally speaking, we expect that \lya\ fluorescence may be also relevant in the explanation of the overall \feiii\ spectrum, considering the large number of radiative transitions possible for the electronic configuration of the Fe$^{++}$. The \feiii\ high ionic fraction in the H{\sc ii} zone and at the transition zone between H{\sc ii} and PIZ are expected to be conditions that make the fluorescent absorption of Lyman continuum photons by \feiii\ ions especially efficient.  In the specific case of the 2080 feature, we suggest that a mechanism of enhancement is due to fluorescence with the Lyman continuum ionization edge. The \feiii\ UV \#48 multiplet is radiatively linked to the multiplets \#118 and \#119\ between the terms z$^{5}P^{\rm o}$  e$^{5}D$ and e$^{5}S$, respectively. The energy difference between the lower level of the lower a$^{5} S$ term of multiplet \#48 and the upper terms of \#118 and 119, is 13.65 and 13.67 eV respectively, slightly above the ionization potential of hydrogen (13.5984 eV). Figure \ref{fig:grotrian} schematically shows the energy levels associated with the spectroscopic terms. Even if the energy is not strictly coincident, the Lyman continuum emission (which is expected to decrease with $\nu^{-3}$) may act as a broadened pseudo-line.}
	
{If this mechanism is indeed operating, we expect \feiii\ emission from multiplets UV $\#$118 and $\#$119 in the range 1595 -- 1610\,\AA.  A faint hump present on the composite spectrum in the predicted range could be due to some \feiii\ emission as well as to the blueshifted component of \heiiuv\ blending with the red side of \civ. Considering the weakness of the features, and the severe blending with \heiiuv\ and \civ, the prediction of \feiii\ emission around 1600\,\AA\ should be verified on high S/N spectra of sources with narrower lines.}

\begin{table}										
\begin{center}	
\tabcolsep=4pt				
\caption{Identification of \feii\ and \feiii\ features \label{tab:feid}   }	
\scriptsize	
\begin{tabular}{c c c c}\\ 					
\hline\hline\noalign{\vskip 0.05cm}     
\multicolumn{1}{c}{Feature} &  \multirow{2}{*}{Identification} & \multirow{2}{*}{Sources where is detected} \\

$[$\AA$ ]$  & & & \\[0.05cm]
\hline
\\[-0.20cm]
$\lambda$1715	&	\feii\ UV \#38	&	J000807, J021606, J084036, J214009, J222753	\\
$\lambda$1785	&	\feii\ UV \#191&	J125659, J101822, J144412	\\
$\lambda$2005	&	\feiii\ UV \#55	&	J233132	\\
$\lambda$2020	&	\feii\ UV  \#83	&	J021606, J233132, J214009	\\
$\lambda$2045	&	\feiii\ UV \#60	&	J222753, J110022, J151258	\\
$\lambda$2080	&	\feiii\ UV \#48	&	J000807, J004241, J021606, J101822, J103527, \\
	& 				&  J125659, J214009, J222753	\\
$\lambda$2093	&	\feiii\ UV \#77	&	J105806	\\
$\lambda$2115	&	\feiii\ UV \#58	&	J220119, J222753, J110022					\\
\hline					
\end{tabular}					
\end{center}					
\end{table}

\end{document}